\documentclass[11pt,a4paper]{article}

\usepackage{hep-title}
\usepackage{epsfig}
\usepackage[usenames,dvipsnames]{xcolor}

\usepackage{amssymb}
\usepackage{amsmath}
\usepackage[utf8]{inputenc}
\usepackage[T1]{fontenc}
\usepackage{graphicx}
\usepackage{float}
\usepackage{rotating}
\usepackage{bigints}
\usepackage{bbold}
\usepackage{enumerate}
\usepackage{tablefootnote}
\usepackage{eucal}
\usepackage{setspace}
\usepackage{lineno}
\usepackage{xspace}
\usepackage{authblk}
\usepackage[comma,square,numbers,sort&compress]{natbib}

\usepackage[hidelinks]{hyperref}
\usepackage{cleveref}
\usepackage{url}

\newcommand{\dndy}{\ensuremath{\ud N/\ud y}\xspace}

\newcommand{\hyt}{$^3_\Lambda \text{H}$}

\newcommand{\pdt}{$N_\text{t}N_\text{p}/N_\text{d}^2$}

\newcommand{\be}{\begin{equation}}
\newcommand{\ee}{\end{equation}}
\newcommand{\ba}{\begin{eqnarray}}
\newcommand{\ea}{\end{eqnarray}}
\newcommand{\ban}{\begin{eqnarray*}}
\newcommand{\ean}{\end{eqnarray*}}

\newcommand{\ud}{\mathrm{d}}
\newcommand{\mub}{\ensuremath{\mu_{B}}\xspace}

\title{Light nucleus and hypernucleus production at RHIC and LHC energies: a comparative study of phenomenological models}

\author[1]{J.~Aichelin}
\author[2]{A.~Andronic}
\author[3]{F.~Bellini}
\author[4]{M.~Bleicher}
\author[5]{K.~Blum}
\author[6]{E.~Bratkovskaya}
\author[6,7]{P.~Braun-Munzinger}
\author[8]{A.~Caliv\`a}
\author[10]{J.~Ditzel}
\author[10]{B.~D\"onigus}
\author[10]{M.~S.~Ege}
\author[6]{H.~Elfner}
\author[4]{S.~Glaessel}
\author[10]{M.V.~Hartung}
\author[9]{C.M.~Ko}
\author[12]{R.~Lea}
\author[13]{M.~Lorenz}
\author[11]{M.~Mahlein}
\author[6]{D.~Mi\'{s}kowiec}
\author[14]{S.~Mrowczynski}
\author[15,23]{C.~Pinto}
\author[15]{M.~Puccio}
\author[16]{K.~Redlich}
\author[11]{B.~Singh}
\author[17]{J.~Stachel}
\author[18]{K.J.~Sun}
\author[19]{B.~Toma\v{s}ik}
\author[20]{V.~Vovchenko}
\author[21]{Z.~Xu}
\author[22]{H.~Zbroszczyk}

\affil[1]{SUBATECH, IMT Atlantique, Universit'e de Nantes, CNRS-IN2P3, Nantes, France}
\affil[2]{Institut f\"ur Kernphysik, Universit\"at M\"unster, Germany}
\affil[3]{Physics Department, University of Bologna and INFN, Bologna, Italy}
\affil[4]{Institute for Theoretical Physics, Goethe University, Frankfurt am Main, Germany}
\affil[5]{Weizmann Institute of Science, Rehovot, Israel}
\affil[6]{GSI Helmholtzzentrum f\"ur Schwerionenforschung, Darmstadt, Germany}
\affil[7]{Research Division and ExtreMe Matter Institute EMMI, Darmstadt, Germany}
\affil[8]{Physics Department, University of Salerno and INFN, Salerno, Italy}
\affil[9]{Cyclotron Institute and Department of Physics and Astronomy, Texas A\&M University, College Station, USA}
\affil[10]{Physics Department, Goethe University, Frankfurt am Main, Germany}
\affil[11]{Physics Department, Technical University of Munich, Munich, Germany}
\affil[12]{Physics Department, University of Brescia and INFN of Pavia, Brescia, Italy}
\affil[13]{Institute of Nuclear Physics, Goethe University, Frankfurt am Main, Germany}
\affil[14]{National Centre for Nuclear Research and Institute of Physics, Warsaw, Poland}
\affil[15]{European Organisation for Nuclear Research (CERN), Geneva, Switzerland}
\affil[16]{Institute of Theoretical Physics, University of Wroclaw, Poland}
\affil[17]{Physikalisches Institut, Ruprecht-Karls-Universit\"at Heidelberg, Heidelberg, Germany}
\affil[18]{Institute of Modern Physics, Fudan University, China}
\affil[19]{Faculty of Nuclear Sciences and Physical Engineering, Czech Technical University in Prague, Prague, Czechia}
\affil[20]{Physics Department, University of Houston, Houston, USA}
\affil[21]{Physics Department, Kent State University, Kent, Ohio, USA}
\affil[22]{Faculty of Physics, Warsaw University of Technology, Warsaw, Poland}
\affil[23]{Laboratoire de Physique des 2 Infinis Irène Joliot-Curie, IJCLab, Université Paris-Saclay, CNRS/IN2P3, Orsay, France}

\begin{document}

%\linenumbers

\maketitle

\begin{abstract}
This document summarizes the work of the EMMI Rapid Reaction Task Force on ``Understanding light (anti)nuclei production at RHIC and LHC'', which took place at GSI on 8--12 April 2024. The meeting brought together experimentalists and theorists to discuss the current understanding of light-nuclei and hypernuclei production in high-energy hadronic collisions, with particular emphasis on statistical hadronization, coalescence, kinetic, and transport approaches. The discussions addressed the experimental status and the ability of different theoretical descriptions to reproduce the available measurements, as well as the connections between nuclear production, femtoscopic correlations, source sizes, nuclear wave functions, and the formation dynamics of loosely bound states. This document reviews the main experimental and theoretical developments presented during the RRTF, summarizes the key questions that emerged from the discussions, and identifies observables and future measurements that may help discriminate among different production mechanisms.
\end{abstract}

\newpage

%%%% Femtoscopy and correlations defs
\newcommand{\onethree}     {$\sqrt{s}~=~13$~Te\kern-.1emV\xspace}

\newcommand{\LedLu}{Lednick\'y\xspace-Lyuboshitz\xspace}
\newcommand{\ks}{\ensuremath{k^*}\xspace}
\newcommand{\rs}{\ensuremath{r^*}\xspace}
\newcommand{\bb}{baryon\mbox{--}baryon}
\newcommand{\abab}{antibaryon\mbox{--}antibaryon}
\newcommand{\corr}[1]{{\color[rgb]{0,1,0}{#1}}}
\newcommand{\MeVc}{\ensuremath{\mathrm{MeV}\kern-0.05em/\kern-0.02em \textit{c}}}
\newcommand{\GeVc}{\ensuremath{\mathrm{GeV}\kern-0.05em/\kern-0.02em \textit{c}}}
\newcommand{\GeVcSq}{\ensuremath{\mathrm{GeV}\kern-0.05em/\kern-0.02em \textit{c}^2}~}
\newcommand{\MeVcSq}{\ensuremath{\mathrm{MeV}\kern-0.05em/\kern-0.02em \textit{c}^2}~}
\newcommand{\fermi}{\ensuremath{\mathrm{fm}}~}
\newcommand{\Led}{Lednick\'y\xspace}
\newcommand{\HALQCD}{HAL QCD\xspace}
\newcommand{\ChargedTrackMultiplicity}{\ensuremath{\num{35}}\,}
\newcommand{\ResolutionEffect}{8}

\newcommand{\pP}{\ensuremath{\mbox{p--p}}~}
\newcommand{\kd}{\ensuremath{\mbox{K$^{+}$--d}}~}
\newcommand{\ApAp}{\ensuremath{\mbox{\ensuremath{\overline{\mbox{p}}}--\ensuremath{\overline{\mbox{p}}}}}\,}
\newcommand{\pd}{\ensuremath{\mbox{p--d}}~}
\newcommand{\ld}{\ensuremath{\mbox{$\Lambda$--d}}~}
\newcommand{\pL}{\ensuremath{\mbox{p--$\Lambda$}}~}
\newcommand{\NL}{\ensuremath{\mbox{N--$\Lambda$}}~}
\newcommand{\NNL}{\ensuremath{\mbox{N--N--$\Lambda$}}~}
\newcommand{\ppL}{\ensuremath{\mbox{p--p--$\Lambda$}}~}
\newcommand{\ppp}{\ensuremath{\mbox{p--p--p}}~}
\newcommand{\ApAL}{\ensuremath{\mbox{\overline{p}--\overline{\Lambda}}}~}
\newcommand{\pAL}{\ensuremath{\mbox{p--$\overline{\Lambda} $}}~}

\newcommand{\pSiZero}{\ensuremath{\mbox{p--$\Sigma^0$}}~}

\newcommand{\rKdeff}{\ensuremath{1.35^{+0.04}_{-0.05}}~} 
\newcommand{\rpdeff}{\ensuremath{1.08\pm0.06}\,~}
\newcommand{\rNNeff}{\ensuremath{1.43^{+0.16}_{-0.16}}\,~}

\tableofcontents
\clearpage
\section{Introduction}
\label{sec:introduction}

The production mechanism of light (anti)nuclei and (anti)hypernuclei in high-energy hadronic collisions has been the subject of intense theoretical and experimental investigation. A particularly intriguing aspect is the production of loosely bound states in ultrarelativistic heavy-ion collisions. Their binding energies, typically of the order of a few MeV or less, are much smaller than the characteristic temperatures of the system created in these collisions, raising fundamental questions about their formation and evolution through the hadronic phase.

Several phenomenological approaches have been developed to describe the production of light nuclei and hypernuclei, including statistical hadronization, coalescence, and microscopic transport models. While these approaches can reproduce several features of the available experimental measurements, they rely on substantially different physical pictures of the formation and subsequent evolution of these states. In particular, they differ in the role attributed to the hadronization and kinetic freeze-out stages, as well as to the destruction and regeneration of loosely bound states during the hadronic evolution. Identifying observables that can discriminate between these scenarios is therefore essential for establishing a consistent description of light-nuclei and hypernuclei production.

To address these questions, the EMMI Rapid Reaction Task Force on ``Understanding light (anti)nuclei production at RHIC and LHC'' was held at GSI on 8--12 April 2024. The meeting brought together experimental and theoretical experts working on light-nuclei and hypernuclei production to critically examine the different phenomenological approaches and confront their predictions with experimental measurements from RHIC and the LHC. Particular emphasis was placed on identifying the assumptions common to the different models, understanding the origin of their differences, and determining which existing or future observables provide the strongest constraints on the underlying production mechanisms.

This document summarizes the studies and discussions carried out in the framework of the Task Force and presents a comparative assessment of the phenomenological approaches considered during the meeting, together with the main conclusions and open questions that emerged from this effort.

\section{Experimental overview}
\label{sec:experimental_overview}

The experimental study of light (anti)nuclei and (anti)hypernuclei production has significantly advanced in recent years, driven by increasingly precise and differential measurements at RHIC and the LHC. The available results span a broad range of collision energies and system sizes and include production yields, transverse-momentum distributions, and measurements of the properties of light nuclear states. In addition, femtoscopic correlation measurements provide complementary information on the particle-emitting source and hadronic interactions relevant to the formation of bound states. This section provides an overview of the main experimental results discussed during the Rapid Reaction Task Force.

\subsection{Hypernuclei results at RHIC and LHC}
\label{subsec:hypertriton}

While quantum chromodynamics (QCD) can be treated perturbatively in the weak-coupling regime, understanding QCD in the strong-coupling regime, and in particular the structure of nuclear matter, remains one of the most important challenges in nuclear physics~\cite{NSACLRP:2015qub}.

As the energy scale increases in nuclear reactions and/or the energy density increases in stars, the strangeness degree of freedom becomes increasingly relevant. Hypernuclei are bound states of ordinary nuclei containing one or more hyperons, i.e., baryons containing strange quarks. Such nuclear systems have provided a natural tool for studying the hyperon--nucleon ($Y$--$N$) interaction since the discovery of the first hypernucleus~\cite{danysz1953delayed} more than seven decades ago. The $Y$--$N$ interaction is expected to be an important ingredient in describing the equation of state of astrophysical objects such as neutron stars~\cite{Lattimer:2004pg}. Since hyperons are unstable in vacuum, direct hyperon--nucleon scattering experiments are challenging to perform, in contrast to, for example, electron scattering experiments used to probe nuclear structure. Hypernuclei therefore remain an ideal laboratory for studying the $Y$--$N$ interaction~\cite{Chrien:1989xeq,Hashimoto:2006aw}.

The hypertriton ($\mathrm{^3_{\Lambda}H}$), the lightest $\Lambda$ hypernucleus, is the most copiously produced hypernucleus in heavy-ion collisions~\cite{Chen:2023mel}. Owing to the additional strangeness degree of freedom and the differences between the $Y$--$N$ and $N$--$N$ interactions, $\mathrm{^3_{\Lambda}H}$ is expected to exhibit properties different from those of ordinary nuclei with the same mass number, such as the triton and $^3$He. Therefore, measurements of the $\mathrm{^3_{\Lambda}H}$ binding energy, excitation energies of particle-bound states, spin, lifetime, and decay branching ratios provide valuable information on the $Y$--$N$ interaction in one of the simplest hypernuclear systems~\cite{Gal:2016boi}.

\medskip

The two main experiments measuring hypernuclei in heavy-ion collisions at relativistic collider energies, ALICE at the LHC and STAR at RHIC, have obtained precise results on the production and properties of light (anti)hypernuclei. 

The ALICE Collaboration used the full Run 2 data sample collected at the LHC to measure the production of (anti)hypertritons in different collision systems, allowing the charged-particle multiplicity dependence of (anti)hypertriton production to be studied.

For the first time, this also enabled a measurement of (anti)hypertriton production in p--Pb collisions~\cite{hypppB}. Comparisons with production models are commonly performed using yield ratios. In particular, the $S_3$ ratio is defined as the double ratio of the hypertriton-to-helium-3 and $\Lambda$-to-proton yields at a given mean charged-particle multiplicity~\cite{Zhang:2009ba,STAR:2010gyg}. Charged-particle multiplicity is correlated with the system size, with larger multiplicities, as observed in Pb--Pb collisions, corresponding to larger systems and lower multiplicities to smaller systems.

\begin{figure}[!hbt]
\centering
\begin{minipage}[b]{0.5\textwidth}
\includegraphics[width=\textwidth]{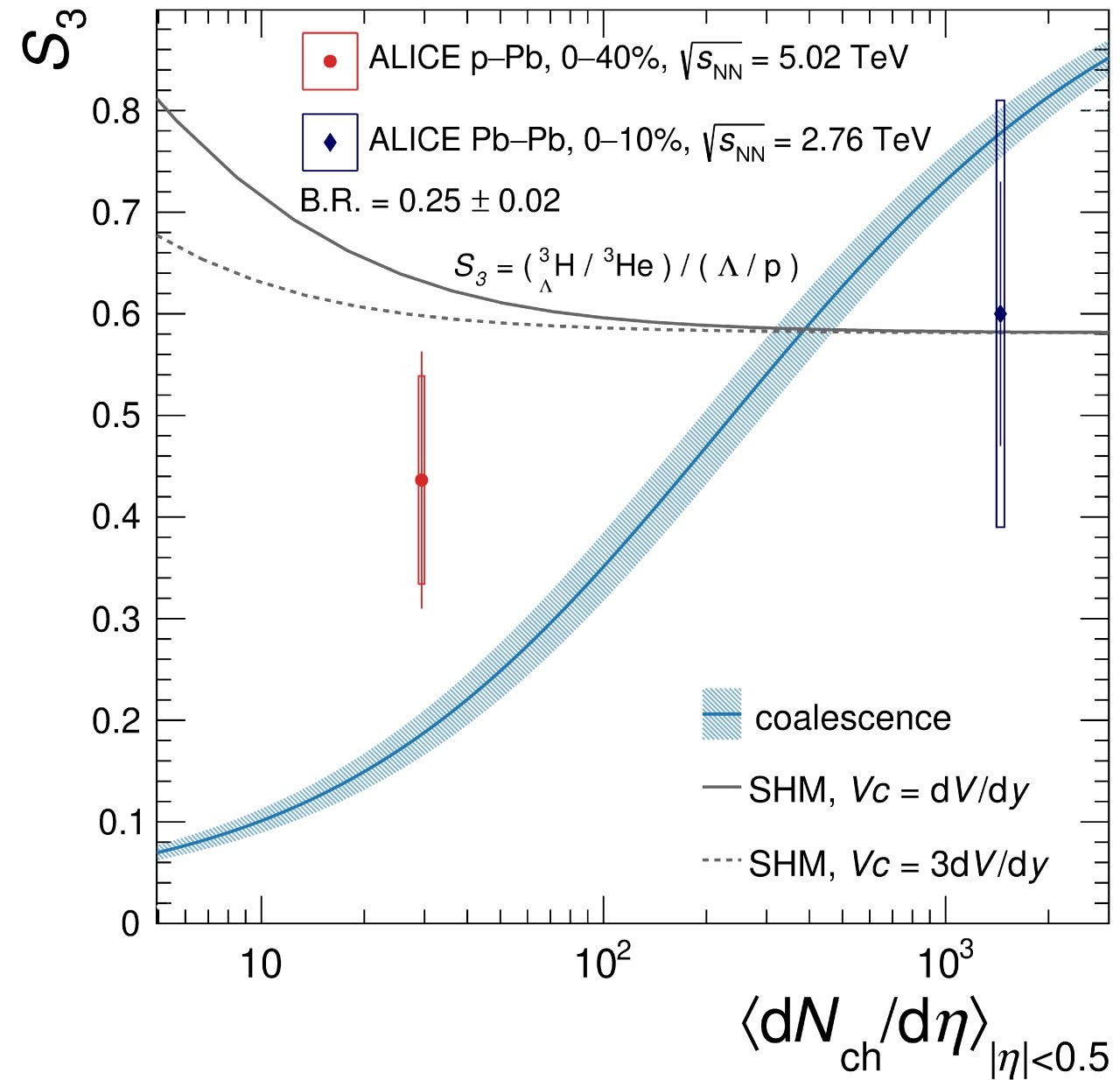}
\end{minipage}
\caption{Measurements of the $S_3$ ratio, defined as the double ratio of the hypertriton-to-helium-3 and $\Lambda$-to-proton yields, as a function of the mean charged-particle multiplicity measured by ALICE at the LHC. The measurements are compared with SHM predictions, including exact quantum-number conservation for small systems, and with coalescence calculations~\cite{hypppB,Sun:2018mqq,Vovchenko:2019kes}.}
\label{fig:3LHprod}
\end{figure}

Figure~\ref{fig:3LHprod} shows the $S_3$ ratio measured by ALICE in p--Pb collisions at $\sqrt{s_{\mathrm{NN}}}=5.02$~TeV~\cite{ALICE:hypertritonpPb502TeV} and in Pb--Pb collisions at $\sqrt{s_{\mathrm{NN}}}=2.76$~TeV~\cite{ALICE:hypertritonPbPb276TeV}, as a function of the mean charged-particle multiplicity. Coalescence models~\cite{Sun:2018mqq,Bellini:2020cbj} predict a decrease of $S_3$ with decreasing system size, whereas the statistical hadronization model (SHM)~\cite{Andronic:2017pug,Vovchenko:2019kes} predicts a milder trend in the opposite direction. However, the current experimental uncertainties are too large for the $S_3$ measurement to provide significant discrimination between the different production models.

\begin{figure}[!hbt]
\centering
\includegraphics[width=0.48\linewidth]{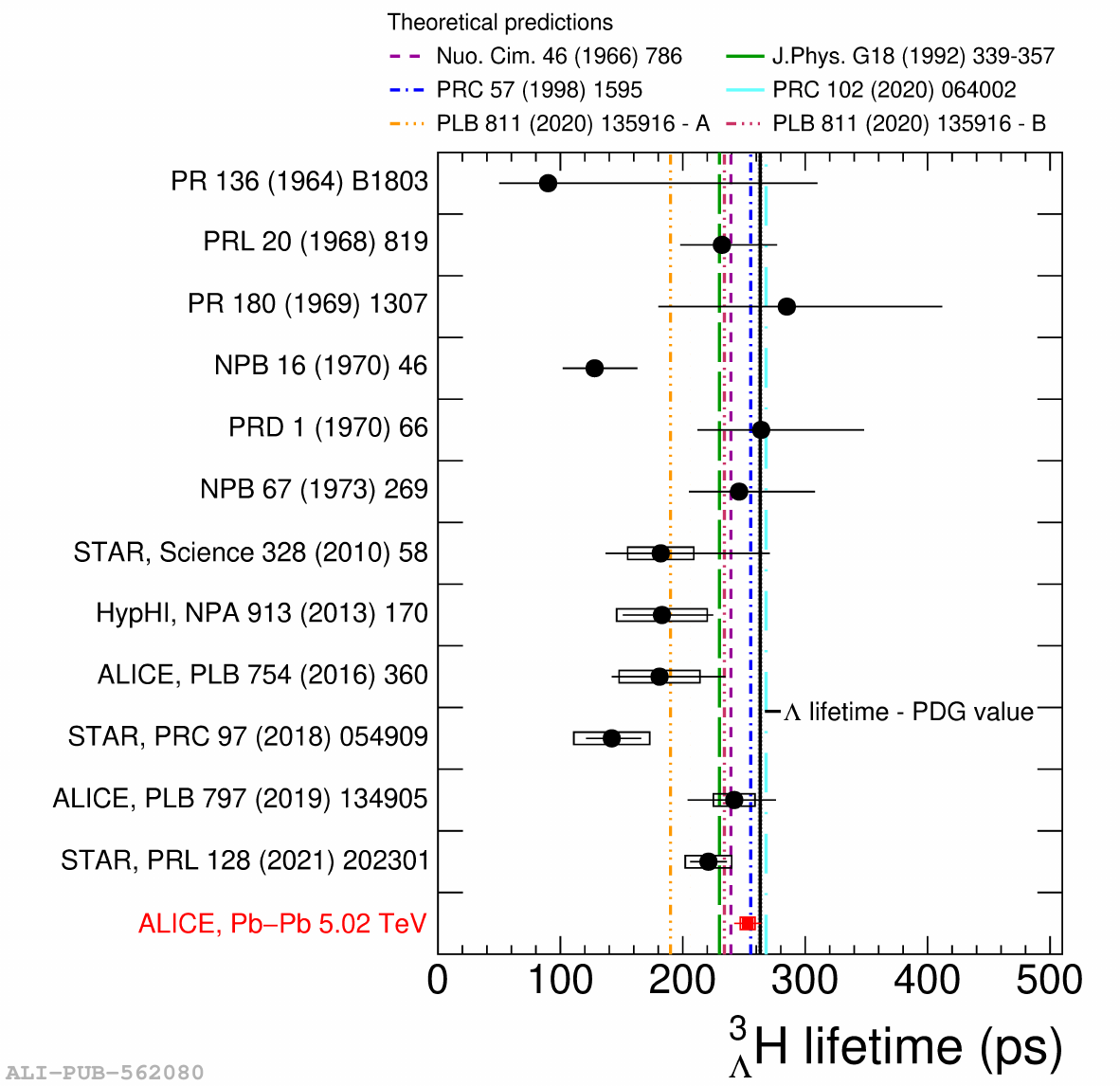}
\includegraphics[width=0.48\linewidth]{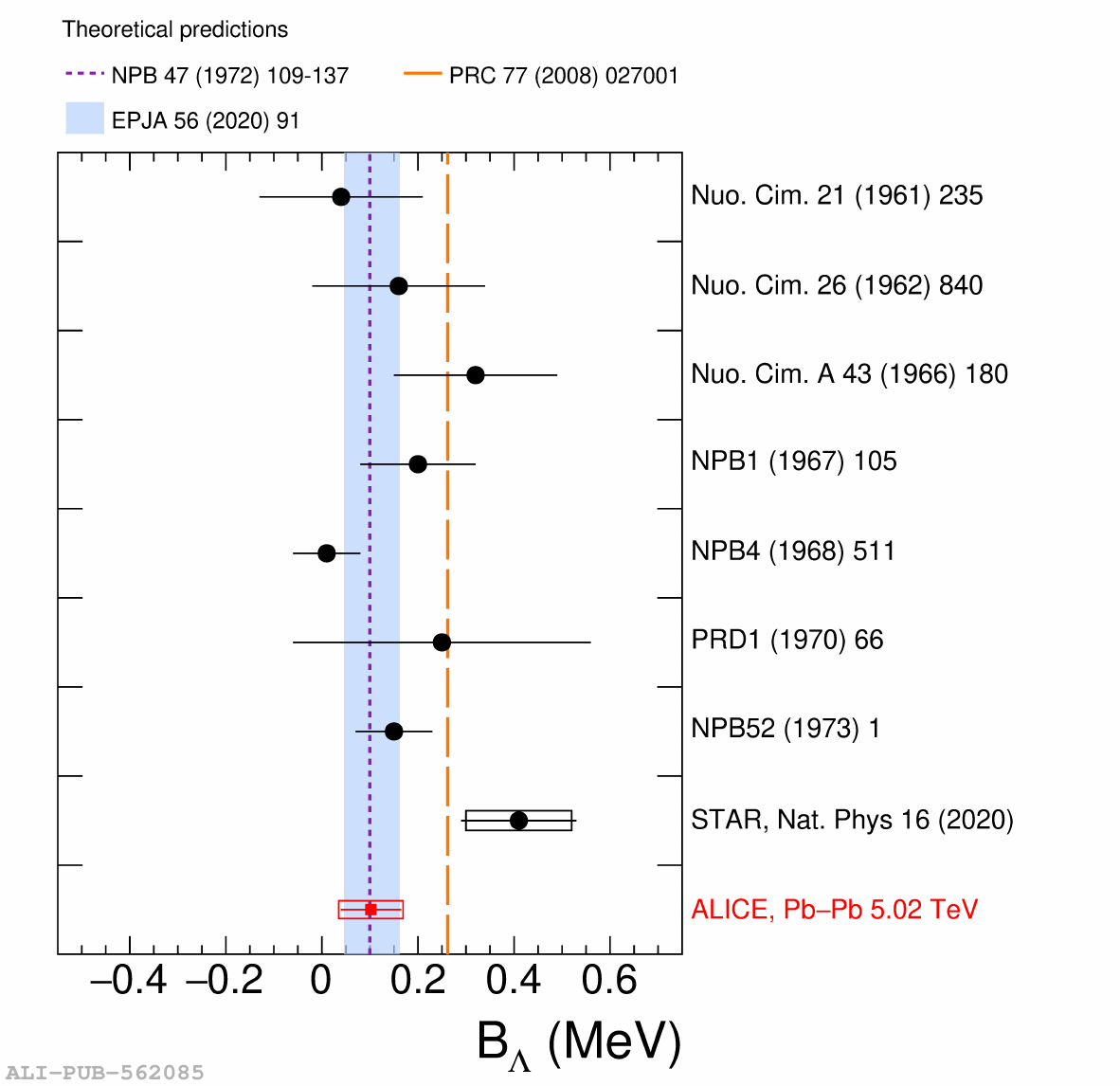}
\caption{Compilation of hypertriton lifetime (left) and $\Lambda$ separation energy (right) measurements, compared with theoretical predictions~\cite{hypLPbPb}.}
\label{fig:3LHlifemass}
\end{figure}

The enhanced production of (anti)hypertritons in Pb--Pb collisions relative to pp collisions also enables precise measurements of the (anti)hypertriton lifetime and binding energy using the ALICE Run 2 data~\cite{hypLPbPb}. These measurements, characterized by small statistical and systematic uncertainties, are compatible with theoretical predictions describing the hypertriton as a weakly bound state. A compilation of the ALICE results together with previous measurements and theoretical predictions is presented in Fig.~\ref{fig:3LHlifemass}.

\begin{figure}[!hbt]
\centering
\includegraphics[width=0.6\linewidth]{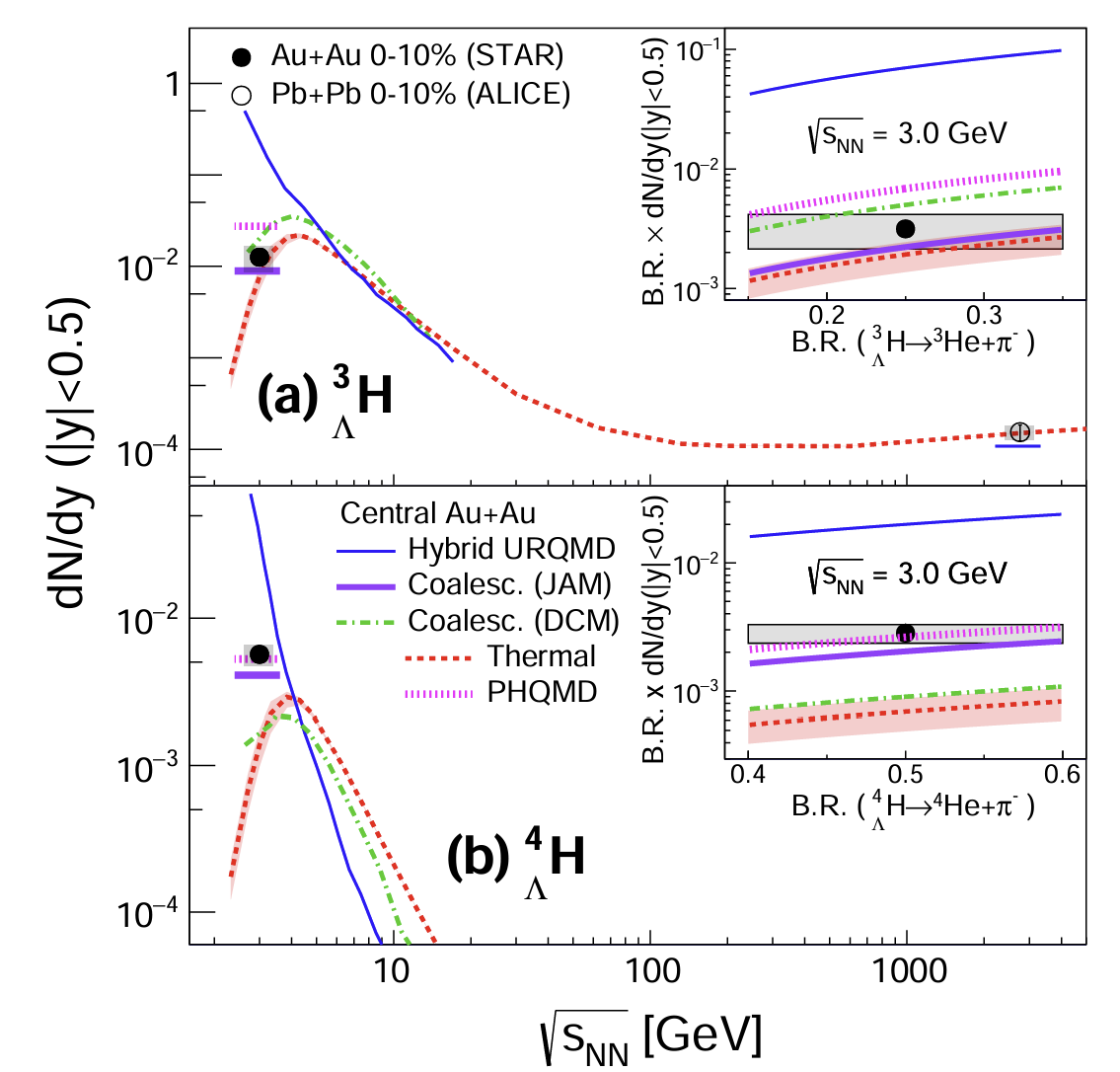}
\caption{Hypertriton (upper panel) and hyperhydrogen-4 (lower panel) production yields as a function of the center-of-mass energy at RHIC, compared with predictions from the SHM and coalescence models~\cite{STAR:2021orx}.}
\label{fig:3LHEngStar}
\end{figure}

At RHIC, the STAR Collaboration studies heavy-ion collisions over a broad center-of-mass energy range, from a few GeV to a few hundred GeV. In the lower-energy regime, enhanced production of hypernuclei is expected, while antiparticle production is strongly suppressed. The large hypernuclei samples available in this energy regime enable precise measurements of their production and properties. In particular, STAR has measured hypertriton production at several center-of-mass energies. The upper panel of Fig.~\ref{fig:3LHEngStar} shows the energy dependence of hypertriton production compared with predictions from the SHM, coalescence approaches, and PHQMD calculations. Within the current experimental uncertainties, all three approaches provide a good description of the measured yields~\cite{STAR:2021orx}.

Beyond the (anti)hypertriton, heavier (anti)hypernuclei can also be studied, such as (anti)hyperhydrogen-4 and (anti)hyperhelium-4. Their production is expected to be suppressed relative to that of the (anti)hypertriton because of their larger masses. However, these nuclei have known excited states that decay electromagnetically to the ground state within about 1~ps, while the ground state subsequently decays weakly~\cite{J-PARCE13:2015uwb}. Feed-down from these excited states is expected to enhance the ground-state yields by approximately a factor of four~\cite{Gal:2016boi,Donigus:2023hiw}.

\begin{figure}[!hbt]
\centering
\includegraphics[width=0.65\linewidth]{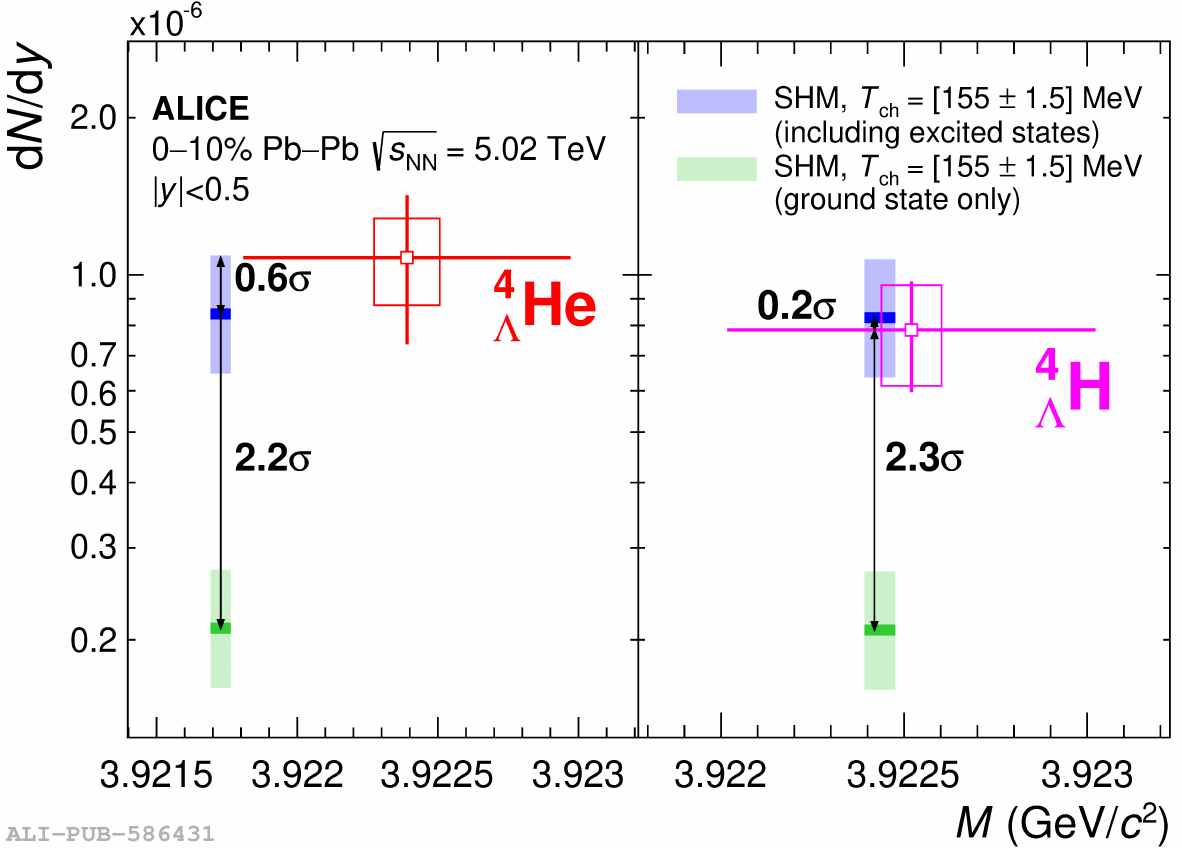}
\caption{Measured integrated yields of (anti)${}^{4}_{\Lambda}$He (left) and (anti)${}^{4}_{\Lambda}$H (right), averaged over particle and antiparticle states. The measured masses are shown on the $x$ axis and compared with the world-average values from the Hypernuclei Database, whose uncertainties are represented by the horizontal widths of the blue and green lines. Statistical and systematic uncertainties of the measurements are shown by bars and boxes, respectively. The measured $\mathrm{d}N/\mathrm{d}y$ values are compared with SHM predictions at $T_{\mathrm{ch}}=155$~MeV considering only the ground state (green) or including feed-down contributions from excited states (blue). The shaded bands represent the variation of the predicted yields obtained by varying $T_{\mathrm{ch}}$ by 1.5~MeV. The arrows indicate the difference between the measured and predicted yields in units of standard deviations, considering only the experimental uncertainties~\cite{ALICE:2024ilx}.}
\label{fig:4LH4LHeALICE}
\end{figure}

The STAR Collaboration discovered antihyperhydrogen-4 by analyzing a large combined dataset in 2021~\cite{STARH4Lbar:2022}, while the first evidence of antihyperhelium-4 was reported by the ALICE Collaboration in 2024~\cite{ALICE:2024ilx}. STAR has measured the production, lifetime, and binding energy of these states~\cite{STAR:2021orx,STAR:2022zrf}. At the LHC, their production is more strongly suppressed because of the lower baryochemical potential. Nevertheless, ALICE recently reported significant invariant-mass signals and measurements of their production yields~\cite{ALICE:2024ilx}. Figure~\ref{fig:4LH4LHeALICE} shows the corresponding ALICE measurements, including the measured masses, which are characterized by relatively large statistical uncertainties.

The measurements are compared with SHM predictions considering either the ground states only or including feed-down contributions from excited states. The uncertainty associated with the chemical freeze-out temperature is evaluated by varying its value by 1.5~MeV. A significant deviation from the ground-state-only prediction is observed, whereas the measured yields are in good agreement with predictions including feed-down from excited states. In this respect, the thermal model can provide a tool complementary to conventional spectroscopy: the inclusion of additional excited states and their feed-down contributions can be tested against the measured ground-state yields, potentially providing sensitivity to states that have not yet been experimentally established. The substantially larger data samples collected during LHC Run 3 will enable more precise measurements of these states. Furthermore, measurements at low charged-particle multiplicities, as already performed for the (anti)hypertriton, will provide additional constraints on the production models.

\begin{figure}[!hbt]
\centering
\includegraphics[width=0.85\linewidth]{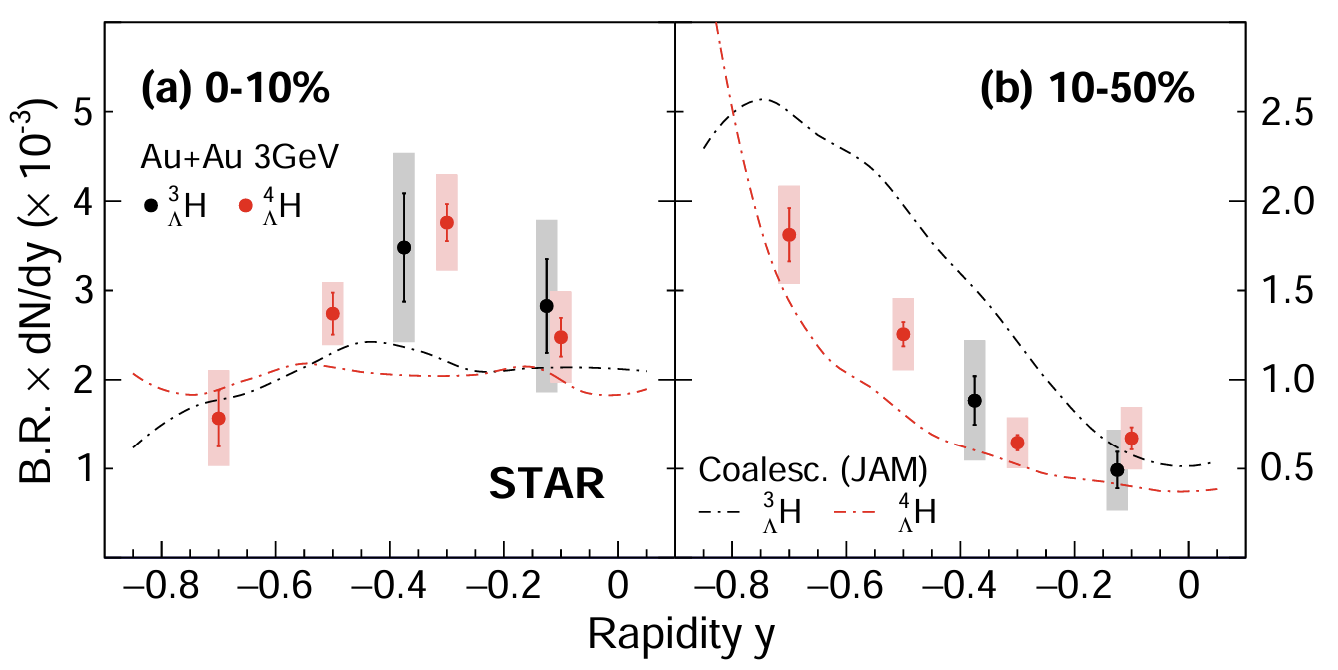}
\caption{Rapidity-differential measurements of hypertriton and hyperhydrogen-4 production at RHIC, compared with predictions from a coalescence model~\cite{STAR:2021orx}.}
\label{fig:3LH4LHRapRHIC}
\end{figure}

As mentioned above, the lower center-of-mass energy regime studied by the STAR Collaboration provides access to high baryon densities and enhanced hypernuclei production relative to LHC energies, enabling more differential studies of these states. The lower panel of Fig.~\ref{fig:3LHEngStar} shows the integrated production yield of hyperhydrogen-4 compared with predictions from different production models~\cite{STAR:2021orx}. The observed deviation from the theoretical predictions is expected to arise from feed-down contributions from excited states. The production yield has also been measured differentially as a function of rapidity, as shown in Fig.~\ref{fig:3LH4LHRapRHIC}, where it is compared with predictions from a coalescence model~\cite{STAR:2021orx}. The measured trend is qualitatively described by the model.

\begin{figure}[!hbt]
\centering
\includegraphics[width=0.6\linewidth]{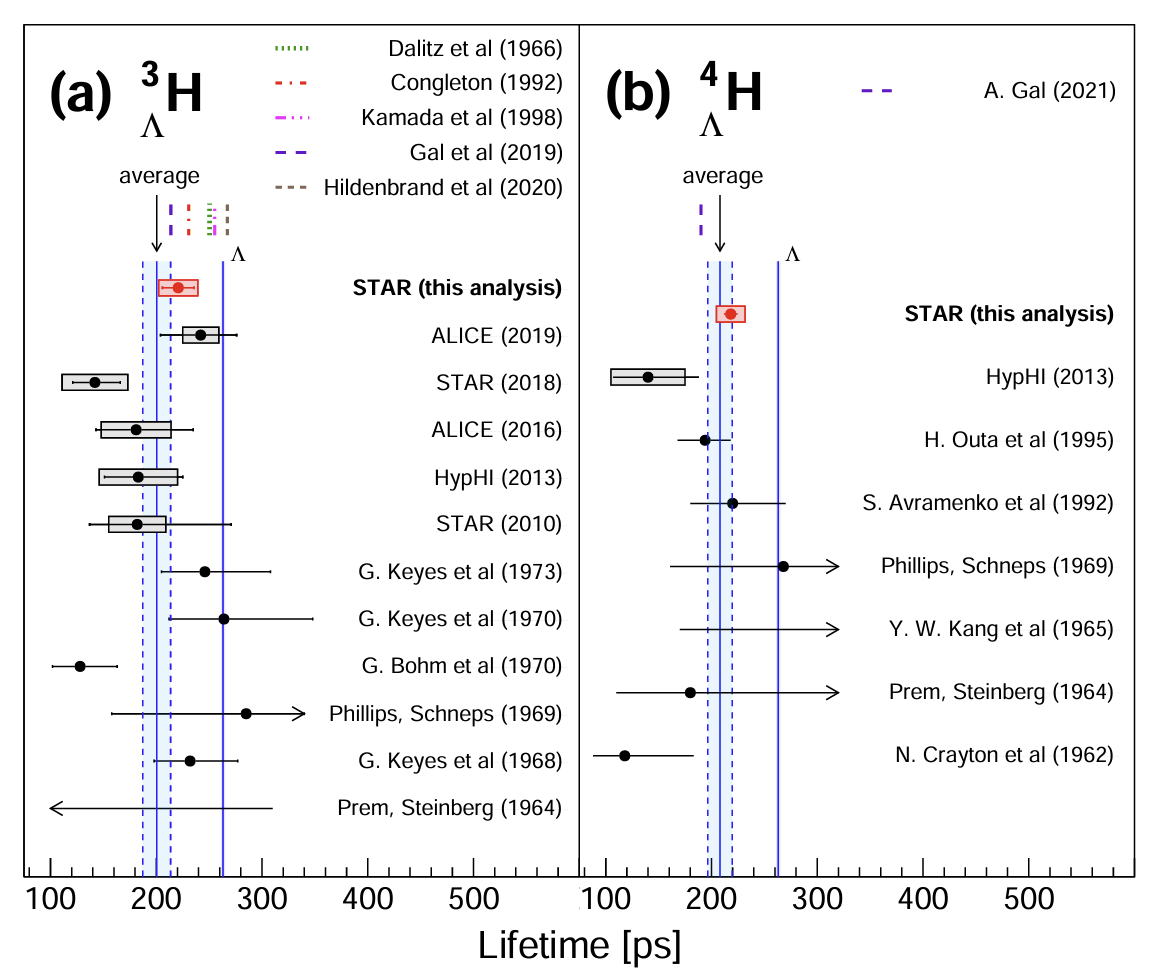}
\caption{Compilation of hypertriton and hyperhydrogen-4 lifetime measurements by the STAR Collaboration, compared with previous measurements and world averages~\cite{STAR:2021orx}.}
\label{fig:4LHSTARLifetime}
\end{figure}

\begin{figure}[!hbt]
\centering
\includegraphics[width=0.48\linewidth]{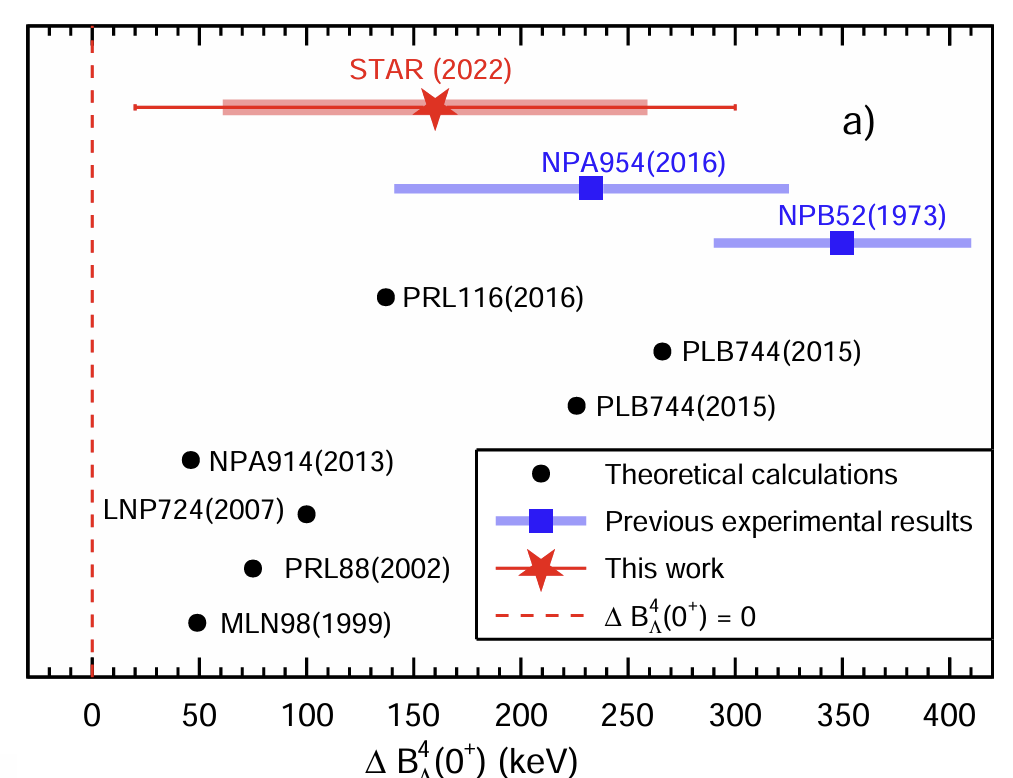}
\includegraphics[width=0.48\linewidth]{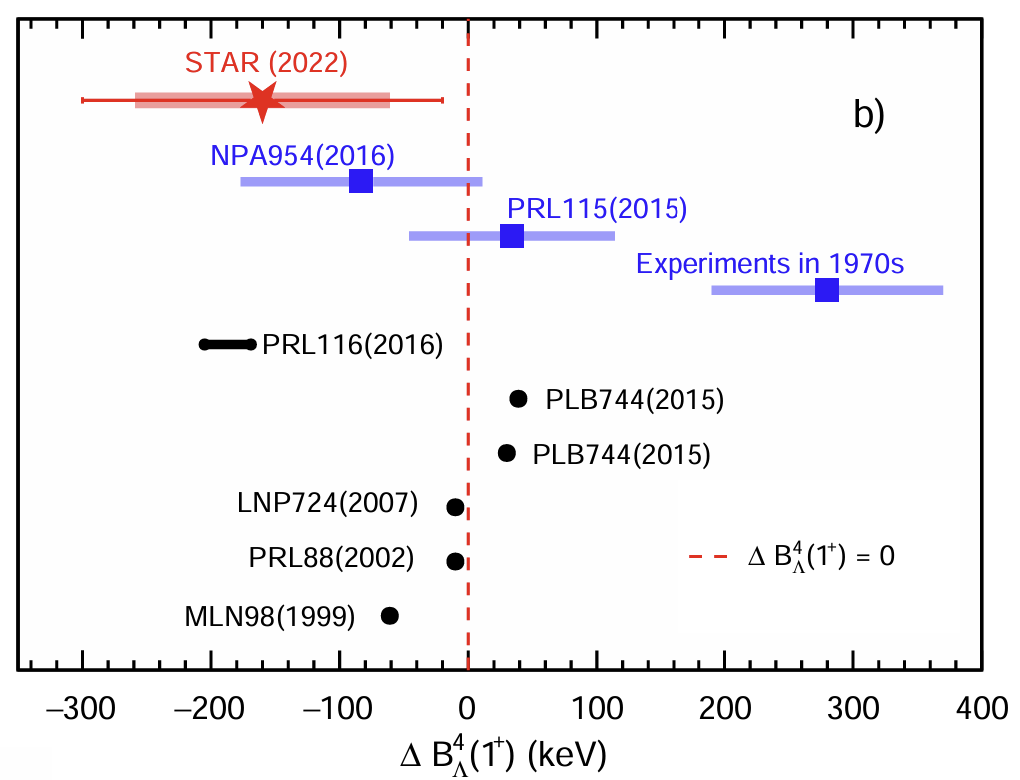}
\caption{Compilation of $\Delta B_{\Lambda}$ measurements for hyperhydrogen-4 and hyperhelium-4, defined as the difference between their measured $\Lambda$ separation energies, compared with previous measurements and world averages~\cite{STAR:2022zrf}.}
\label{fig:4LHSTARBL}
\end{figure}

In addition, the lifetimes of the hypertriton and hyperhydrogen-4 have been measured with high precision. The results are shown in Fig.~\ref{fig:4LHSTARLifetime} and compared with previous measurements and world averages~\cite{STAR:2021orx}. The measured values are consistent with the current averages within uncertainties. The $\Lambda$ separation energies of hyperhydrogen-4 and hyperhelium-4 have also been measured, and their difference is presented in Fig.~\ref{fig:4LHSTARBL}. Although the uncertainties remain sizable in the previous published data~\cite{STAR:2022zrf}, the analysis of additional data has now been completed after the RRTF meeting, and the improved precision reveals a symmetric energy-level splitting that resolves the longstanding discrepancy between theoretical models and experimental observations~\cite{STAR:2026hds,Gazda:2015qyt}.

\medskip

The prospects for further investigations of these systems are very promising. Larger data samples are being collected at the LHC, while several datasets remain to be analyzed at RHIC. In parallel, advances in tracking and reconstruction techniques are expected to improve reconstruction efficiencies and, consequently, increase the number of available hypernuclear candidates. This will open new opportunities for studying double-strange hypernuclei and heavier states with $A>4$. Looking further ahead, the proposed ALICE 3 experiment will substantially extend the reach of hypernuclear measurements at the LHC, enabling studies of exotic nuclear states and heavier hypernuclei, including ${}^{4}_{\Lambda}\mathrm{He}$ and ${}^{5}_{\Lambda}\mathrm{He}$, and potentially extending the experimental reach to hypernuclei with $A=6$~\cite{ALICE3LoI}.

These developments will enable more differential studies of hypernuclei production as a function of charged-particle multiplicity and center-of-mass energy, together with more precise determinations of their properties.

Since the RRTF meeting, the HADES Collaboration has reported the first
observation of $\mathrm{^{3}_{\Lambda}H}$ and $\mathrm{^{4}_{\Lambda}H}$ in
Ag+Ag collisions at $\sqrt{s_{NN}}=2.55$~GeV~\cite{HADES:2026hypernuclei}.
The measured rapidity distributions are centered around midrapidity, and the
yield of $\mathrm{^{4}_{\Lambda}H}$ is found to be equal to or larger than that
of $\mathrm{^{3}_{\Lambda}H}$. This ordering differs from that observed by
STAR at $\sqrt{s_{NN}}=3$~GeV and is consistent with a scenario including
feed-down from excited hypernuclear states. These results provide an additional
low-energy benchmark for systematic comparisons of hypernuclear production
models.

\subsection{Measurements of (anti)nuclei at RHIC}
\label{subsec:nuclei_rhic}

Relativistic heavy-ion collisions at RHIC~\cite{Chen:2018tnh} produce a hot and dense phase of matter containing approximately equal numbers of quarks and antiquarks. The large detector acceptance and particle-identification capabilities provide favorable conditions for the experimental study of (anti)nuclei. Studies of (anti)nuclei production with the STAR detector began with the first RHIC operation in 2000~\cite{Adler:2001uy} and subsequently led to the discovery of the first antimatter hypernucleus~\cite{STAR:2010gyg} and of the heaviest stable antinucleus observed to date~\cite{Agakishiev:2011ib}. With the upgrade of the micro-vertex detector, STAR was able to perform precision measurements of the masses and binding energies of (anti)hypernuclei, as well as their differences~\cite{STAR:2019wjm}. More recently, STAR has performed searches for heavier antimatter hypernuclei using existing datasets and data collected in Au--Au collisions at $\sqrt{s_{\mathrm{NN}}}=200$~GeV~\cite{STARH4Lbar:2022}. The $A=4$ antimatter hydrogen hypernucleus (${}^{4}_{\Lambda}\overline{\mathrm{H}}$) was discovered~\cite{STAR:2023fbc}. This provides a valuable opportunity to further investigate interactions among antimatter particles~\cite{STAR:2015kha} and to measure mass differences between heavier (hyper)nuclei and (hyper)antinuclei. Measurements of (hyper)antinucleus production at RHIC will provide valuable input for understanding their production mechanisms~\cite{Chen:2018tnh}, as well as the associated backgrounds in astrophysical and cosmological searches.

In addition, STAR has requested a high-statistics fixed-target (FXT) data sample of two billion events at $\sqrt{s_{\mathrm{NN}}}=3$~GeV to search for double hypernuclei~\cite{STAR2020BUR}. Efforts have also been devoted to searches for muonic atoms~\cite{KFXinPhD,Wang:2024njn,Wang:2026ftb}, with tentative evidence observed at low statistical significance. Renewed interest in these searches, together with new analyses of high-statistics datasets, has motivated further studies. A possible next step in the search for hypernuclei unique to relativistic heavy-ion collisions is the investigation of charmed hypernuclei~\cite{Dover:1977jw}. In particular, the feasibility of searching for a specific charmed mesonic nuclear bound state, $\alpha$--$D^-$, composed of an alpha particle and a $D^-$ meson bound through the Coulomb interaction, has been explored. Such a state would be stable against strong decay owing to the large binding energy of the alpha particle, whereas a proton--$D^-$ system would strongly decay into a neutron and a $D^0$. The production of such exotic states is expected to be experimentally accessible at midrapidity at FAIR and at forward rapidity at RHIC and the LHC. Rate estimates are based on existing measurements at similar rapidities~\cite{E864:2000auv} and on a coalescence model for charmed mesons~\cite{xuzb2024EMMI}.

The production of nuclear clusters is sensitive both to the dynamics of the system evolution and to the cluster-formation mechanism~\cite{Sun:2022xjr,Andronic:2026nev,Braaten:2024cke,Wang:2026rcf}. RHIC has successfully measured the yields and flow of nuclei and hypernuclei in both collider and fixed-target modes~\cite{STAR:2021orx,STAR:2022fnj}. The baryon chemical potential, and correspondingly the baryon density, decreases dramatically as the center-of-mass energy increases from about 3~GeV in the RHIC fixed-target program to several TeV at the LHC. At the same time, the yields of nuclear clusters vary by several orders of magnitude. While this demonstrates the sensitivity of nuclear production to the system dynamics, it also makes quantitative comparisons with models over such a large dynamic range challenging.

\begin{figure}[!hbt]
\centering
\begin{minipage}[b]{0.7\textwidth}
	\includegraphics[width=\textwidth]{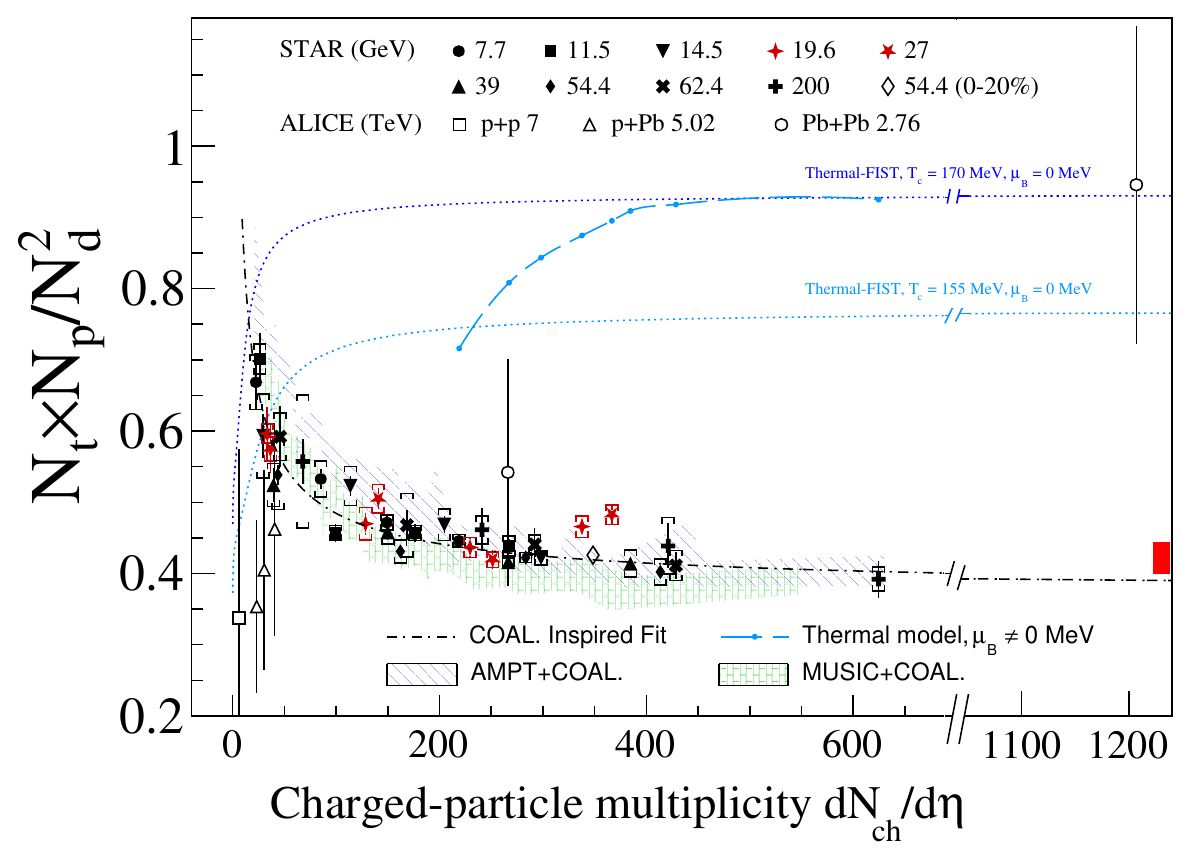}
\end{minipage}
\caption{Yield ratio $N_{\mathrm{t}}N_{\mathrm{p}}/N_{\mathrm{d}}^2$ as a function of charged-particle multiplicity, using measurements from RHIC to LHC energies. The data are compared with predictions from thermal and coalescence models. Thermal-model predictions for the most central collisions at each RHIC Beam Energy Scan energy are also shown (blue points connected by a dashed line). Figure adapted from Ref.~\cite{STAR:2022hbp}, with additional ALICE data points (see Sec.~\ref{subsec:nuclei_lhc}).}
\label{fig:AGS2RHIC2LHC4TritonVSNch}
\end{figure}

It was proposed in Ref.~\cite{Sun:2017xrx} that yield ratios of nuclear clusters, such as $tp/d^2$, can effectively cancel much of the ``trivial'' dependence on thermodynamic quantities such as density and temperature. This provides sensitivity to smaller, non-trivial effects, such as density fluctuations, over a broad range of collision energies. Such ratios are also sensitive to the nuclear production mechanism. Figure~\ref{fig:AGS2RHIC2LHC4TritonVSNch} shows the yield ratio $N_{\mathrm{t}}N_{\mathrm{p}}/N_{\mathrm{d}}^2$, where $N_{\mathrm{t}}$, $N_{\mathrm{p}}$, and $N_{\mathrm{d}}$ denote the triton, proton, and deuteron yields, respectively, as a function of charged-particle multiplicity from the RHIC Beam Energy Scan I to LHC energies. The data show a characteristic increase toward smaller systems~\cite{STAR:2022hbp}. Within the coalescence model, this behavior is interpreted as an effect of the overlap with the nuclear wave function, whereas it is absent in the thermal model. Both thermal and coalescence models predict only a weak multiplicity dependence for large systems, but with limiting values that differ by approximately a factor of two. The ALICE measurements appear to indicate a slight increasing trend toward high multiplicities. Future measurements with improved precision will be crucial for determining the limiting value of this ratio in large systems and further discriminating between the different production mechanisms.

\subsection{Measurements of (anti)nuclei at the LHC}
\label{subsec:nuclei_lhc}

At the Large Hadron Collider (LHC), the nearly vanishing baryochemical potential provides favorable conditions for the study of both nuclei and antinuclei production. The ALICE Collaboration has performed systematic measurements across a broad range of collision systems and center-of-mass energies, from pp collisions at $\sqrt{s}=900$~GeV to Pb--Pb collisions at multi-TeV energies~\cite{muB,Adam:2015pna,3HePbPb,Acharya:2017dmc,Acharya:2017fvb,Acharya:2017bso,deuteron_pp7TeV,Acharya:2019rys,Acharya:2019ttn,Acharya:2019xmu,Acharya:2020sfy,Acharya:2020lus,deuteron_jets,nuclei_pp_13TeV,nuclei_pp_5TeV,hypertriton_pPb_ALICE,ALICE:2022boh,ALICE:2022pbb,ALICE:2022jmr,ALICE:2024say,ALICE:2024yin}. (Anti)nuclei up to $A=4$ have been measured in Pb--Pb collisions, and up to $A=3$ in pp, p--Pb, and Xe--Xe collisions. The large variety of collision systems and charged-particle multiplicities accessible at the LHC enables systematic studies of nuclear production over a broad range of system sizes.

The multiplicity dependence of light (anti)nuclei production provides important constraints on their formation mechanisms. The deuteron-to-proton yield ratio exhibits a smooth increase with charged-particle multiplicity, approaching a nearly constant value in central heavy-ion collisions. This trend is qualitatively described by both statistical hadronization and coalescence models, making it difficult to discriminate between the different production scenarios using $A=2$ nuclei alone. The $^{3}$He-to-proton yield ratio provides a more sensitive observable, with model predictions differing by approximately 40\% in the multiplicity range corresponding to heavy-ion collisions. However, the current experimental uncertainties do not yet allow these predictions to be distinguished. Measurements of yield ratios involving nuclei with $A=4$ remain relatively scarce, limiting their ability to provide additional constraints on the different production scenarios. Hypernuclei, on the other hand, offer particularly promising observables for discriminating among models, owing to the significant differences in spatial extension between loosely bound hypernuclei and ordinary nuclei of comparable mass. In particular, their production yields are sensitive to the interplay between the spatial extent of the nuclear wave function and the size of the particle-emitting source, as further discussed in Sect.~\ref{ss_looselybs}.

Complementary information is provided by transverse-momentum-differential measurements, whose interpretation within coalescence models is sensitive to the nuclear wave function and the size of the baryon-emitting source. In particular, the deuteron transverse-momentum spectrum measured in high-multiplicity pp collisions at $\sqrt{s}=13$~TeV can be described by Wigner-function-based coalescence calculations employing a realistic deuteron wave function, such as Argonne v18, together with experimentally constrained source sizes~\cite{Mahlein:2024pur,Mahlein:2023fmx}, as further discussed in Sect.~\ref{ss:tomcca}.

Additional insights into the formation mechanism are provided by measurements of deuteron production in jets, where the reduced phase-space separation between nucleons can enhance their coalescence probability. The ALICE Collaboration has investigated this production in pp collisions at $\sqrt{s}=13$~TeV~\cite{ALICE:2022ugx} and, more recently, in p--Pb collisions at $\sqrt{s_{\rm NN}}=5.02$~TeV~\cite{ALICE:2026vqn}. In pp collisions, the coalescence parameter $B_2$ measured in jets is approximately one order of magnitude larger than that in the underlying event. An even larger enhancement is observed in p--Pb collisions, providing further evidence of the sensitivity of nuclear formation to local nucleon phase-space correlations and the characteristics of the particle-emitting source. These results are discussed in more detail in Sect.~\ref{dinjets}.

The rapidity dependence of antideuteron production provides complementary constraints on the coalescence mechanism, with direct implications for astrophysical applications. In pp collisions at $\sqrt{s}=13$~TeV, ALICE has measured the antideuteron yields and the coalescence parameter $B_2$ up to $|y|<0.7$~\cite{ALICE:2024yin}. The coalescence parameter is found to be approximately independent of rapidity within the measured range. These measurements provide an important experimental reference for extrapolating antinuclei production towards forward rapidities, where direct experimental constraints remain limited. Such extrapolations are particularly relevant for calculations of secondary antinuclei production in cosmic-ray interactions with the interstellar medium, which constitute an important background for indirect dark-matter searches. In particular, model calculations~\cite{Blum:2023lmr} indicate that a large fraction of the secondary cosmic-ray antideuteron flux originates from production at rapidities $0.5<|y|<1.5$, highlighting the importance of future measurements beyond midrapidity and at lower collision energies.

More recently, measurements of (anti)nuclei and (anti)hypernuclei at the LHC have been extended by the LHCb Collaboration~\cite{LHCb:2008vvz}, which has reported measurements of (anti)$^3$He~\cite{LHCb:2023tlg} and (anti)hypertriton production~\cite{LHCb-CONF-2023-002}.

\subsection{Correlation measurements}
\label{subsec:correlations}

The study of strong interactions among hadrons and their role in the formation of bound states, such as light nuclei, remains an active topic in nuclear physics. The strong interaction between baryons gives rise to residual interactions~\cite{Lednicky:1981su,Braaten:2024cke,Wang:2026rcf}, an important example being the hyperon--nucleon ($Y$--$N$) interaction. Our understanding of these interactions relies largely on effective theories and scattering experiments, the latter being particularly challenging for unstable particles because of their short lifetimes. Traditionally, hadron--hadron interactions have been investigated through low-energy scattering experiments using stable and unstable beams. For nucleon--nucleon ($N$--$N$) scattering, these measurements have provided an extensive dataset, comprising approximately 8,000 data points for scattering observables and phase shifts. In contrast, direct three-to-three scattering measurements for three-hadron systems are currently not feasible because of experimental limitations. In the strange sector, the unstable nature of hyperons and strange mesons also results in limited statistics for reactions involving these particles, limiting our knowledge of the corresponding strong interactions. Consequently, the limited experimental information on hyperon--nucleon and three-hadron interactions represents an important challenge for understanding composite systems such as light (anti)hypernuclei.

Hadronic interactions can also be investigated using femtoscopy, which exploits momentum-space correlations between particles produced in high-energy collisions. Historically, the femtoscopic technique can be traced back to the Hanbury Brown and Twiss (HBT) method, originally developed in the 1950s using photon correlations to determine the angular size of stars~\cite{Brown1954}. This concept was subsequently applied to pairs of identical particles produced in elementary and heavy-ion collisions and has become a powerful tool for investigating the space-time structure of the particle-emitting source.

Momentum-space femtoscopic correlations can be studied for a pair of particles of interest and provide access to their final-state interaction (FSI) through correlations in their relative momentum. Experimentally, the correlation function is defined as

$$
C(\ks)=\xi(\ks)\otimes\frac{N_{\mathrm{same}}(\ks)}{N_{\mathrm{mixed}}(\ks)},
$$

where $\xi(\ks)$ denotes corrections for experimental effects, $N_{\mathrm{same}}(\ks)$ is the distribution of particle pairs in a given $\ks$ interval obtained by combining particles produced in the same collision, and $N_{\mathrm{mixed}}(\ks)$ is the corresponding reference distribution of uncorrelated pairs obtained by combining particles from different collisions (mixed events).

The theoretical correlation function can be expressed as~\cite{Pratt:1986cc,Lisa:2005dd}

$$
C(\ks)=\int d^3\rs\,S(\rs)\left|\psi(\vec{\ks},\vec{\rs})\right|^2,
$$

where $S(\rs)$ describes the distribution of the relative separation $\rs$ between the emitted particles and therefore characterizes the particle-emitting source. The function $\psi(\vec{\ks},\vec{\rs})$ represents the wave function of the relative motion of the particle pair, including the FSI, while $\ks$ is the reduced relative momentum of the pair,

$$
\ks=\frac{|\vec{p}^{\,*}_2-\vec{p}^{\,*}_1|}{2}.
$$

The asterisk indicates that the momenta are evaluated in the pair rest frame, where $\vec{p}^{,*}_1=-\vec{p}^{,*}_2$. For three-particle correlations, an analogous formalism can be employed using the so-called hypermomentum

$$
Q_3=\sqrt{-q_{12}^{2}+q_{23}^{2}+q_{31}^{2}},
$$

which is constructed from the relative momenta $q_{ij}$ of the three two-particle pairs~\cite{Acharya2023}.

The particle-emitting source is commonly studied through correlation measurements of pion or kaon pairs, for which quantum statistics together with the Coulomb interaction determine the characteristic shape of the correlation function. Over the last three decades, femtoscopic measurements have provided detailed information on particle emission in experiments spanning a broad range of collision energies, from SIS, AGS, and SPS to RHIC and the LHC. The extensive datasets collected in collider and fixed-target experiments have enabled studies of the three-dimensional structure and evolution of the particle-emitting source, providing information on the kinetic freeze-out of different hadron species. Typical source radii measured at RHIC and LHC energies range from approximately 2 to 8~fm.

In practice, one-dimensional femtoscopic correlation functions are frequently used to constrain hadron--hadron interactions. In this case, the particle-emitting source describes the distribution of the relative separation of the two particles at kinetic freeze-out and is commonly parameterized by a Gaussian probability density distribution characterized by a source radius $r_0$.

\subsubsection{Particle emission source}
\label{subsec:particle_emission_source}

\begin{figure}[!hbt]
\centering
\includegraphics[width=0.6\linewidth]{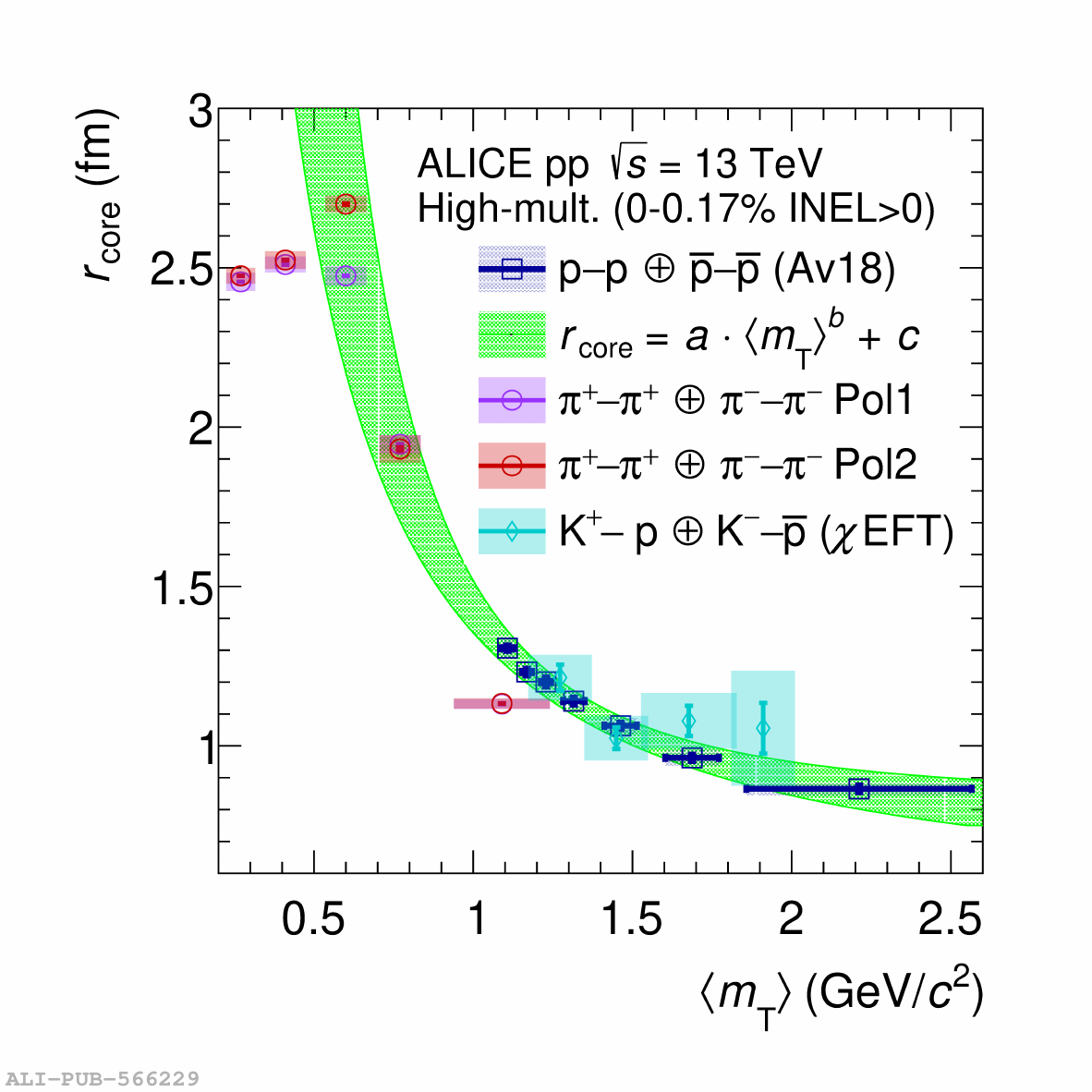}
\caption{Two-particle source radius as a function of $m_{\mathrm{T}}$ for meson--meson, meson--baryon, and baryon--baryon correlations in pp collisions at $\sqrt{s}=13$~TeV~\cite{sourceBaryonBaryon}. The green band represents a parametrization of the $m_{\mathrm{T}}$ dependence of the p--p source radius, together with its $3\sigma$ uncertainty band~\cite{ALICE:SourceMax}.}
\label{fig:alicesourcefig}
\end{figure}

Over the last decades, both the ALICE and STAR Collaborations have exploited femtoscopic correlations to constrain the particle-emitting source and investigate the underlying strong interactions. In particular, ALICE has used pp collisions at the LHC, where typical source radii of about $1$--$2$~fm provide sensitivity to the short-range strong interaction between hadrons~\cite{sourceBaryonBaryon,ALICE:SourceMax}. In pp collisions at $\sqrt{s}=13$~TeV, the source size has been constrained using several particle pairs for which the interaction is relatively well known from scattering experiments, including baryon--baryon (p--p), meson--baryon ($\mathrm{K}^{\pm}$--p), and meson--meson ($\pi^{\pm}$--$\pi^{\pm}$) pairs~\cite{sourceBaryonBaryon,ALICE:SourceMax}. The extracted source radii $r_0$, shown as a function of the transverse mass of the pair ($m_{\mathrm{T}}$) in Fig.~\ref{fig:alicesourcefig}, exhibit a common scaling behavior for the different hadron pairs.

This common $m_{\mathrm{T}}$ scaling can then be used to constrain the source size for hadron pairs whose interaction is less well known, with the resulting constraints subject to the uncertainties of the underlying measurements and scaling relation. Using this approach, the ALICE Collaboration has investigated several hadron--hadron systems, including pairs with strangeness down to $S=-3$, which were previously inaccessible to traditional scattering experiments~\cite{ALICE:nature2022}.

\subsubsection{Hyperon--nucleon interactions}
\label{subsec:hyperon_nucleon_interactions}

One of the major challenges in hadronic physics is to constrain the strong interaction involving hyperons, since relatively little experimental information is available on nucleon--hyperon ($N$--$Y$) and hyperon--hyperon ($Y$--$Y$) interactions. The existence of hypernuclei itself provides evidence for an attractive interaction between nucleons and hyperons. In the context of light (anti)hypernuclei formation, the nucleon--$\Lambda$ ($N$--$\Lambda$) interaction is particularly important for understanding systems such as the hypertriton. Scattering data for the $N$--$\Lambda$ interaction are scarce, especially in the low-momentum region. Femtoscopic measurements therefore provide an important complementary approach.

\begin{figure}[!hbt]
\centering
\includegraphics[width=0.9\linewidth]{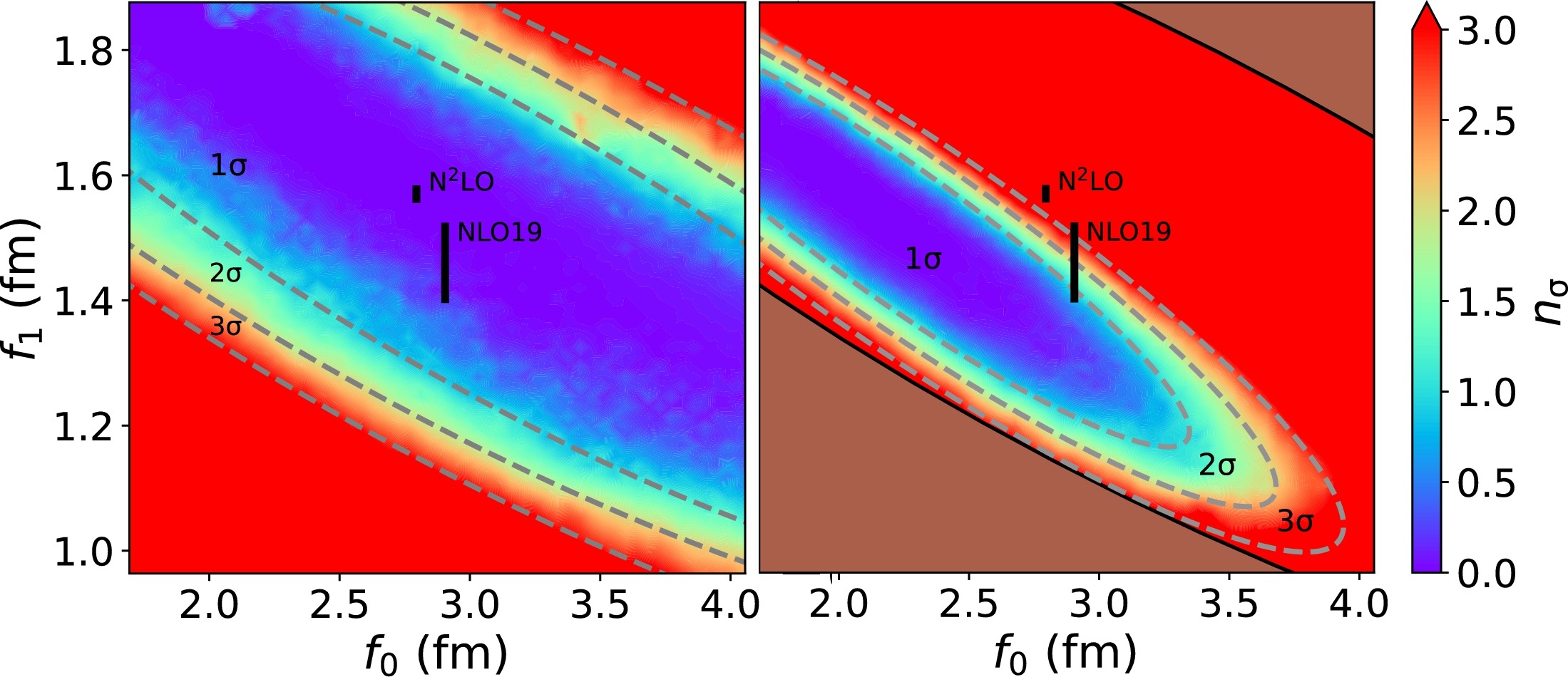}
\caption{Constraints on the singlet and triplet p--$\Lambda$ scattering lengths obtained from scattering cross-section data alone (left) and from a combined analysis of scattering and femtoscopic correlation data (right)~\cite{MIHAYLOV2024138550}.}
\label{fig:PLambdaScatteringAnalysis}
\end{figure}

Recent measurements of p--$\Lambda$ correlations in pp collisions at the LHC have reached a precision at the percent level~\cite{Acharya2022}. A combined analysis of femtoscopic correlation functions and available scattering data by Mihaylov et al.~\cite{MIHAYLOV2024138550} provided new constraints on the p--$\Lambda$ interaction. Figure~\ref{fig:PLambdaScatteringAnalysis} shows the resulting constraints on the singlet and triplet scattering lengths, $f_0$ and $f_1$, respectively.

As shown in Fig.~\ref{fig:PLambdaScatteringAnalysis}, the constraints obtained using scattering cross-section measurements alone are significantly improved when femtoscopic p--$\Lambda$ correlations in pp collisions are included. In particular, current state-of-the-art chiral effective field theory calculations at N$^2$LO, which are constrained by the available scattering data, lie outside the $3\sigma$ contour of the combined analysis~\cite{MIHAYLOV2024138550}. These measurements therefore provide new constraints for chiral EFT calculations of the $N$--$\Lambda$ interaction, with potential implications for theoretical predictions of the hypertriton wave function.

Complementary studies of d--$\Lambda$ correlations are being pursued by the STAR Collaboration~\cite{Hu2024}. Within the Lednick'y--Lyuboshitz (L--L) approach~\cite{Lednicky:1981su}, two-particle correlations can be used to constrain both the source radius and the final-state interaction parameters, namely the scattering length $f_0$ and effective range $d_0$. For non-identical particles, different spin configurations contribute to the correlation function. The p--$\Lambda$ system can be in singlet (S) or triplet (T) spin states, whereas the d--$\Lambda$ system can be in doublet (D) or quartet (Q) states, with different values of $f_0$ and $d_0$ associated with each spin channel.

STAR analyses of d--$\Lambda$ correlations have been performed~\cite{Hu2024}. In the d--$\Lambda$ analysis, contamination from hypertriton decays is explicitly taken into account. Based on the measured hypertriton yields, approximately $4$--$8\%$ of the entries at $k^*<100$~MeV/$c$ are estimated to originate from hypertriton decays. The d--$\Lambda$ correlation functions are analyzed simultaneously in different centrality classes, using common strong-interaction parameters while allowing the source radii to vary. This approach makes it possible to disentangle the particle-emitting source from the final-state interaction. The extracted d--$\Lambda$ source radius is slightly smaller than that obtained for p--$\Lambda$, consistent with the larger mean transverse mass of the d--$\Lambda$ pair.

\begin{figure}[!hbt]
    \centering
    \includegraphics[width=0.6\textwidth]
    {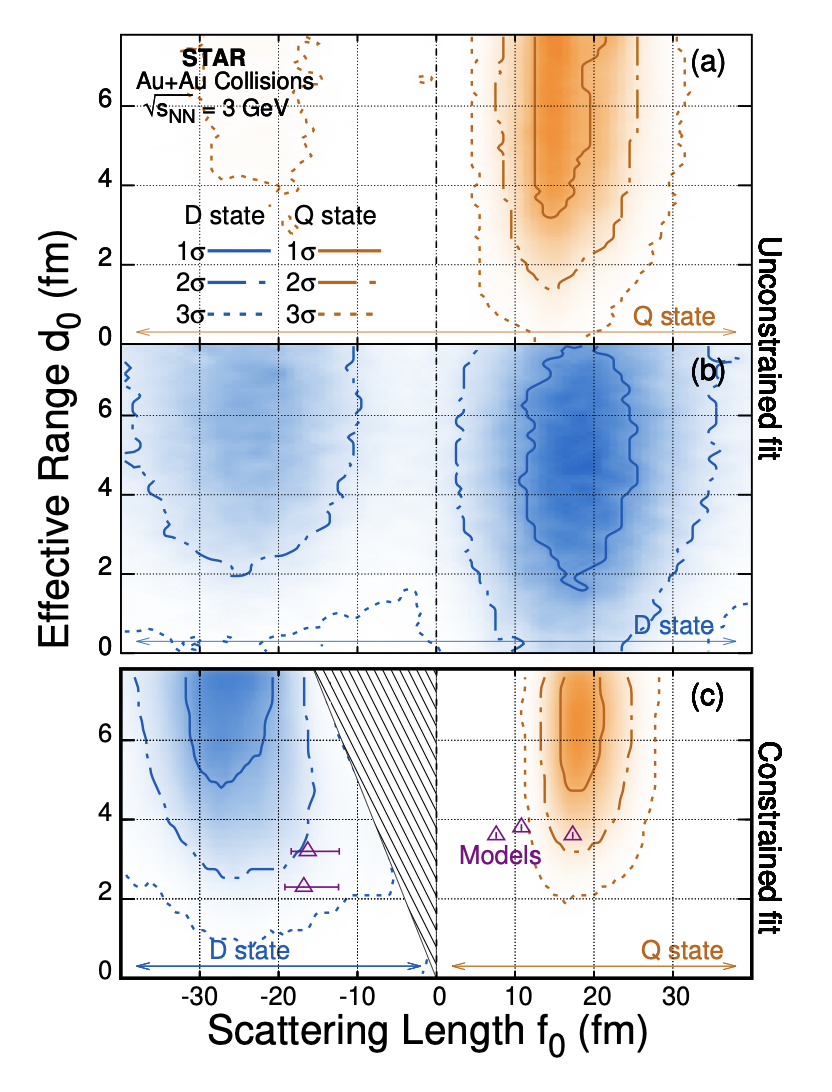}
    \caption{Constraints on the scattering length $f_0$ and effective range $d_0$ for d--$\Lambda$ systems. The contours represent the regions allowed by the femtoscopic measurements, while the markers indicate theoretical predictions for the different spin states. Figure adapted from Ref.~\cite{Hu2024}.}
    \label{fig:STARLambdaCorrelations}
\end{figure}

The corresponding constraints on $f_0$ and $d_0$ for the d--$\Lambda$ systems are shown in Fig.~\ref{fig:STARLambdaCorrelations}. The theoretical predictions for the doublet and quartet d--$\Lambda$ channels can be directly compared with the regions allowed by the femtoscopic measurements. In particular, the negative value of $f_0$ in the doublet d--$\Lambda$ channel is associated with the presence of the hypertriton bound state. Using the extracted $f_0^{\mathrm{D}}$ and $d_0^{\mathrm{D}}$ parameters together with the Bethe formula within the effective-range expansion (ERE), the $\Lambda$ separation energy of the hypertriton is estimated to be $B_{\Lambda}=0.04^{+0.12}_{-0.03}$ ~MeV at $95\%$ confidence level (constrained-fit results) and $B_{\Lambda}=0.03^{+0.25}_{-0.02}$ ~MeV at $95\%$ confidence level (unconstrained-fit results). This range is consistent with the current world average~\cite{Eckert2022}.

A similar strategy is employed for the p--$\Lambda$ correlation measurement shown in Fig.~\ref{fig:STARLambdaCorrelations}. Three centrality classes are analyzed simultaneously, assuming common strong-interaction parameters but different source radii. Since the scattering lengths in the singlet and triplet channels are expected to be similar, a spin-averaged description is employed, yielding $f_0=2.32^{+0.12}_{-0.11}$~fm and $d_0=3.5^{+2.7}_{-1.3}$~fm.

\subsubsection{Hadron--deuteron and deuteron--deuteron correlations}
\label{subsec:hadron_deuteron_correlations}

In recent years, two-particle hadron--deuteron correlation measurements by the ALICE and STAR Collaborations have provided new insights into the formation of light (anti)nuclei.

\begin{figure}[!hbt]
\centering
\includegraphics[width=0.9\linewidth]{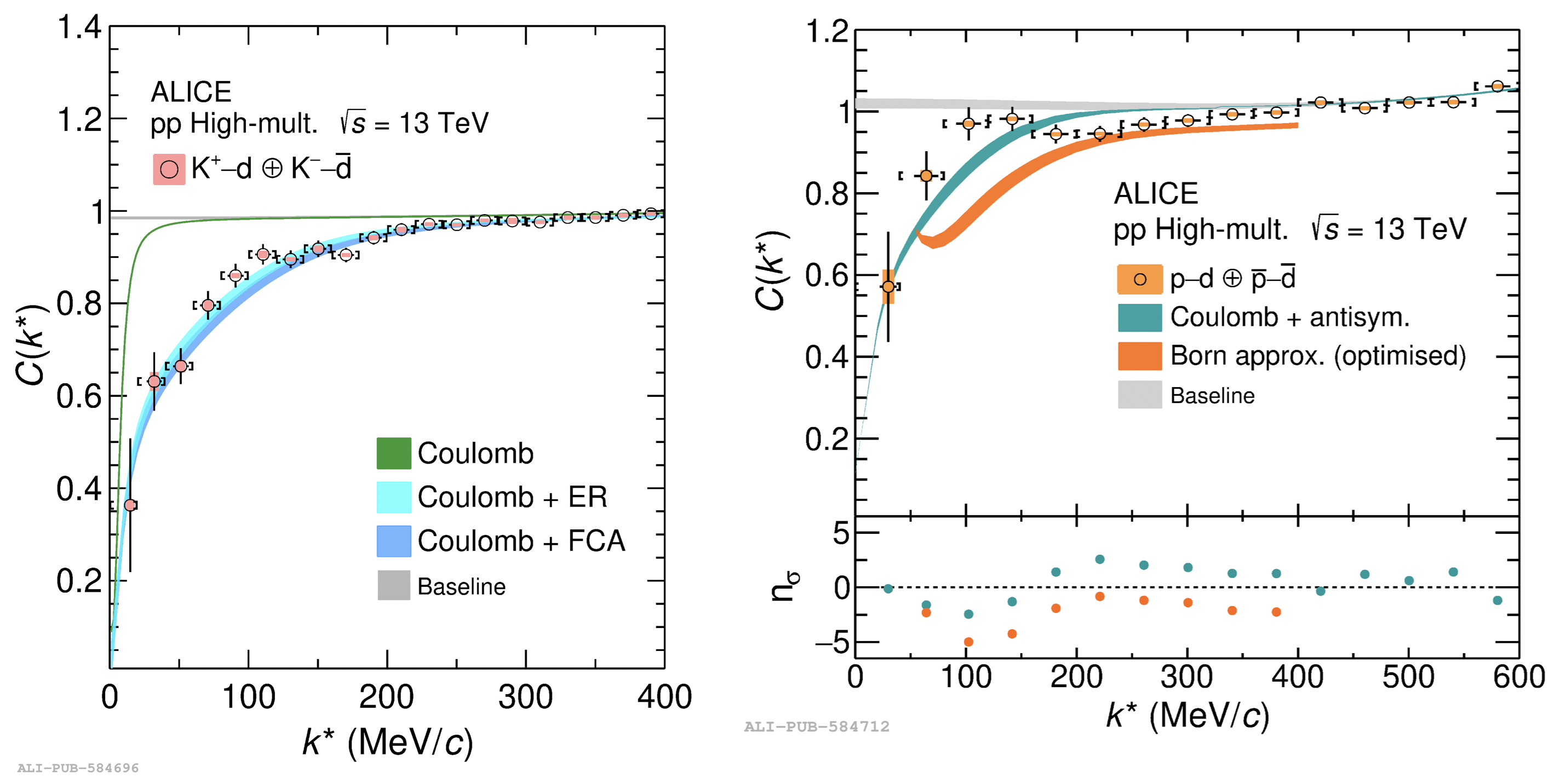}
\caption{Measured K--d (left) and p--d (right) correlation functions in high-multiplicity pp collisions at $\sqrt{s}=13$~TeV at the LHC, compared with theoretical calculations~\cite{pdPaper}.}
\label{fig:CFkdpd}
\end{figure}

Measurements of K--d and p--d correlations by ALICE in high-multiplicity pp collisions at $\sqrt{s}=13$~TeV at the LHC provide important information on deuteron formation in small collision systems~\cite{pdPaper}. The measured K--d correlation function can be described by effective two-body calculations based on the Lednick'y--Lyuboshitz model, in which the deuteron is treated as a point-like particle. In this approach, the emission source size is constrained using the $m_{\mathrm{T}}$ scaling observed for other hadron pairs and the average transverse mass of the K--d pairs. The resulting effective source radius is $r_{\mathrm{eff}}^{K^+\text{--}d}=1.35^{+0.04}_{-0.05}$~fm, corresponding to a most probable relative distance between the kaon and deuteron of about 2~fm. The interaction parameters entering the K--d calculation are constrained by both K--N interactions and available K--d scattering data. The comparison between the measured K--d correlation function and the effective two-body calculation is shown in the left panel of Fig.~\ref{fig:CFkdpd}.

An important observation of this analysis is that deuterons follow the same $m_{\mathrm{T}}$ scaling of the source radius observed for other hadrons in pp collisions. Furthermore, the K--d correlation measurement indicates that deuterons are produced at characteristic distances of about 2--3~fm from kaons. In contrast, the measured p--d correlation function cannot be described by an effective two-body calculation using source sizes constrained by femtoscopy and interaction parameters obtained from p--d scattering data~\cite{pdPaper}. This discrepancy points to the importance of the internal structure of the deuteron in the p--d system, in particular Pauli blocking between the external proton and the proton inside the deuteron, as well as short-range strong-interaction effects.

The difference between the two systems can be expressed explicitly in terms of their constituent structure. For K--d correlations, the relevant system can be written as

$$
K + d(pn) \longrightarrow K+d,
$$

or, at the constituent level,

$$
K+(p+n)\longrightarrow K+d.
$$

Since the kaon is distinguishable from both nucleons forming the deuteron, the deuteron formation probability can approximately factorize from the K--d interaction, allowing an effective two-body description.

For p--d correlations, instead, the underlying system is

$$
p+d(pn)\longrightarrow p+d,
$$

corresponding at the constituent level to

$$
p+(p+n)\longrightarrow p+d.
$$

In this case, the external proton is identical to the proton that eventually becomes part of the deuteron. The full wave function must therefore account for the exchange of the two protons, together with Pauli blocking and short-range nucleon--nucleon interactions. A genuine three-body description of the underlying p--p--n system is consequently required. In such an approach, the three nucleons are initially produced as a p--p--n system and dynamically evolve toward the asymptotic p--d final state~\cite{theorypd}. This picture follows the description proposed by S.~Mrówczyński, in which the deuteron is not assumed to exist as a fully formed object at the initial stage but emerges dynamically from the constituent nucleons~\cite{coalescence6}.

Complementary information is provided by femtoscopic measurements in heavy-ion collisions at lower energies, where the enhanced production of light nuclei at high baryon density provides large samples for correlation studies. The STAR Collaboration has performed the first measurements of p--d and d--d correlation functions in different centrality classes at midrapidity in Au--Au collisions at $\sqrt{s_{\mathrm{NN}}}=3$~GeV. The measured correlation functions exhibit a characteristic depletion at low $k^*$, reflecting the repulsive final-state interactions in these systems.

To investigate the sensitivity of these measurements to the deuteron production mechanism, the data are compared with calculations based on the Simulating Many Accelerated Strongly-interacting Hadrons (SMASH) transport model~\cite{smash}. The calculations are performed in cascade mode, in which particles propagate freely between successive interactions. Two scenarios for deuteron production are considered. In the first, deuterons are produced dynamically through hadronic reactions, such as

$$
p+n+\pi \leftrightarrow d+\pi.
$$

In the second, protons and neutrons are produced during the system evolution and subsequently form deuterons through a coalescence prescription~\cite{Wigner}.

\begin{figure}[!hbt]
\centering
\includegraphics[width=\linewidth]{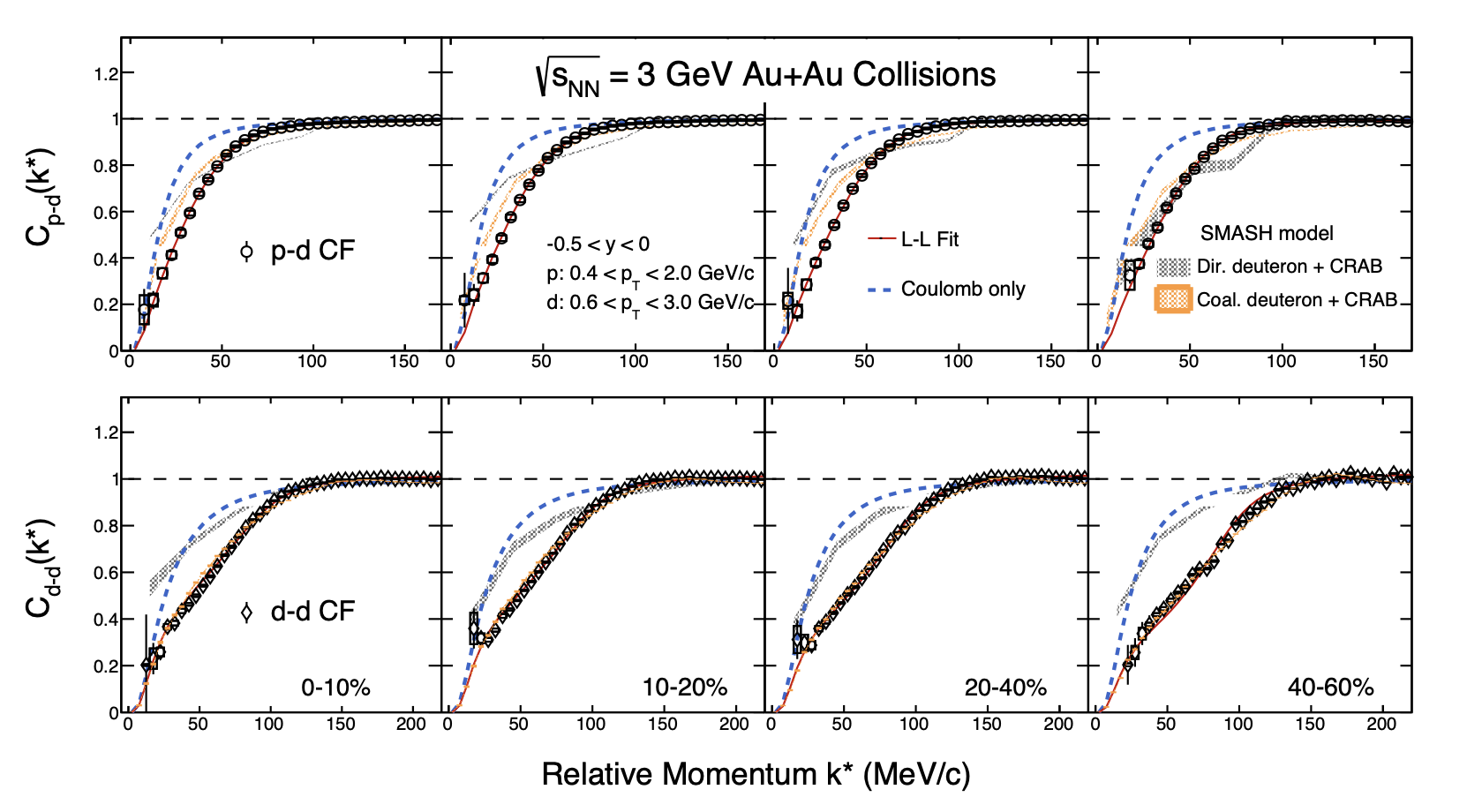}
\caption{Proton--deuteron (top) and deuteron--deuteron (bottom) correlation functions measured in different centrality classes of Au--Au collisions at $\sqrt{s_{\mathrm{NN}}}=3$~GeV. The colored bands represent SMASH+CRAB predictions for scenarios in which deuterons are produced through hadronic reactions or through nucleon coalescence~\cite{STAR2025LightNucleiFemtoscopy}.}
\label{fig:pp-dd}
\end{figure}

Since SMASH does not intrinsically include femtoscopic two-particle correlations, the correlation functions are calculated using the Correlation Afterburner (CRAB) package~\cite{crab}. For d--d correlations, both quantum-statistical effects and final-state interactions, including Coulomb and strong interactions, are taken into account, whereas for p--d correlations only final-state interactions contribute. The resulting predictions for the two deuteron-production scenarios are compared with the measurements in Fig.~\ref{fig:pp-dd}.

The measured correlation functions are better reproduced by the calculations in which deuterons are formed through nucleon coalescence. In contrast, the scenario based on direct deuteron production through hadronic reactions reproduces only qualitatively the overall behavior of the data. These results therefore provide additional evidence that the internal structure and dynamical formation of the deuteron play an important role in femtoscopic correlations and favor a description in which deuterons emerge from the coalescence of nucleons.

\section{The statistical hadronization model}
\label{sec:theory_overview}

The statistical hadronization model (SHM) provides a successful framework for describing the production of hadrons and light (hyper)nuclei in heavy-ion collisions over a broad range of collision energies~\cite{Reichert:2022mek}. In this approach, cluster abundances are determined by the thermodynamic conditions at chemical freeze-out, treating light nuclei and hypernuclei on the same footing as other hadronic species. The main ingredients of the statistical hadronization model and its application to cluster production are discussed in the following sections.

\subsection{Grand-canonical SHM}
\label{subsec:thermal_grandcanonical}

In the grand-canonical formulation of SHM, the system produced in heavy-ion collisions is described at chemical freeze-out as a hadron-resonance gas in thermal and chemical equilibrium~\cite{Andronic:2017pug,Andronic:2026cdu}. Its thermodynamic state is characterized by the chemical freeze-out temperature $T_{\rm CF}$, the volume $V$, and the chemical potentials associated with the conserved quantum numbers. The abundances of the different hadron species are then determined by these thermodynamic parameters and by the corresponding particle properties.

Chemical freeze-out denotes the stage of the collision evolution at which inelastic interactions cease and the primordial abundances of hadron species are fixed. Subsequent changes in the measured particle yields arise from the decays of unstable hadrons and resonances. The thermodynamic parameters at chemical freeze-out can therefore be determined by comparing the particle yields predicted by the SHM with experimental measurements.

At high collision energies, where the baryochemical potential $\mub$ is small, the chemical freeze-out temperature is found to be close to the pseudocritical temperature ($T_c$) of the QCD crossover. At vanishing $\mub$, Lattice QCD (LQCD) calculations yield $T_c = 156.5\pm1.5$~MeV~\cite{Bazavov:2018mes} and $T_c = 158.0\pm0.6$~MeV~\cite{Borsanyi:2020fev}. Recent LQCD calculations also indicate a mild decrease of $T_c$ with increasing $\mub$ for $\mub \lesssim 300$~MeV~\cite{Bonati:2018nut,Bazavov:2018mes,Borsanyi:2020fev}. Within this range, the transition remains a crossover~\cite{Aoki:2006we}.

Below the QCD transition temperature, the relevant degrees of freedom are hadrons rather than quarks and gluons. In this regime, QCD thermodynamics can be successfully described within the hadron resonance gas (HRG) framework, in which the interacting hadronic system is approximated as a gas of hadrons and resonances. In particular, the QCD partition function is well reproduced by the HRG model for temperatures below $T_c$ (see, e.g., \cite{Bazavov:2017dus} and references therein).

The grand-canonical partition function for hadron species $i$ is given by

\begin{equation}
\ln Z_i =
\frac{V g_i}{2\pi^2}
\int_0^\infty
\pm p^2 \mathrm{d}p
\ln\left[1 \pm \exp\left(-\frac{E_i-\mu_i}{T}\right)\right]^{\pm 1},
\end{equation}

where the $+$ sign applies to fermions and the $-$ sign to bosons. Here, $V$ is the system volume, $g_i=2J_i+1$ is the spin degeneracy factor, $T$ is the temperature, and $E_i=\sqrt{p^2+m_i^2}$ is the energy of a particle of mass $m_i$ and momentum $p$.

The chemical potential of species $i$ is determined by its conserved quantum numbers,
\begin{equation}
\mu_i = \mu_B B_i + \mu_{I_3} I_{3i} + \mu_S S_i + \mu_C C_i,
\end{equation}
where $\mu_B$, $\mu_{I_3}$, $\mu_S$, and $\mu_C$ are the chemical potentials associated with baryon number, isospin, strangeness, and charm, respectively. Their values are constrained by the corresponding conserved quantum numbers of the colliding system. In particular, the relative baryon and isospin content is fixed by that of the initial nuclei, while overall strangeness and charm neutrality require
\begin{equation}
\sum_i n_i S_i = 0,
\qquad
\sum_i n_i C_i = 0.
\end{equation}

The calculation requires knowledge of the hadronic spectrum, which is conventionally taken from the states listed by the PDG. Resonances play an important role in the HRG description: their decays contribute to the final particle yields through feed-down, while their inclusion in the partition function effectively accounts for attractive hadronic interactions.

Excited nuclear states can be incorporated in the same framework. In particular, excited states of $^4$He and their decay contributions have been considered in Refs.~\cite{Hahn:1986mb,Shuryak:2019ikv}. Their impact on the predicted light-nuclei yields is expected to be most relevant at low collision energies, corresponding to large \mub~\cite{Vovchenko:2020dmv}.

Repulsive interactions were traditionally modeled using an ``excluded-volume'' prescription~\cite{Rischke:1991ke}, in which a common excluded-volume radius $r_0 \leq 0.3$~fm is assigned to all particle species, including light nuclei~\cite{Andronic:2005yp}. For such values of $r_0$, the correction reduces the particle densities, while the extracted thermal parameters $T$ and \mub are only weakly affected.
More elaborate prescriptions, such as bag-model mass scaling or the assignment of different radii to mesons and baryons, can significantly modify hadron-yield ratios and lead to extracted chemical freeze-out temperatures well above $T_c$~\cite{Vovchenko:2015cbk,Vovchenko:2016ebv}. The phenomenological excluded-volume approach should therefore be applied with care, particularly when the resulting corrections are large.
A more systematic treatment of hadronic interactions is provided by the S-matrix formulation of statistical mechanics, which incorporates experimentally constrained scattering information and, importantly, includes both resonant and non-resonant contributions~\cite{Andronic:2018qqt}. In its current implementation, this approach is restricted to meson--meson and meson--baryon interactions in the strange and non-strange sectors~\cite{Fernandez-Ramirez:2018vzu}, while contributions from multi-pion--nucleon interactions are estimated using lattice-QCD calculations~\cite{Lo:2017lym}.
The excluded-volume approach can nevertheless be used to model non-resonant repulsive interactions in the baryon--baryon sector~\cite{Satarov:2016peb}. At $\mub=0$, comparisons with lattice-QCD susceptibilities constrain the corresponding excluded-volume parameter to approximately $b=1$~fm$^3$~\cite{Vovchenko:2017xad,Karthein:2021cmb}.
The treatment of interactions involving light nuclei within the HRG framework is considerably less developed. A recent study based on the Beth--Uhlenbeck approach represents a first step toward incorporating such interactions more systematically~\cite{Donigus:2022haq}.

\begin{figure}[!hbt]
\centering
\includegraphics[width=0.8\linewidth]{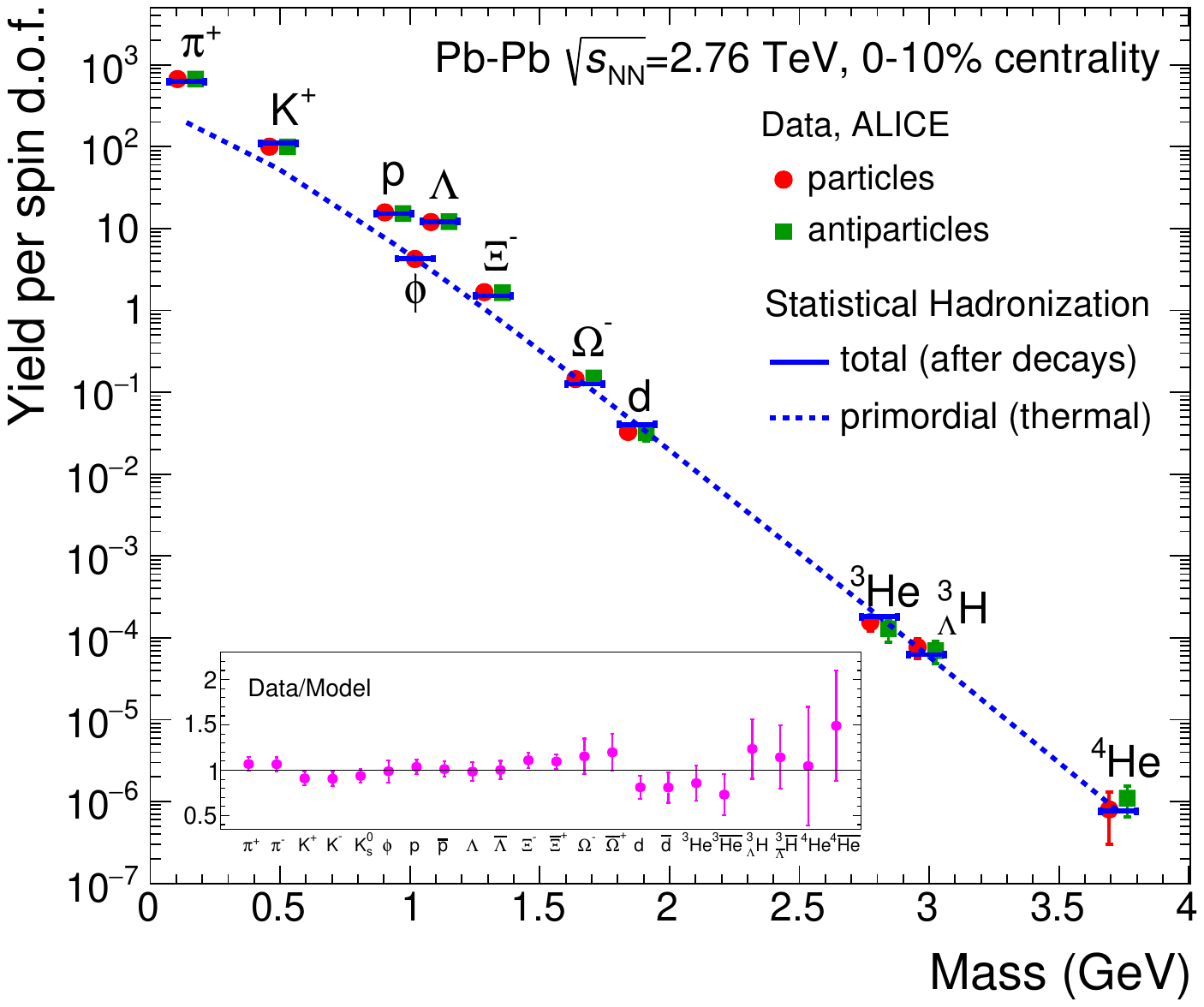}
\caption{Hadron yields $\ud N/\ud y$ measured in central Pb--Pb collisions at the LHC, compared with the best-fit SHM predictions~\cite{Andronic:2017pug}. The yields are divided by the spin degeneracy factor $(2J+1)$ and presented as a function of particle mass. The SHM ``total'' yields include feed-down contributions from resonance decays, whereas the ``primordial'' yields correspond to those before strong and electromagnetic decays. For the $\Lambda$ hyperon, the total yield also includes the contribution from the electromagnetic decay $\Sigma^0\rightarrow\Lambda\gamma$, which cannot be experimentally resolved. The inset shows the ratio of the measured yields to the model predictions; statistical and systematic uncertainties of the data are added in quadrature.}
\label{fig:Fit}
\end{figure}

In practice, the chemical freeze-out parameters $T_{CF}$, \mub, and $V$ are determined by fitting the particle yields measured experimentally. For the 0--10\% most central Pb--Pb collisions at $\sqrt{s_{\rm NN}}=2.76$~TeV, a fit to the ALICE particle yields measured at midrapidity~\cite{Acharya:2017bso} gives $T_{CF}=156.6\pm1.7$~MeV, $\mu_B=0.7\pm3.8$~MeV, and $V=4175\pm380$~fm$^3$~\cite{Andronic:2018qqt}, as shown in Fig.~\ref{fig:Fit}. The volume refers to a rapidity interval of one unit centered at midrapidity. The quoted uncertainties reflect the experimental uncertainties and do not include systematic uncertainties associated with the model implementation.  With more recent ALICE data an order of magnitude improvement in the precision of $\mub =0.7\pm0.45$ MeV was demonstrated~\cite{ALICE:2023ulv}, implying that the central region at LHC energies is essentially baryon-free.

Consistent results are obtained with different SHM implementations~\cite{Wheaton:2004qb,Torrieri:2004zz,Vovchenko:2019pjl}, provided that the same particle lists, input data, and model assumptions are used~\cite{ALICE:2022wpn}.

The SHM provides a very good description of the measured particle yields over approximately nine orders of magnitude in abundance, including strange and non-strange mesons, baryons, multi-strange hyperons, as well as light nuclei, hypernuclei, and their antiparticles. The previously observed over-prediction of the proton and antiproton yields, corresponding to a deviation of $2.7\sigma$, is resolved when hadronic interactions are treated within the S-matrix approach~\cite{Andronic:2018qqt}. In this implementation, the excluded-volume correction is not applied.

The inclusion of approximately 500 additional hadronic states predicted by LQCD and quark models, beyond those listed by the PDG, has been shown to worsen the quality of the fit (see, however, Ref.~\cite{Alba:2020jir}). This effect is no longer observed when the S-matrix treatment of hadronic interactions is employed~\cite{Andronic:2020iyg}, a further indication that the S-matrix approach is a consistent one.

As evident in Fig.~\ref{fig:Fit}, the yields of the lightest measured mesons and baryons ($\pi$, $K$, p, and $\Lambda$) receive substantial feed-down contributions from resonance decays; for pions, for example, these contributions account for about 70\% of the total yield.

In contrast, feed-down contributions to light-nucleus yields are generally much smaller, although contributions from excited nuclear states may become relevant under specific conditions, as discussed above. Owing to their large masses, the thermal yields of light nuclei depend strongly on the temperature. Consequently, their measured abundances provide a sensitive constraint on the chemical freeze-out temperature. At $\sqrt{s_{\rm NN}}=2.76$~TeV, a fit to the light-nucleus yields gives $T_{\rm nuclei}=159\pm5$~MeV, consistent with the value of $T_{CF}$ obtained from the global fit to hadron yields.

A similar analysis using the more precise Run 2 data from Pb--Pb collisions at $\sqrt{s_{\rm NN}}=5.02$~TeV~\cite{ALICE:2022veq,ALICE:2023qyl} yields a lower temperature, $T_{\rm nuclei}=147\pm2$~MeV, with $\chi^2/\mathrm{dof}=12.9/3$. The relatively poor fit quality is driven primarily by the lower-than-expected yields of the $A=3$ nuclei ($^3$He and t), indicating that their yields cannot be simultaneously described with those of the lighter nuclei using a single thermal temperature. A similar suppression of $A=3$ nuclei relative to thermal-model expectations is also observed at lower collision energies at RHIC.

The rapidity densities of light (anti)nuclei and hypernuclei at LHC energies were predicted within the SHM based on the systematics of hadron production measured at lower collision energies~\cite{Andronic:2010qu}. The successful description of these loosely bound states in SHM is particularly striking given the large difference between their binding energies and the chemical freeze-out temperature. For example, the deuteron binding energy is only about 2.2~MeV, compared with $T_{CF}\approx T_c\approx157$~MeV~\cite{BraunMunzinger:1994iq}.

An important aspect in interpreting this apparent hierarchy of energy scales is the formation time of the bound state, which was at the center of extensive discussions within the RRTF. Owing to their small binding energies, loosely bound nuclei are expected to have formation times that can exceed the characteristic lifetime of the evolving hadronic system. The physical bound state may therefore form only after its constituents have escaped the dense interacting medium, avoiding the apparent difficulty of a fragile object surviving at temperatures far above its binding energy.

While the abundances of light nuclei are well described by the SHM, their microscopic formation mechanism remains an open question (see, e.g., Refs.~\cite{Braun-Munzinger:2018hat,Donigus:2020ctf}), particularly in relation to the assumptions underlying the thermal description~\cite{Cai:2019jtk}. One possibility, proposed already in Ref.~\cite{Chapline:1978kg}, is that the relevant quantum numbers are fixed at QGP hadronization in compact, colorless configurations that subsequently evolve into the observed nuclear bound states.

\begin{figure}[!hbt]
\centering
\includegraphics[width=0.6\linewidth]{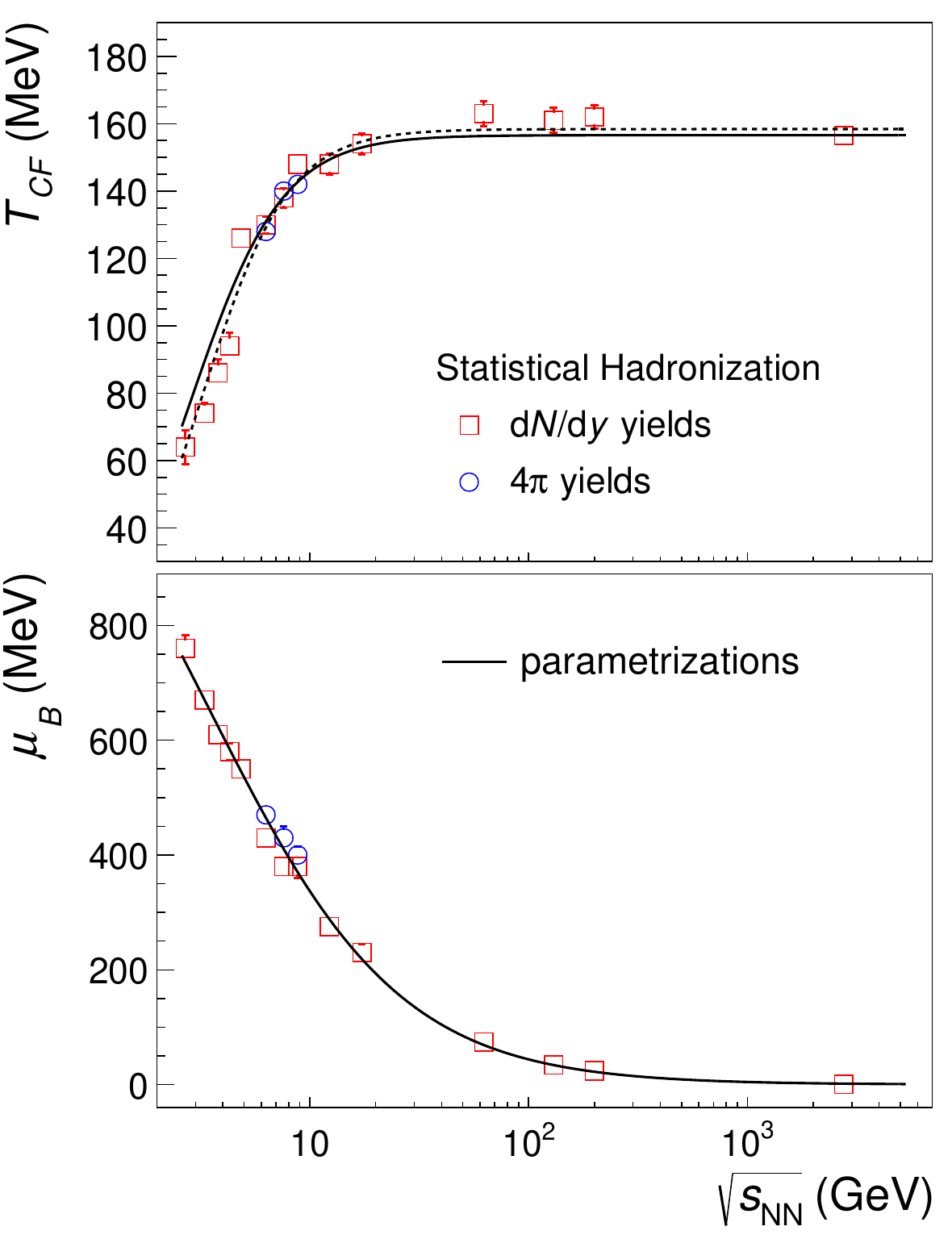}
\caption{Energy dependence of the chemical freeze-out parameters $T_{CF}$ and \mub. The results are obtained from SHM analyses of hadron yields (at midrapidity, \dndy, and in full phase space, $4\pi$) for central collisions at different collision energies. The parametrization for the temperature is slightly updated for the low-energy domain to account for the current strangeness canonical treatment, see~\cite{Andronic:2026cdu}; the previous parametrization~\cite{Andronic:2017pug} is shown for comparison as the dashed line.}
\label{fig:edep}
\end{figure}

The description of particle production in SHM is successful not only at LHC energies but also over the broad range of lower collision energies explored at RHIC, SPS, and AGS. As the collision energy decreases, the net baryon density becomes increasingly large, leading to non-zero values of the baryochemical potential and to growing differences between particle and antiparticle yields.

Furthermore, in central collisions at energies below $\sqrt{s_{\rm NN}}\approx6$~GeV, strange-particle production becomes sufficiently rare that the average multiplicity of strange hadrons can fall significantly below unity. In this regime, strangeness conservation must be enforced exactly, requiring a canonical rather than grand-canonical treatment of this quantum number~\cite{Hagedorn:1984uy,Hamieh:2000tk}. Similar considerations apply to smaller systems, such as peripheral nucleus--nucleus and elementary collisions, where the small number of strange particles can likewise make canonical effects important.

While \mub decreases smoothly with increasing collision energy, the energy dependence of $T_{CF}$ exhibits a characteristic saturation behavior, as illustrated in Fig.~\ref{fig:edep}. The chemical freeze-out temperature increases from about 50~MeV at low collision energies, reaching a plateau for $\sqrt{s_{\rm NN}}\gtrsim20$~GeV. 
The saturation of $T_{CF}$ at high collision energies supports the interpretation that chemical freeze-out occurs close to the QCD hadronization temperature~\cite{BraunMunzinger:1998cg,Stock:1999hm,BraunMunzinger:2003zz,Andronic:2008gu}, implying a close connection between experimentally extracted chemical freeze-out conditions in central heavy-ion collisions and the QCD phase boundary. 
Hagedorn noted in 1965~\cite{Hagedorn:1965st} that hadronic matter cannot be heated beyond a certain limit due to the divergence of the number of states, but the saturation observed here indicates a different boundary, i.e. the chiral phase transition.

The collision-energy dependence of the chemical freeze-out parameters shown in Fig.~\ref{fig:edep} is described by the empirical parametrizations
\begin{equation}
T_{CF} =
\frac{T_{CF}^{\rm lim}}
{1+\exp\left(2.20-\ln\left(\sqrt{s_{\rm NN}}\right)/0.48\right)},
\end{equation}
and
\begin{equation}
\mu_B =
\frac{a}{1+0.288\sqrt{s_{\rm NN}}},
\end{equation}
where $\sqrt{s_{\rm NN}}$ is in GeV. The 'limiting temperature' $T_{CF}^{lim}=156.6\pm 1.7$ MeV is fixed to the temperature obtained from the fit of the LHC data and $a=1307.5$ MeV.

\begin{figure}[htb]
\begin{tabular}{cc}  \begin{minipage}{.48\textwidth}
\includegraphics[width=.95\textwidth,height=.85\textwidth]{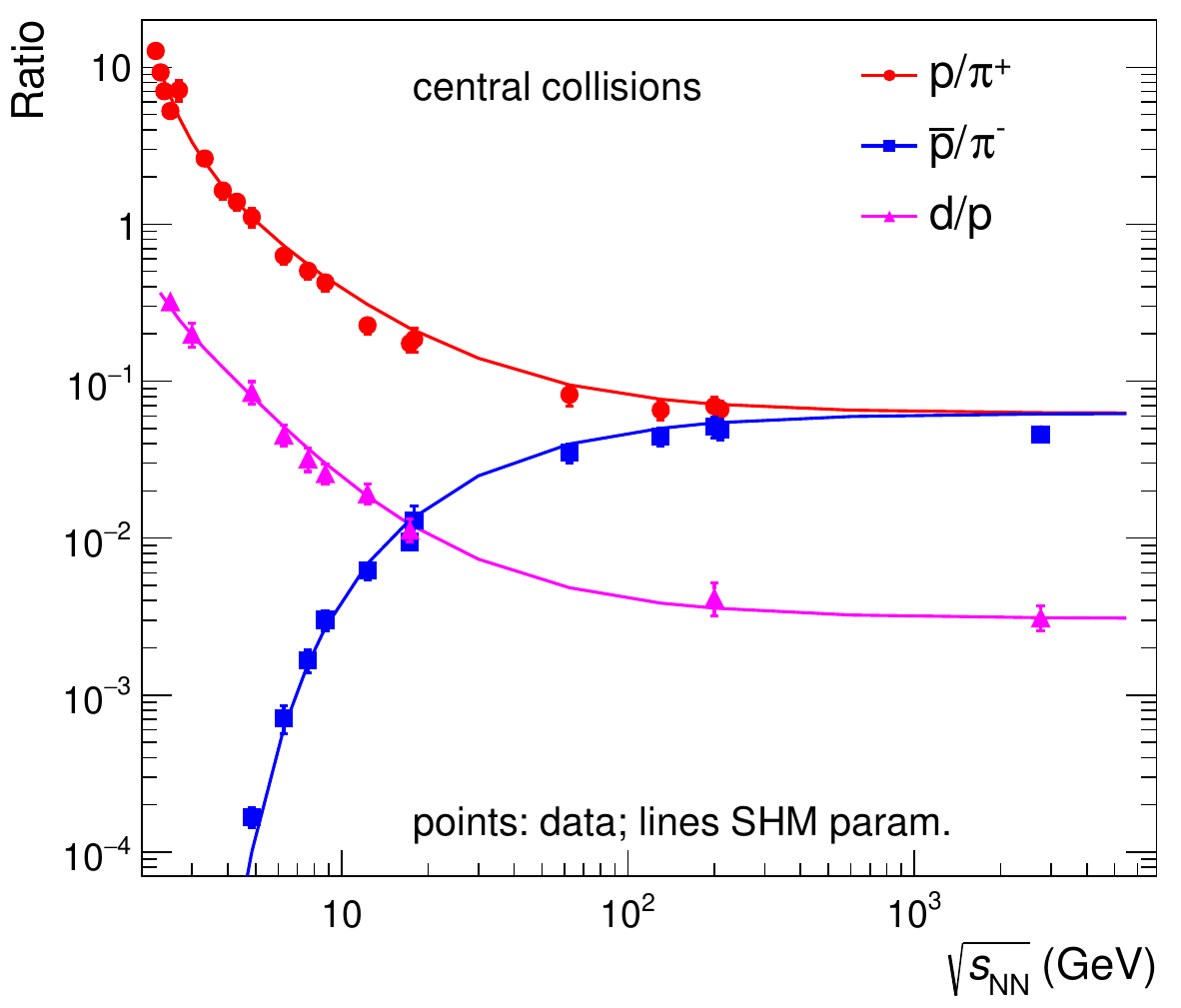} 
\end{minipage} & \begin{minipage}{.48\textwidth}
\includegraphics[width=.95\textwidth,height=.85\textwidth]{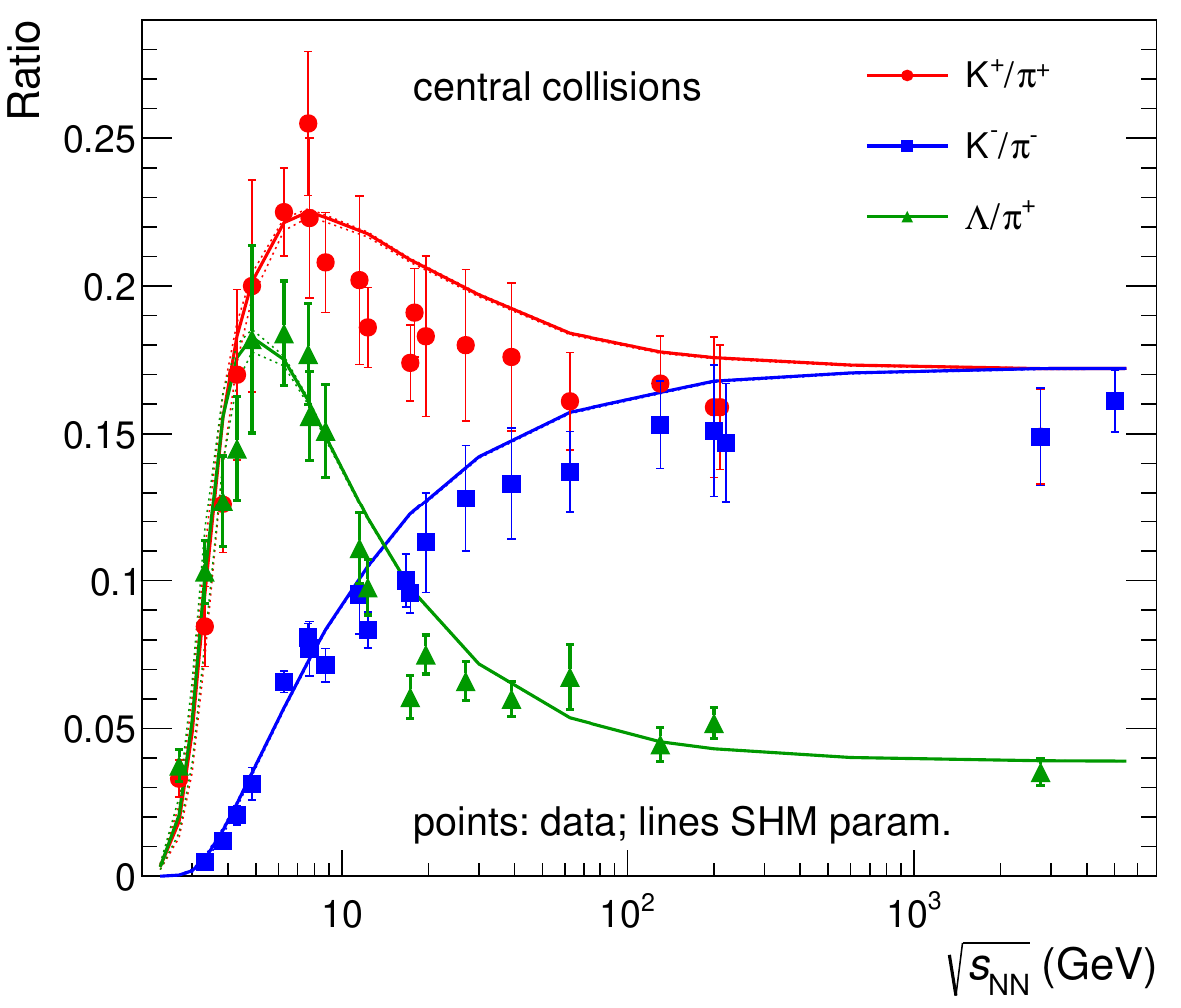}
\end{minipage} 
\end{tabular}
\caption{Collision energy dependence of the relative abundance of selected hadron species. The data, compiled in ref.~\cite{Andronic:2026cdu}, with the error bars represent the total statistical and systematic uncertainties, are compared to SHM predictions for the parametrized dependence of $T_{CF}$ and $\mu_B$ with energy. The dotted band, visible at the lower energies, quantifies the $\pm30$\% uncertainty of the strangeness-canonical volume $V_{cs}$.}
\label{fig:edep-ratios}
\end{figure}

The quality of the thermal description over a broad range of collision energies is illustrated in Fig.~\ref{fig:edep-ratios}, which shows the excitation functions of selected particle-yield ratios together with the corresponding SHM predictions obtained using the parametrized energy dependence of the chemical freeze-out parameters. In particular, the maxima observed in the $K^+/\pi^+$ and $\Lambda/\pi^+$ ratios, which occur at slightly different collision energies, arise within the SHM from the interplay between the energy dependence of $T_{CF}$ and $\mu_B$ and the constraints imposed by strangeness conservation~\cite{Andronic:2008gu}.

\begin{figure}[htb]
%\hspace{.4cm} 
\includegraphics[width=.48\textwidth]{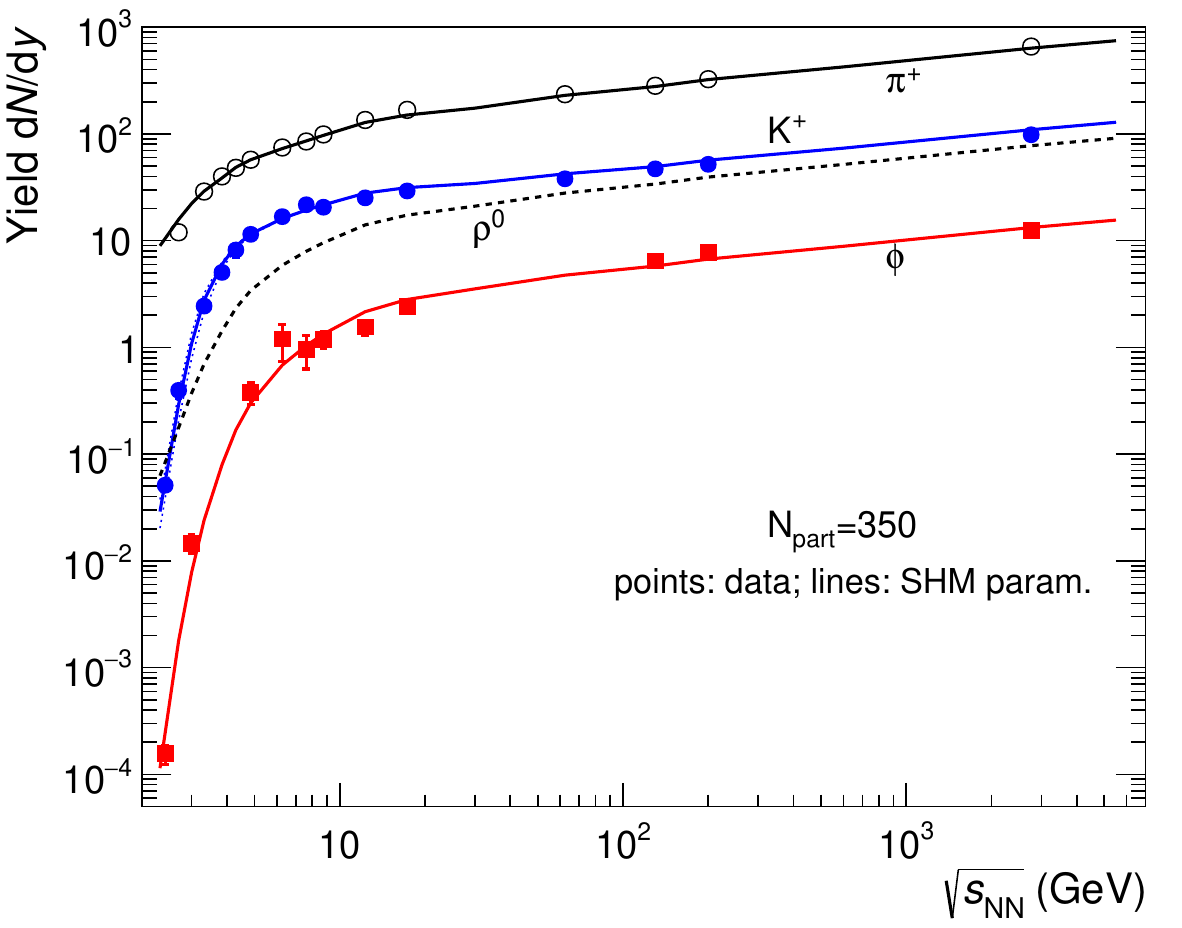} \hspace{.3cm} 
\includegraphics[width=.48\textwidth]
{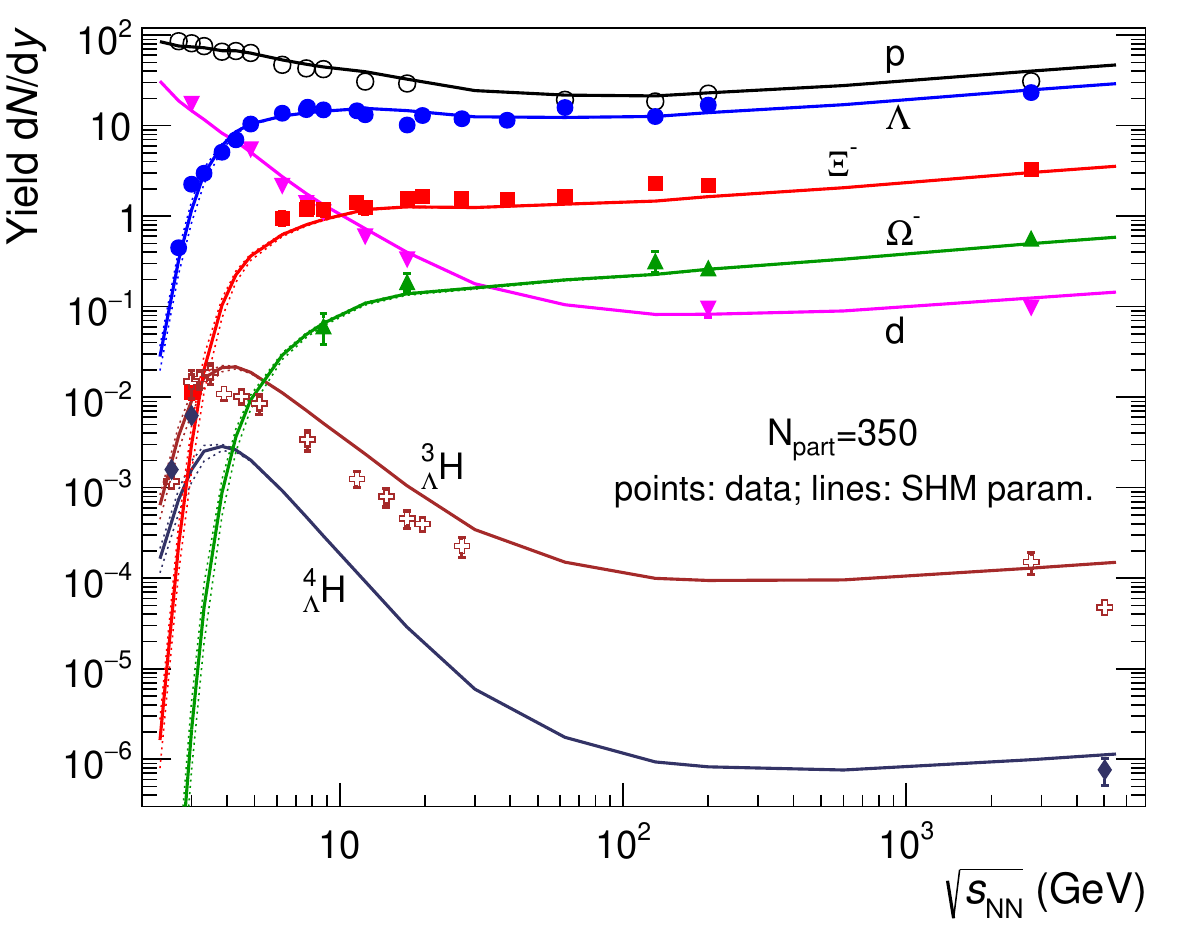}
\caption{As Fig.~\ref{fig:edep-ratios}, but for yields of selected mesons (left) and baryon and hypernuclei (right) species. Note the performed $N_{part}$-scaling of the data to correspond to $N_{part}=350$. In the left panel we show for comparison the predicted yields of the $\rho^0$ mesons.} \label{fig:edep-yields}
\end{figure}

The remarkable description of the absolute yields seen in Fig.~\ref{fig:edep-yields} implies a universal hadron production mechanism at chemical freeze-out, despite vastly different fireball properties at the lowest and highest ends of the collision range.
The production of hypernuclei, which is significantly enhanced at lower energies as a result of the large baryo-chemical potential, shows interesting features. The recent hypertriton data from HADES in Ag-Ag collisions at 2.55 GeV~\cite{HADES:2026hypernuclei}, STAR in central Au-Au collisions over a broad range of energies in the BES program~\cite{STAR:2026khp}, and ALICE in central Pb-Pb collisions at the LHC~\cite{ALICE:2015oer,ALICE:2024koa} are described by the model with a mixed degree of success, good at the lowest energies and overestimated otherwise.
The $^4_\Lambda$H hypernucleus data~\cite{STAR:2021orx,ALICE:2024djx,HADES:2026hypernuclei} is underpredicted at low energies and well described at the LHC. This lack of consistency between data and model description is under scrutiny. We note in this context that the efficiency for the detection of loosely bound states such as the hyper-triton is notoriously difficult to determine with precision.

We note that the S-matrix correction, which at the LHC energies leads to a decrease of the proton yield in the SHM by about 17\%, is not applied for the curves in Fig.~\ref{fig:edep-ratios} and Fig.~\ref{fig:edep-yields}.

\begin{figure}[!hbt]
\centering
\includegraphics[width=0.65\linewidth]{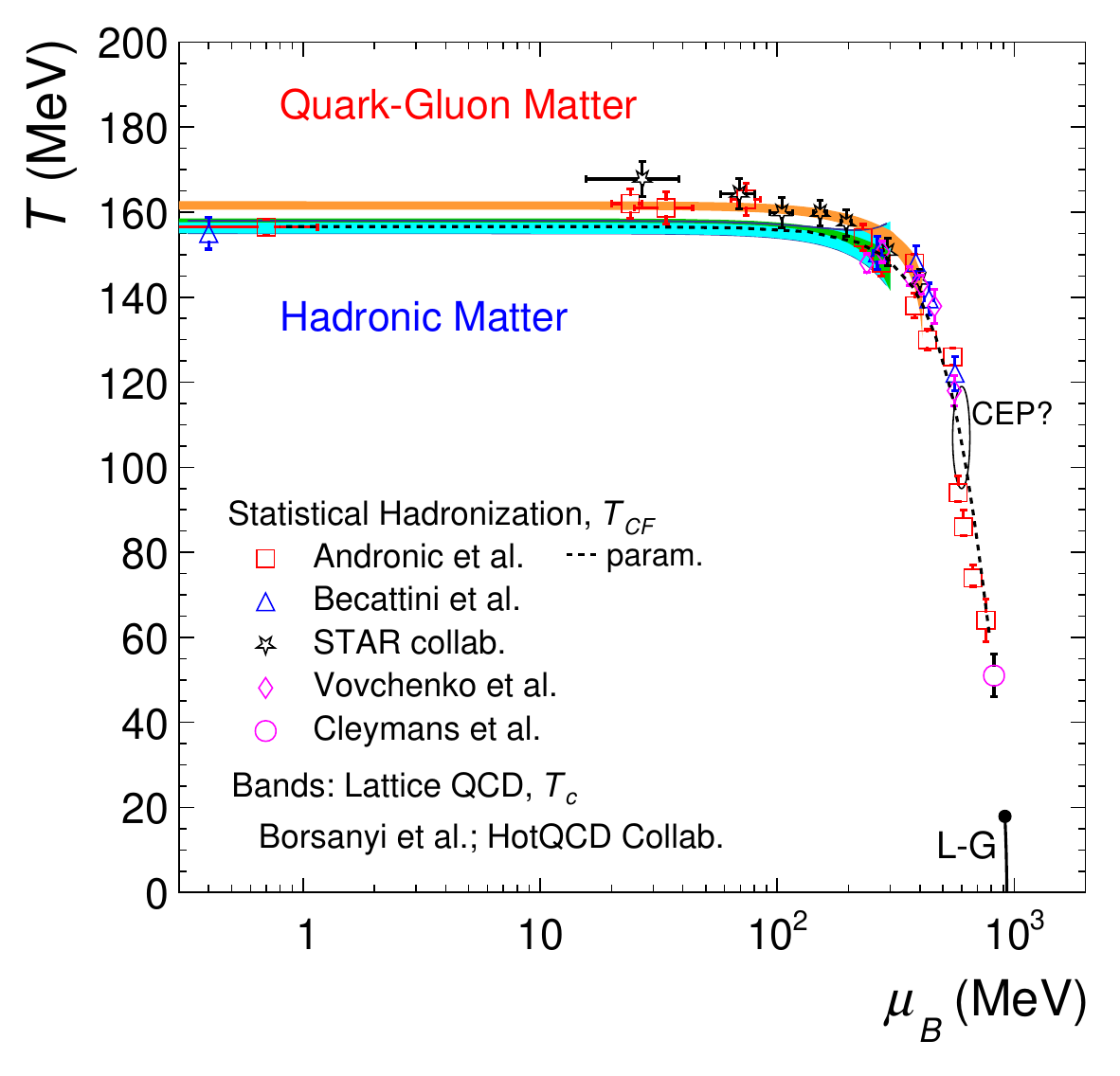}
\caption{Phase diagram of strongly interacting matter constructed from chemical freeze-out points for central collisions at different energies. Points are extracted from experimental datasets in our own (squares) and other similar SHM analyses~\cite{Cleymans:1998yb,Vovchenko:2015idt,Becattini:2016xct,Adamczyk:2017iwn}. They are compared with predictions from LQCD shown as bands. The turquoise band represents the chiral transition as computed in Refs.~\cite{Bazavov:2018mes,Borsanyi:2020fev}. The orange band corresponds to the deconfinement transition~\cite{Borsanyi:2024xrx}, in this case not yet extrapolated to the continuum. For the location of a possible critical endpoint (CEP), see Ref.~\cite{Fischer:2026uni}. For the liquid-gas phase transition (L-G), see Ref.~\cite{Kaiser:2026msy}.}
\label{fig:t-mu}
\end{figure}

The pair of chemical freeze-out parameters, $(T_{CF},\mub)$, provided through the SHM fit to data can be used to construct the phenomenological phase diagram shown in Fig.~\ref{fig:t-mu}. At low temperatures, the extracted freeze-out points approach the region corresponding to ground-state nuclear matter, where $\mub\simeq m_N$. As discussed in Ref.~\cite{Floerchinger:2012xd}, however, this limiting behavior does not necessarily indicate the presence of a phase transition.

The low-temperature, high-\mub region remains theoretically challenging and is not currently accessible to first-principles lattice-QCD calculations. In contrast, at high collision energies, where $T$ is large and \mub is small, the extracted chemical freeze-out conditions can be directly compared with lattice-QCD predictions for the QCD crossover, providing a connection between experimentally measured particle yields and the QCD phase boundary.

\newpage

\subsection{Hadronic phase and Saha equation}
\label{subsec:Saha}

The yields of short-lived resonances, such as $\text{K}^*$ and $\rho_0$, are significantly over-predicted by thermal models~\cite{ALICE:2014jbq,ALICE:2018qdv}, indicating the presence of additional mechanisms that modify their observable yields during the hadronic phase.
Furthermore, blast-wave analyses of the transverse-momentum spectra of hadrons and light nuclei indicate a kinetic freeze-out temperature of $T_{\rm kin} \sim 100$~MeV in the most central Pb--Pb collisions at the LHC~\cite{ALICE:2023qyl}, significantly below the chemical freeze-out temperature of $T_{\rm ch} \sim 150$--$160$~MeV.
The extracted value depends on the particle species included in the fit, with Ref.~\cite{ALICE:2023qyl} reporting temperatures in the range $T_{\rm kin}=85$--$132$~MeV for different particle selections.

\begin{figure}[htb]
\centering
\includegraphics[width=0.48\textwidth]{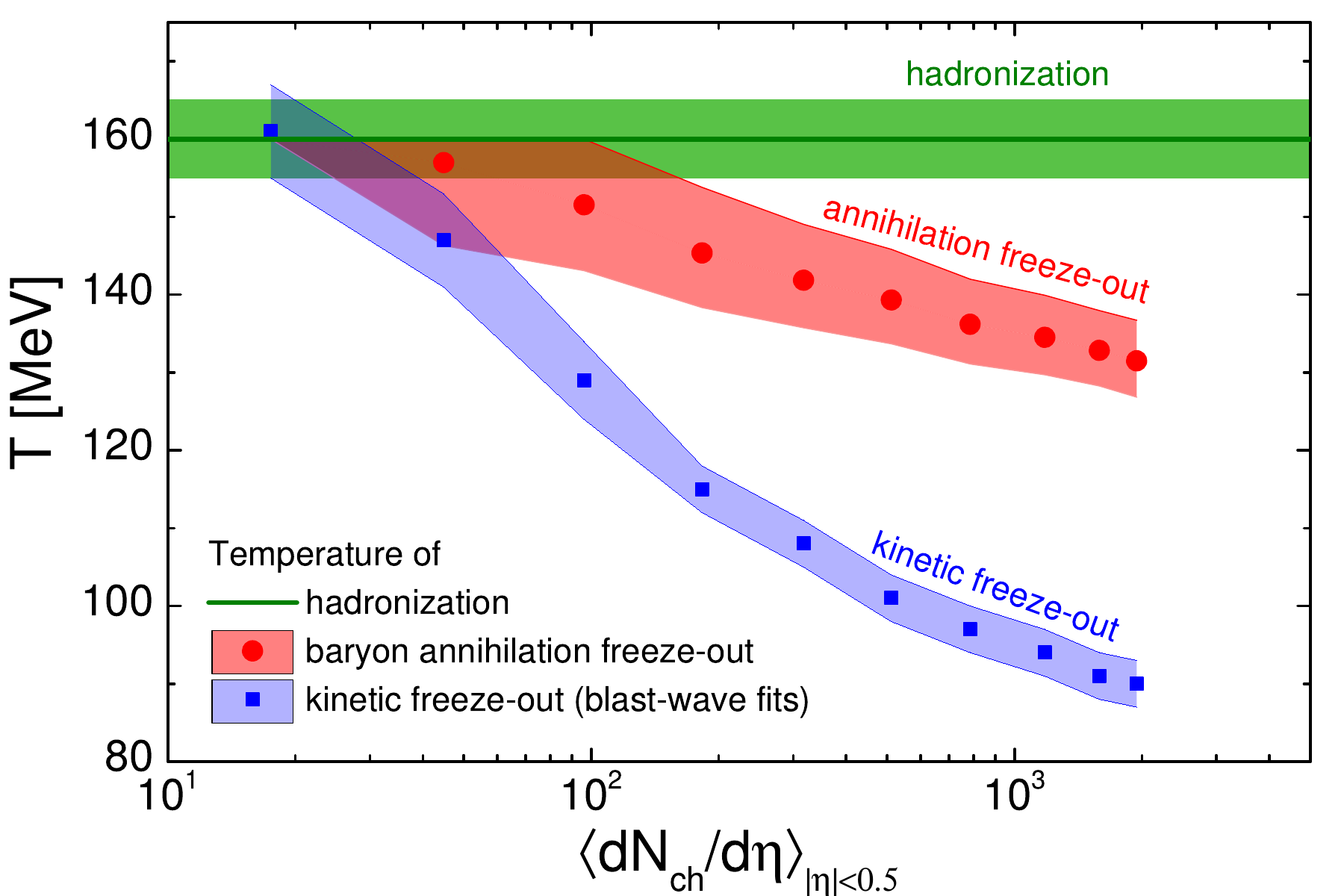}
\includegraphics[width=0.48\textwidth]{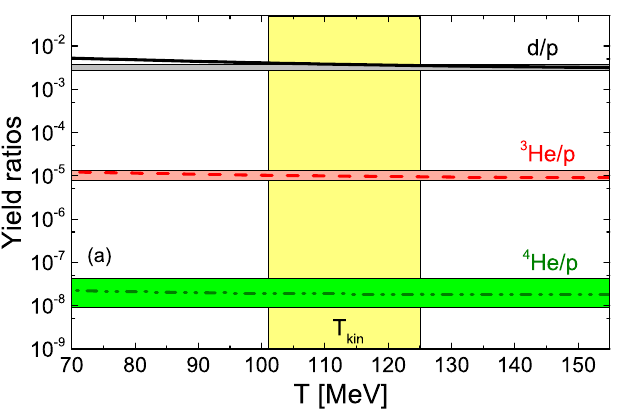}
\caption{
Left panel: From Ref.~\cite{Vovchenko:2022xil}. Charged-particle multiplicity dependence of the hadronization temperature, $T_{\rm had}=160\pm5$~MeV (green line and band), approximately corresponding to conventional chemical freeze-out; the kinetic freeze-out temperature extracted from blast-wave fits (blue symbols and band)~\cite{ALICE:2019hno}; and the baryon annihilation freeze-out temperature (red symbols and band), extracted from the centrality dependence of the $p/\pi$ ratio in Pb--Pb collisions at $\sqrt{s_{\rm NN}}=5.02$~TeV.
Right panel: From Ref.~\cite{Vovchenko:2019aoz}. Temperature dependence of the light-nucleus-to-proton yield ratios at $T<T_{\rm ch}$, calculated using the Saha-equation approach within a hadron resonance gas (HRG) in partial chemical equilibrium (PCE). The horizontal bands represent measurements by the ALICE Collaboration in the most central Pb--Pb collisions at $\sqrt{s_{\rm NN}}=2.76$~TeV~\cite{Adam:2015vda,Acharya:2017bso}.
}
\label{fig:Sahaeq}
\end{figure}

In the thermal-model description of light-nuclei production, this implies that, although their abundances are fixed at chemical freeze-out, light nuclei continue to interact with the hadronic medium until kinetic freeze-out (left panel of Fig.~\ref{fig:Sahaeq}).
Given their small binding energies, however, such interactions would be expected to readily dissociate light nuclei, apparently challenging the assumption that their yields remain unchanged between chemical and kinetic freeze-out.

The suppression of short-lived resonance yields and the low kinetic freeze-out temperature indicate the presence of a sufficiently long-lived hadronic phase, during which the expanding system remains in kinetic, but not chemical, equilibrium after chemical freeze-out and until kinetic freeze-out at $T_{\rm kin}<T_{\rm ch}$.
An extension of the thermal model that accounts for this hadronic phase is based on the hadron resonance gas in partial chemical equilibrium (HRG-PCE)~\cite{Bebie:1991ij,Huovinen:2007xh}.

The HRG-PCE model describes the hadronic phase between $T_{\rm ch}$ and $T_{\rm kin}$ as a gas of hadrons and resonances in which inelastic reactions that change the abundances of stable hadrons are frozen out, while elastic reactions (e.g. $\pi\pi \leftrightarrow \pi\pi$) and pseudo-elastic reactions involving short-lived resonances (e.g. $\pi\pi \leftrightarrow \rho$, $\pi K \leftrightarrow K^*$, and $\pi N \leftrightarrow \Delta$) remain in equilibrium~\cite{Bebie:1991ij,Huovinen:2007xh}.
Entropy is assumed to be conserved in this model, as are the total abundances of stable hadrons, including contributions from feed-down of short-lived resonances.
For instance, the total pion and nucleon numbers,
$N^{\rm tot}_{\pi} = N_{\pi} + 2N_{\rho} + 3N_{\omega} + \ldots$ and
$N^{\rm tot}_{N} = N_{N} + N_{\Delta} + N_{N^*} + \ldots$,
respectively, are conserved.
Effective chemical potentials, such as $\mu_\pi$ and $\mu_N$, can therefore be introduced to enforce these conservation laws.

The chemical potentials of short-lived resonances are not independent but are related to the effective chemical potentials of their decay products through the condition of relative chemical equilibrium between decay and regeneration reactions. For example, $\mu_\rho = 2\mu_\pi$ and $\mu_\Delta = \mu_N + \mu_\pi$.
The chemical potentials at a given temperature $T$ are determined by solving the system of conservation equations

\begin{equation}
\frac{n_{i}^{\rm tot}}{s} =
\frac{n_i^{\rm tot,ch}}{s^{\rm ch}},
\qquad i \in \pi, K, N, \ldots ,
\end{equation}

where $n_i^{\rm tot}$ denotes the total density of stable hadron species $i$, including feed-down contributions, and $s$ is the entropy density; quantities evaluated at chemical freeze-out are denoted by the superscript ``ch''.

By construction, HRG-PCE reduces to the conventional SHM with respect to the final abundances of stable hadrons. The yields of short-lived resonances, however, are not conserved and continue to evolve during the hadronic phase. In Ref.~\cite{Motornenko:2019jha}, this feature was used to estimate the temperature at which short-lived resonance yields freeze out by fitting measurements of the $K^{*0}$ and $\rho_0$ yields. The resulting temperature was found to be consistent, within uncertainties, with $T_{\rm kin}$ extracted from blast-wave fits.

Light nuclei can be incorporated into the HRG-PCE framework by assuming nuclear statistical equilibrium, in which their chemical potentials are determined by those of their constituents. For instance, the deuteron chemical potential satisfies $\mu_d=\mu_p+\mu_n$, where $\mu_p$ and $\mu_n$ are determined from the PCE equations~\cite{Xu:2018jff}.
This treatment, known as the Saha equation approach~\cite{Vovchenko:2019aoz}, is analogous to the description of nucleosynthesis in the early Universe.

Results obtained by applying the Saha equation approach to Pb--Pb collisions at the LHC are shown in the right panel of Fig.~\ref{fig:Sahaeq}.
Compared with the standard SHM, the temperature dependence of the light-nucleus-to-proton yield ratios is considerably weaker. As a result, a broad range of temperatures below $T_{\rm ch}$ is compatible with the experimental data, including temperatures close to the kinetic freeze-out temperature $T_{\rm kin}$.
It should be emphasized, however, that the light-nucleus yields themselves do not remain constant during the hadronic evolution in the Saha equation approach. Their quantitative evolution also depends on the baryonic resonance spectrum employed in the calculation~\cite{Vovchenko:2019aoz}.

The assumption of nuclear statistical equilibrium employed in the Saha equation approach thus allows the observed light-nucleus yields to be established at temperatures significantly below $T_{\rm ch} \sim 150$--$160$~MeV.
A possible microscopic mechanism for maintaining such equilibrium is provided by nuclear formation and breakup reactions catalyzed by a third particle, $X + A \leftrightarrow X + \sum_i A_i$.

Pion-catalyzed deuteron formation was studied using hadronic transport in Ref.~\cite{Oliinychenko:2018ugs}. The study showed that the reaction $\pi + d \leftrightarrow \pi + p + n$ maintains approximate relative chemical equilibrium during the hadronic phase at LHC energies, with deuteron formation and breakup rates remaining approximately balanced due to the large $\pi d$ cross section.
In this picture, deuterons present around $T_{\rm ch}$ are rapidly dissociated and continuously regenerated throughout the hadronic evolution, with the finally observed deuterons predominantly formed at later stages of the hadronic phase.
This framework has subsequently been extended to heavier light nuclei, such as $^3$He and $^4$He, using rate-equation calculations~\cite{Neidig:2021bal}.

A conceptual issue with the microscopic description of light-nucleus production is related to their formation time. A naive estimate of the deuteron formation time based on the uncertainty principle yields a timescale of order 100~fm/$c$, considerably longer than the lifetime of the hadronic phase. This raises questions about kinetic approaches that employ the vacuum properties of nuclei and treat them as point-like degrees of freedom throughout the hadronic evolution.

A more complete treatment of this problem may require an open-quantum-system description of light-nucleus formation. Recent studies based on the time-dependent Schr\"odinger equation~\cite{Rais:2022gfg} and Kadanoff--Baym equations~\cite{Neidig:2023kid} suggest that the relevant formation timescale is not necessarily set by the uncertainty-principle estimate, but may instead be governed by the momentum-transfer scale of reactions involving loosely bound states such as light nuclei.

\subsubsection{Baryon annihilation}
\label{sec:annihilation}

Baryon--antibaryon annihilation played an important role in the evolution of the early Universe as the primordial matter cooled below the QCD transition temperature, $T \lesssim 160$~MeV. Such reactions may also be relevant during the hadronic phase of heavy-ion collisions.

Historically, the proton yield measured in Pb--Pb collisions at the LHC was found to be lower than predictions of the statistical hadronization model (SHM)~\cite{ALICE:2013mez}. This discrepancy, commonly referred to as the ``proton anomaly'', was initially interpreted as possible evidence for baryon--antibaryon annihilation during the hadronic phase~\cite{Becattini:2012sq,Steinheimer:2012rd}. However, the discrepancy can be resolved within the statistical hadronization framework by employing an S-matrix treatment of pion--nucleon interactions, which accounts for resonance widths and non-resonant contributions and leads to a reduction of the predicted proton yield.

\begin{figure}[!hbt]
\centering
\includegraphics[width=0.7\linewidth]{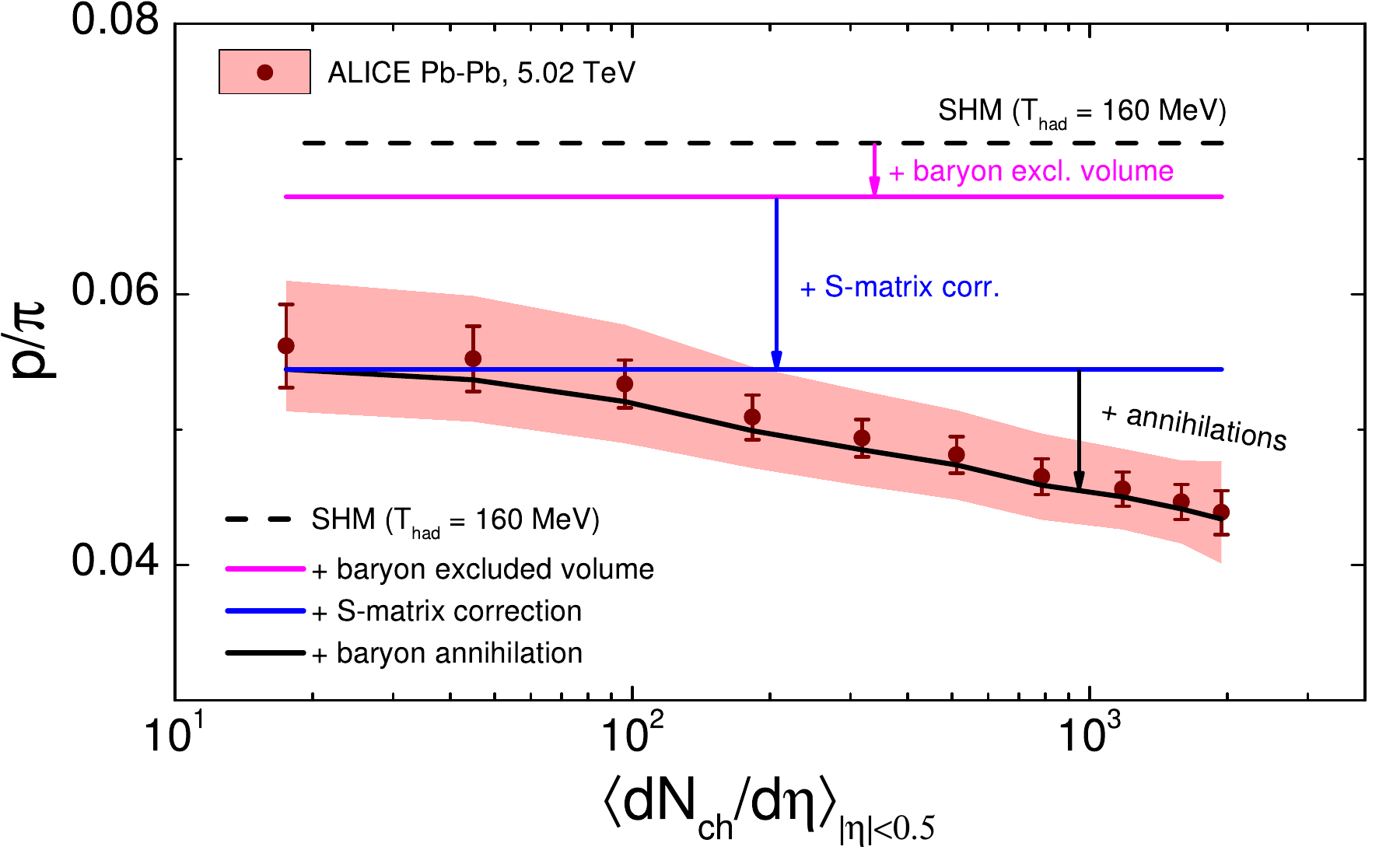}
\caption{Adapted from Ref.~\cite{Vovchenko:2022xil}. Charged-particle multiplicity dependence of the $p/\pi$ ratio in Pb--Pb collisions at $\sqrt{s_{\rm NN}}=5.02$~TeV. The calculations show the standard SHM at $T_{\rm had}=160$~MeV (dashed black line), the SHM including baryon--baryon repulsive interactions (magenta line), the additional inclusion of S-matrix corrections (solid blue line)~\cite{Andronic:2018qqt}, and the further inclusion of $N\bar{N} \leftrightarrow 5\pi$ annihilation and regeneration reactions during the hadronic phase (solid black line). The symbols represent measurements by the ALICE Collaboration~\cite{ALICE:2019hno}.}
\label{fig:ppiAnn}
\end{figure}

A distinct observation is the centrality dependence of the $p/\pi$ ratio measured in Run 2~\cite{ALICE:2019hno}, which decreases from peripheral to central Pb--Pb collisions (Fig.~\ref{fig:ppiAnn}). While corrections to the thermal description, such as those obtained within the S-matrix approach, modify the overall proton yield, they do not by themselves account for this centrality dependence. The latter may therefore provide sensitivity to baryon--antibaryon annihilation during the hadronic phase, whose effect is expected to increase with the duration of the hadronic evolution. If baryon--antibaryon annihilation has a sizable effect on proton yields, analogous processes may also affect the abundances of light nuclei and antinuclei.

Baryon--antibaryon annihilation can be incorporated into the thermal framework through the generalized PCE approach described in Ref.~\cite{Vovchenko:2022xil}.
Consider, for instance, the nucleon--antinucleon annihilation and regeneration reaction
\begin{equation}
N\bar{N} \leftrightarrow 5\pi.
\end{equation}
When this reaction is active, neither the total number of (anti)nucleons nor the total number of pions is separately conserved. However, the combination
\begin{equation}
N^{\rm tot}_{\rm ann}
= \frac{N_N^{\rm tot}+N_{\bar{N}}^{\rm tot}}{2}
+ \frac{N_\pi^{\rm tot}}{5}
\end{equation}
is conserved and therefore plays the role of a generalized conserved quantity, together with the net nucleon number,
\begin{equation}
N_N^{\rm net,tot}=N_N^{\rm tot}-N_{\bar{N}}^{\rm tot}.
\end{equation}
Furthermore, assuming that the reaction $N\bar{N} \leftrightarrow 5\pi$ remains in relative chemical equilibrium through detailed balance leads to the relation
\begin{equation}
\mu_N+\mu_{\bar{N}}=5\mu_\pi.
\end{equation}

Incorporating these generalized conservation laws and chemical-equilibrium conditions into the PCE framework allows the effect of baryon--antibaryon annihilation and regeneration on hadron yields during the hadronic phase to be evaluated.
In Ref.~\cite{Vovchenko:2022xil}, this framework was used to extract the annihilation freeze-out temperature $T_{\rm ann}$ as a function of centrality by describing the measured centrality dependence of the $p/\pi$ ratio~\cite{ALICE:2019hno}. The extracted values are shown by the red symbols and band in the left panel of Fig.~\ref{fig:Sahaeq}.

When combined with an improved description of the thermal baseline, including an S-matrix treatment of pion--nucleon interactions (or, alternatively, finite resonance-width effects~\cite{Vovchenko:2018fmh}) and excluded-volume corrections in the baryon--baryon sector, the generalized PCE framework provides a description of the measured $p/\pi$ ratio in Pb--Pb collisions at the LHC over the full centrality range (Fig.~\ref{fig:ppiAnn}).

\begin{figure}[t]
\centering
\includegraphics[width=0.32\textwidth]{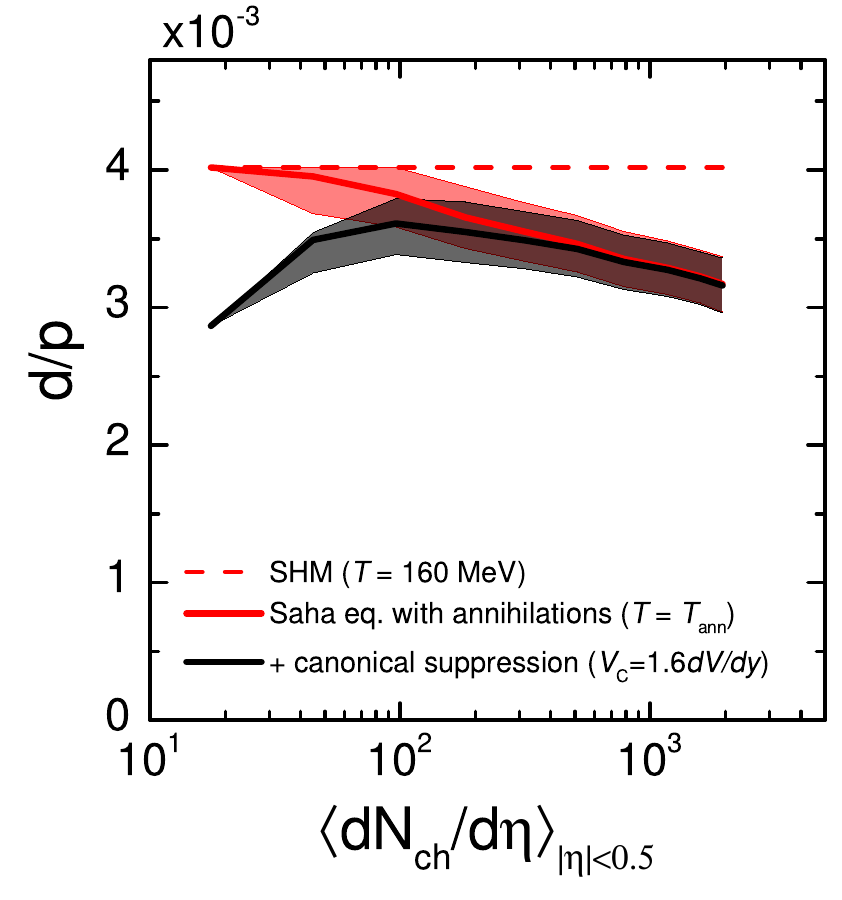}
\includegraphics[width=0.32\textwidth]{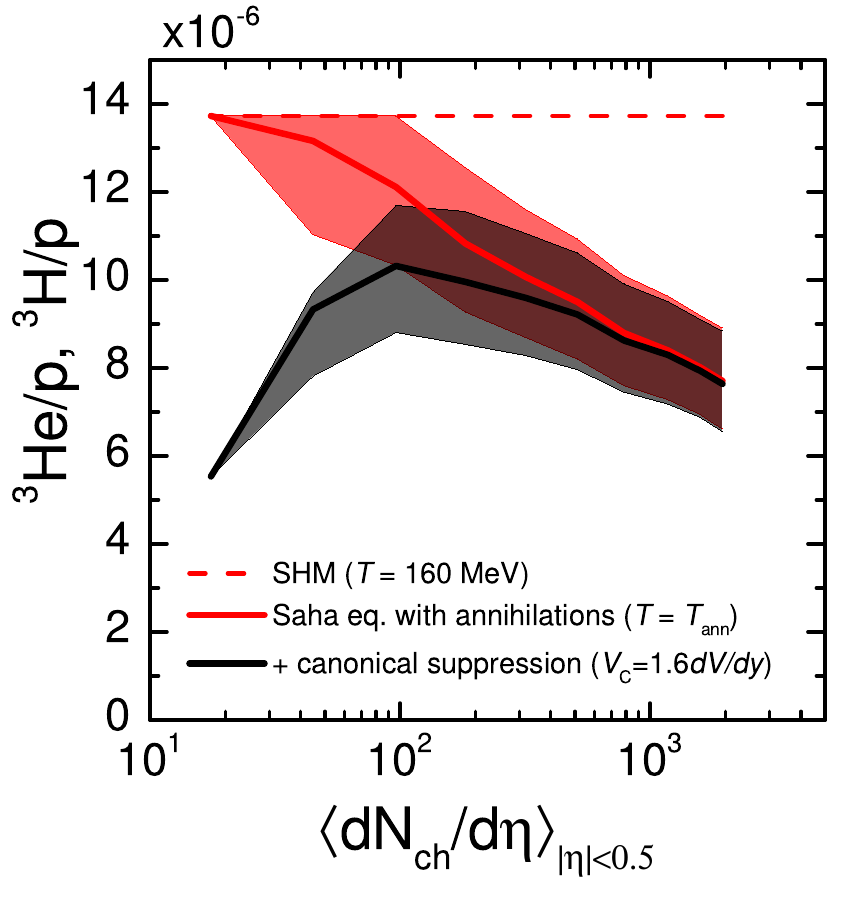}
\includegraphics[width=0.32\textwidth]{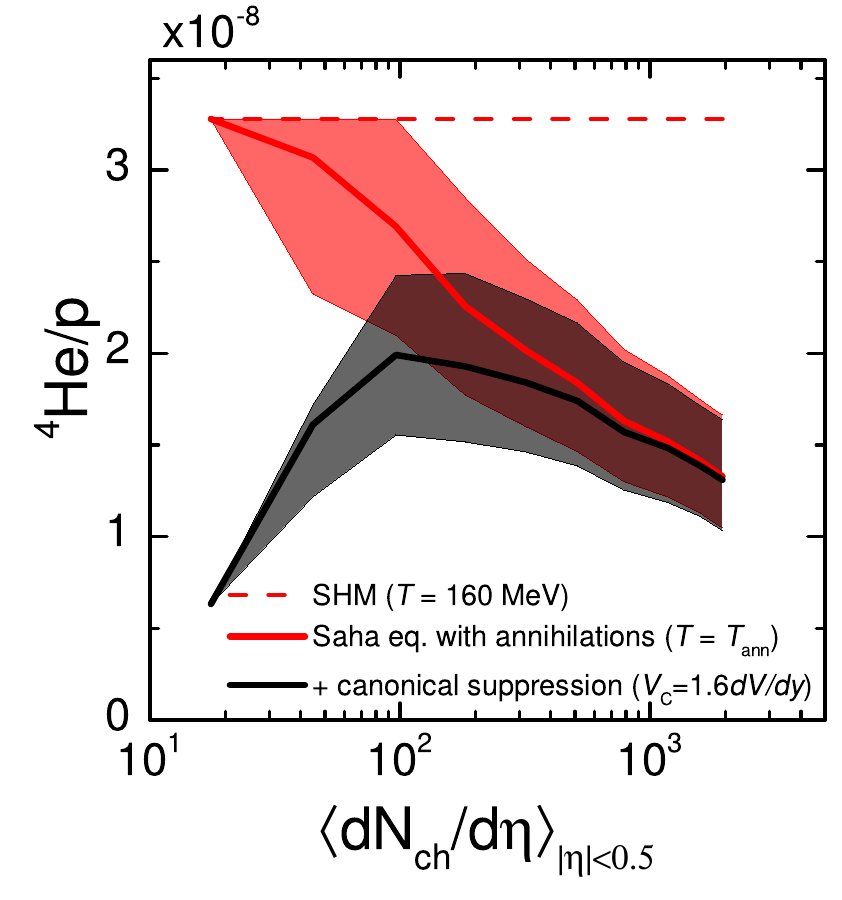}
\caption{
From Ref.~\cite{Vovchenko:2022xil}.
Charged-particle multiplicity dependence of the light-nucleus-to-proton yield ratios calculated within the extended SHM using the Saha equation approach, including baryon--antibaryon annihilation (red lines) and the additional effect of canonical suppression (black lines).
}
\label{fig:SahaAnn}
\end{figure}

The effect of baryon--antibaryon annihilation on light-nucleus production can be evaluated by using the nucleon chemical potentials obtained within the generalized PCE framework as input to the Saha equation.
Figure~\ref{fig:SahaAnn} shows the $\mathrm{d}/p$, $^3$He/$p$, and $^4$He/$p$ yield ratios as a function of charged-particle multiplicity, evaluated using the Saha equation at $T=T_{\rm ann}$ (red bands).
The calculations predict a suppression of light-nucleus yields relative to protons in central collisions due to baryon--antibaryon annihilation, with an increasingly pronounced effect for heavier nuclei.
The magnitude of the annihilation effect obtained within this framework is consistent with earlier studies based on a UrQMD afterburner combined with coalescence~\cite{Sombun:2018yqh,Reichert:2022mek} and on rate-equation approaches~\cite{Neidig:2021bal}.

When this effect is combined with the expected canonical suppression of light-nucleus production in peripheral collisions, a non-monotonic multiplicity dependence of the light-nucleus-to-proton yield ratios is predicted.
Run 2 measurements at the LHC~\cite{ALICE:2022veq} show indications of a non-monotonic multiplicity dependence for the $\mathrm{d}/p$ and $^3$He/$p$ ratios, qualitatively compatible with this interplay between canonical suppression and annihilation effects.
However, the available measurement of the $^4$He/$p$ ratio, which is currently limited to central collisions, does not support the sizable suppression of the $^4$He yield predicted from baryon--antibaryon annihilation in this framework.

\subsection{Canonical thermal model}
\label{subsec:thermal_canonical}

The simplest formulation of the thermal model employs the grand-canonical ensemble, characterized by the temperature $T$, the volume $V$, and chemical potentials associated with the conserved charges, such as the baryochemical potential $\mu_B$. These parameters are determined at each collision energy by fitting measured particle yields. In the grand-canonical ensemble, conservation of baryon number $B$, electric charge $Q$, and strangeness $S$ is imposed only on average through the corresponding chemical potentials.

In the canonical formulation of the thermal model, also referred to as the Canonical Statistical Model (CSM), the conserved charges $(B,Q,S)$ are instead fixed exactly~\cite{Rafelski:1980gk,Hagedorn:1984uy}. Due to the equivalence of statistical ensembles in the thermodynamic limit, the grand-canonical and canonical formulations yield the same predictions for particle abundances in sufficiently large systems, such as central heavy-ion collisions. In this regime, the grand-canonical formulation is generally preferred because of its technical simplicity.

The canonical formulation of the thermal model becomes relevant in two main scenarios:

\begin{enumerate}

\item When the system is sufficiently small that the thermodynamic limit is not applicable.
As a rule of thumb, canonical effects become important when the mean abundance of particles carrying a given conserved charge is of order unity or smaller.
This situation can occur, for instance, in pp and p--Pb collisions, as well as in heavy-ion collisions at lower energies, where the production of strange particles is limited.
For light nuclei, the most relevant effects are those associated with exact baryon-number conservation in pp and p--Pb collisions at the LHC~\cite{Vovchenko:2018fiy}, and with exact strangeness conservation at low collision energies ($\sqrt{s_{\rm NN}}\lesssim5$~GeV)~\cite{Hamieh:2000tk}, which is particularly relevant for hypernucleus production~\cite{Reichert:2022mek}.

\item When the thermal model is used to describe fluctuations and correlations of particle multiplicities, including those involving light (anti)nuclei and (anti)hypernuclei.
While different statistical ensembles become equivalent for average thermodynamic quantities in the thermodynamic limit, fluctuations and correlations of conserved charges can retain a strong dependence on the implementation of conservation laws~\cite{Begun:2004gs}.
The total net baryon number provides a simple illustration~\cite{Bzdak:2012an}.
In the grand-canonical ensemble, baryon and antibaryon numbers fluctuate independently in the ideal-gas limit, leading to a Skellam distribution of their difference\footnote{The Skellam distribution describes the probability distribution of the difference between two independent Poisson-distributed random variables.}.
In the canonical ensemble, by contrast, the total net baryon number is fixed exactly and therefore does not fluctuate.
\end{enumerate}

The canonical partition function is given by~\cite{Becattini:1995if,Becattini:1997rv}
\begin{equation}
\label{eq:Zcan}
Z(B,Q,S) =
\int \limits_{-\pi}^{\pi}
\frac{d \phi_B}{2\pi}
\int \limits_{-\pi}^{\pi}
\frac{d \phi_Q}{2\pi}
\int \limits_{-\pi}^{\pi}
\frac{d \phi_S}{2\pi}~
e^{-i  (B \phi_B + Q \phi_Q + S \phi_S)}
\exp\left[
\sum_{j \in \text{species}} \sum_{n=1}^{\infty}
z_j^n ,
e^{i , n , (B_j \phi_B + Q_j \phi_Q + S_j \phi_S)}
\right].
\end{equation}
The sum over $j$ runs over all particle species included in the model, including light nuclei, while the sum over $n$ accounts for quantum-statistical effects.
The quantities $B_j$, $Q_j$, and $S_j$ denote the baryon number, electric charge, and strangeness, respectively, carried by particle species $j$. The quantity $z_j^n$ is the corresponding single-particle partition-function contribution,
\begin{equation}
\label{eq:zj}
z_j^n = (\mp 1)^{n-1} \, V_c \, \int dm \, \rho_j(m) \,
d_j \frac{m^2 T}{2\pi^2 n^2} \, K_2(n , m/T).
\end{equation}
Here, $d_j$ is the degeneracy factor of particle species $j$, with the minus sign applying to fermions and the plus sign to bosons.
The integration over the mass distribution $\rho_j(m)$ allows finite resonance widths to be incorporated, for instance through a Breit--Wigner prescription~\cite{Vovchenko:2018fmh}.
The parameter $V_c$ denotes the canonical correlation volume over which the conserved charges $(B,Q,S)$ are enforced exactly. In general, $V_c$ need not coincide with the acceptance volume $V_A$ over which particle yields or fluctuations are measured.

The mean multiplicities of individual particle species can be calculated by introducing auxiliary fugacities into the partition function in Eq.~\eqref{eq:Zcan} and taking the corresponding derivatives with respect to these fugacities (see Refs.~\cite{Becattini:1995if,Becattini:1997rv} for details):
\begin{equation}
\label{eq:Ncanon}
\langle N_j^{\rm prim} \rangle =
\frac{V_A}{V_c}
\sum_{n=1}^{\infty}
\frac{Z(B-nB_j,Q-nQ_j,S-nS_j)}
{Z(B,Q,S)}
\, n\,z_j^n~.
\end{equation}
The prefactor $V_A/V_c$ accounts for the difference between the acceptance volume and the canonical correlation volume. The ratios
\begin{equation}
\frac{Z(B-nB_j,Q-nQ_j,S-nS_j)}{Z(B,Q,S)}
\end{equation}
are the canonical chemical factors and encode the effect of exact conservation of the charges $(B,Q,S)$ on the production of particle species $j$.

In the thermodynamic limit, these factors approach the corresponding grand-canonical fugacity factors,
\begin{equation}
\frac{Z(B-nB_j,Q-nQ_j,S-nS_j)}{Z(B,Q,S)}
\rightarrow
\exp\left[
\frac{n(\mu_B B_j+\mu_Q Q_j+\mu_S S_j)}{T}
\right],
\end{equation}
and the canonical results therefore approach the grand-canonical limit.
As in Eq.~\eqref{eq:Zcan}, the sum over $n$ accounts for quantum-statistical effects. In practice, terms with $n\geq2$ are generally negligible for light-nucleus yields because of their large masses relative to the temperature.

Away from the thermodynamic limit, the canonical chemical factors must be evaluated from the partition function in Eq.~\eqref{eq:Zcan}. In the general case, this requires numerical evaluation of the corresponding integrals, although analytical expressions can be obtained in certain limits using properties of Bessel functions~\cite{Braun-Munzinger:2001alt}.

In addition to analytical calculations, the canonical thermal model can be implemented using Monte Carlo event sampling. In this approach, the multiplicities of all particle species are sampled from independent Poisson distributions based on their grand-canonical mean values, and rejection sampling is subsequently applied to retain only configurations that satisfy the exact conservation laws~\cite{Becattini:2004rq}.

The sampled particle multiplicities can be combined with a hydrodynamics-inspired collective-flow profile, such as that of the blast-wave model~\cite{Schnedermann:1993ws}, to generate realistic momentum distributions and study $p_{\rm T}$-differential observables, including fluctuations and correlations. Further details of this procedure can be found in Refs.~\cite{Vovchenko:2020kwg,Vovchenko:2022syc}.

State-of-the-art implementations of the canonical thermal model, in both analytical and Monte Carlo formulations, are available in the open-source \texttt{Thermal-FIST} package~\cite{Vovchenko:2019pjl}.

\subsubsection{Multiplicity dependence of light-nucleus yields at the LHC}

The CSM can be applied to systems produced in collisions at the LHC by imposing vanishing net conserved charges, $B=Q=S=0$, and specifying the chemical freeze-out temperature $T_{\rm ch}$ and the canonical correlation volume $V_c$, with the latter related to the measured charged-particle multiplicity.

In the simplest implementation, the chemical freeze-out temperature is fixed to $T_{\rm ch}=155$~MeV for all multiplicities, while $V_c$ is assumed to scale proportionally with $dN_{\rm ch}/d\eta$. The multiplicity dependence of canonical effects is therefore entirely governed by the variation of $V_c$, with the strongest effects expected at low multiplicities and in the smallest systems.

\begin{figure}[t]
\centering
\includegraphics[width=0.7\textwidth]{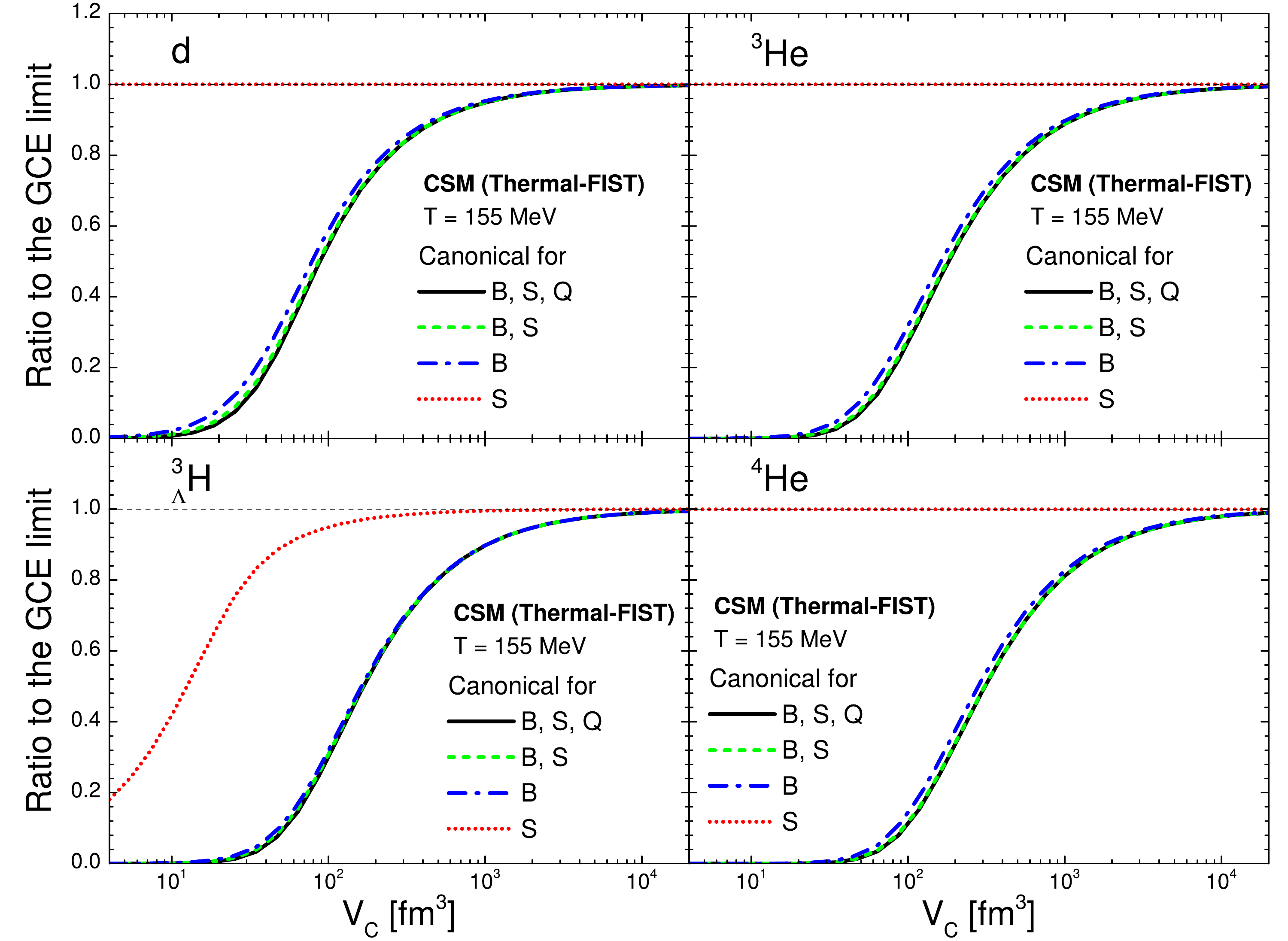}
\caption{
Dependence of the yields of d, $^3$He, $^3_\Lambda$H, and $^4$He on the canonical correlation volume $V_C$, calculated within the CSM at $T=155$~MeV and vanishing net conserved charges~\cite{Vovchenko:2018fiy}. The different lines correspond to different sets of conserved charges treated canonically. The yields are normalized to their grand-canonical (infinite-volume) limits.
}
\label{fig:CSMBQS}
\end{figure}

Figure~\ref{fig:CSMBQS} shows the relative change in the yields of d, $^3$He, $^3_\Lambda$H, and $^4$He as a function of the canonical correlation volume $V_c$ for different choices of the conserved charges treated canonically.
In the full scenario (black lines), all three conserved charges, $B$, $Q$, and $S$, are treated canonically ($BQS$-canonical ensemble). In the other three scenarios ($BS$-, $B$-, and $S$-canonical), only the indicated charges are treated canonically, while the remaining ones are treated grand-canonically.
Comparison of these scenarios allows the relative importance of exact conservation of the different charges to be assessed.

At LHC energies, the canonical suppression of the considered light (anti)nuclei and (anti)hypernuclei is dominated by exact baryon-number conservation.
This results from two main factors: (i) light nuclei carry multiple units of baryon number, making them particularly sensitive to exact baryon-number conservation, and (ii) the abundance of particles carrying baryon number is smaller than that of particles carrying electric charge or strangeness, enhancing the corresponding canonical effects.
Consequently, within the CSM, the multiplicity dependence of light-nucleus yields is primarily governed by the number of baryons contained within the canonical correlation volume.

This mechanism is conceptually distinct from coalescence, where the multiplicity dependence arises from the interplay between the size of the particle-emitting source and the spatial extent of the light-nucleus wave function~\cite{Sun:2018mqq}.

To relate CSM predictions to different collision systems at the LHC, the canonical correlation volume $V_c$ must be connected to the acceptance volume $V_A$, which is taken here to correspond to the volume per unit rapidity, $V_A=dV/dy$.
Analyses of hadron yields in pp, p--Pb, and Pb--Pb collisions within the CSM suggest values in the range $V_c=(1$--$3)\,dV/dy$~\cite{Vovchenko:2019kes}, while an analysis of antiproton--antideuteron correlations yields $V_c\simeq1.6 dV/dy$~\cite{ALICE:2022xiu}.

\begin{figure}[t]
\centering
\includegraphics[width=0.33\textwidth]{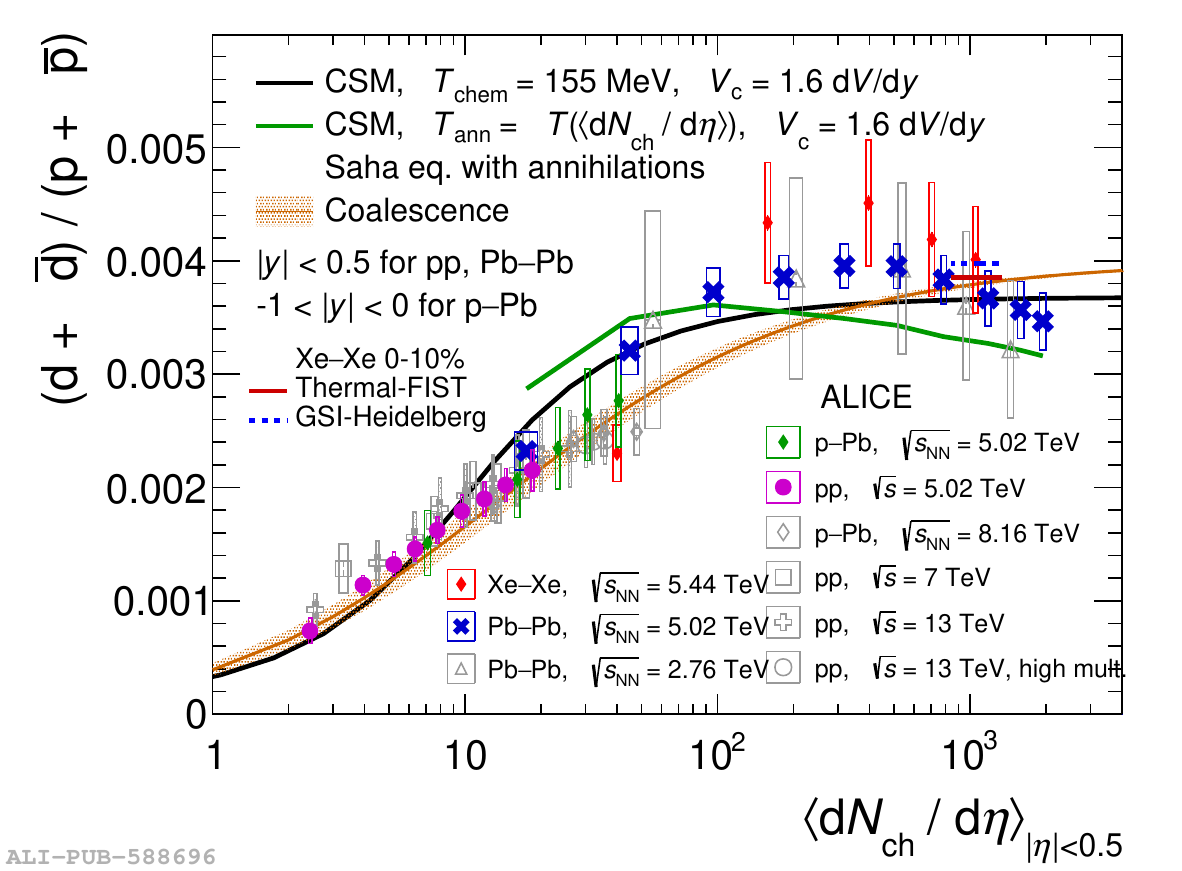}
\includegraphics[width=0.33\textwidth]{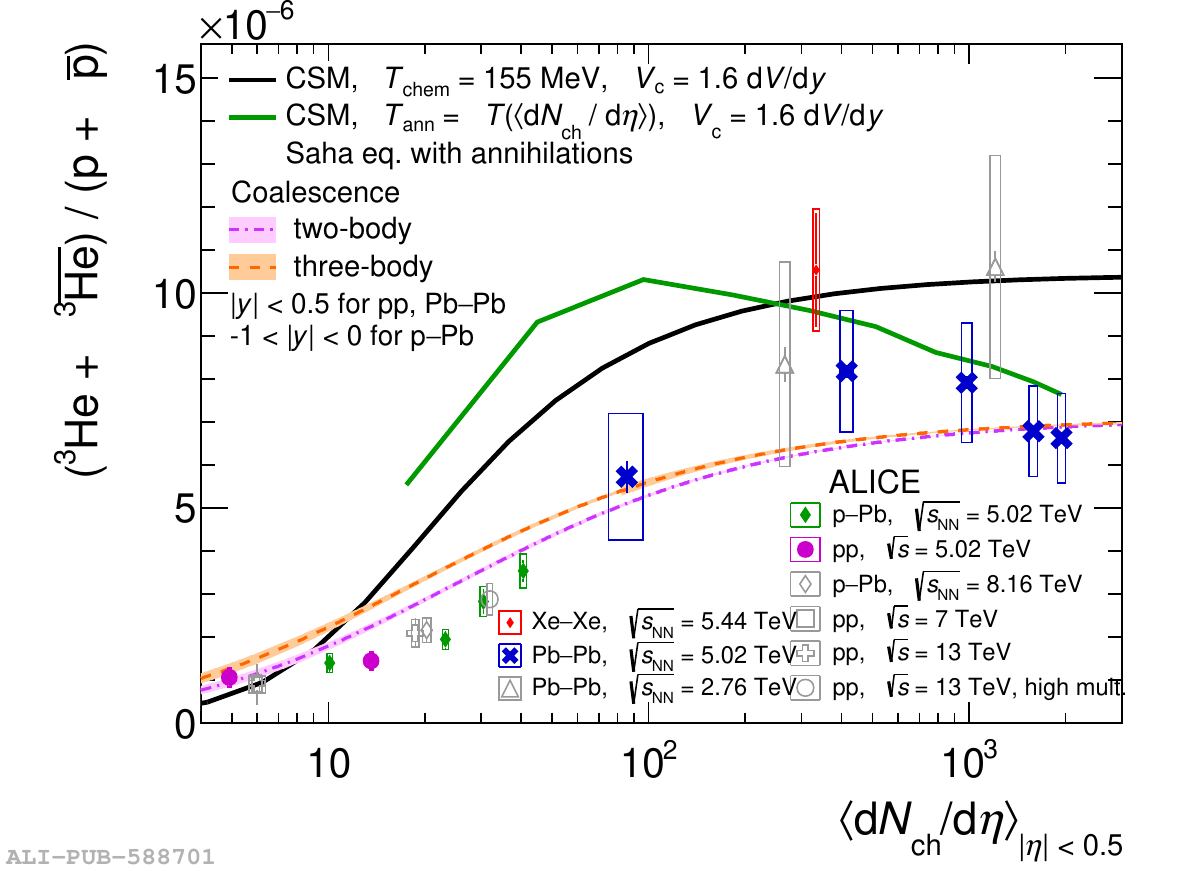}
\includegraphics[width=0.28\textwidth]{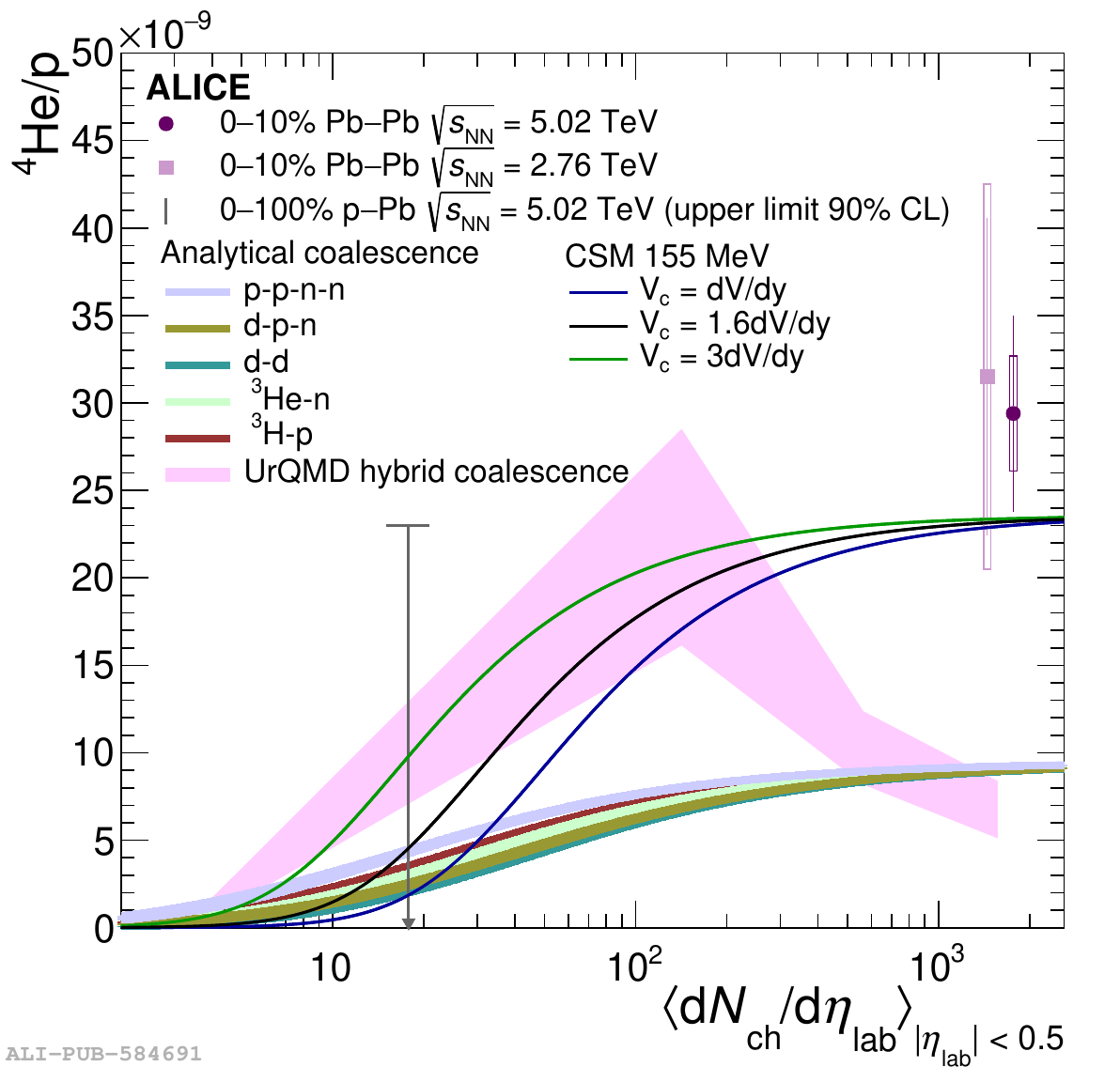}
\caption{
Multiplicity dependence of the $\mathrm{d}/p$ (left), $^3$He/$p$ (middle), and $^4$He/$p$ (right) yield ratios measured in pp, p--Pb, and Pb--Pb collisions at the LHC. The left and middle panels are taken from Ref.~\cite{ALICE:2024say}, while the right panel is taken from Ref.~\cite{ALICE:2023qyl}. The measurements are compared with predictions from different models.
}
\label{fig:nucleiToProtonYieldVsMult}
\end{figure}

Figure~\ref{fig:nucleiToProtonYieldVsMult} shows the multiplicity dependence of the $\mathrm{d}/p$, $^3$He/$p$, and $^4$He/$p$ yield ratios calculated within the CSM using $V_c=1.6\,dV/dy$. The predictions are compared with measurements by the ALICE Collaboration in pp, p--Pb, and Pb--Pb collisions. The CSM provides a reasonable description of the measured $\mathrm{d}/p$ and $^4$He/$p$ ratios, but significantly over-predicts the $^3$He/$p$ ratio.

A further challenge for the CSM is to achieve a simultaneous description of the $p/\pi$ ratio and light-nucleus-to-proton yield ratios. In particular, for $V_c=1.6\,dV/dy$, the CSM predicts a sizable suppression of the $p/\pi$ ratio in pp and p--Pb collisions that is not observed in the data~\cite{Vovchenko:2019kes}.

In Ref.~\cite{Sharma:2022poi}, this tension was addressed within a baryon-canonical formulation of the CSM by treating $V_c$ as a free parameter and determining it from the data at each multiplicity. In this approach, the ratio $V_c/V_A$ is allowed to vary with multiplicity, leading to a reasonable simultaneous description of the pion, proton, and deuteron yields in pp and p--Pb collisions. However, the yields of $^3$He and $^3_\Lambda$H remain over-predicted by approximately a factor of two.

A recent study~\cite{Andronic:2026nev} investigated nuclear-size corrections to the canonical SHM by applying asymptotic nuclear wave functions at chemical freeze-out. Within this prescription, the additional suppression generally worsens the description of the measured multiplicity dependence of light-(hyper)nucleus yields, particularly in small collision systems, compared with the canonical SHM without such corrections. This finding may be consistent with a picture in which (hyper)nucleus production is seeded by initially compact multiquark configurations that subsequently evolve into the physical bound states. Other recent developments, including those based on Contact theory~\cite{Bethe:1935zbl,Braaten:2004rn}, similarly require source-conditioned compact configurations at system freeze-out~\cite{Braaten:2024cke,Wang:2026rcf}.

\subsubsection{Hypernuclei and the \texorpdfstring{$S_3$}{S3} ratio}

Hypernuclei are incorporated into the thermal model analogously to ordinary nuclei, by including them as explicit degrees of freedom in the partition function. However, owing to their nonzero strangeness, hypernuclei are additionally affected by exact strangeness conservation in the CSM. The magnitude of this effect depends strongly on the collision energy. As shown in Fig.~\ref{fig:CSMBQS}, at LHC energies the effect of strangeness conservation on the $^3_\Lambda$H yield is negligible compared with that of baryon-number conservation.

A useful observable in this context is the so-called double ratio $S_3$~\cite{Zhang:2009ba,STAR:2010gyg}:

\begin{equation}
S_3 = \frac{N_{^3_\Lambda\text{H}} / N_{^3\text{He}}}{N_\Lambda / N_p}~.
\end{equation}

\begin{figure}[t]
\centering
\includegraphics[width=0.44\textwidth]{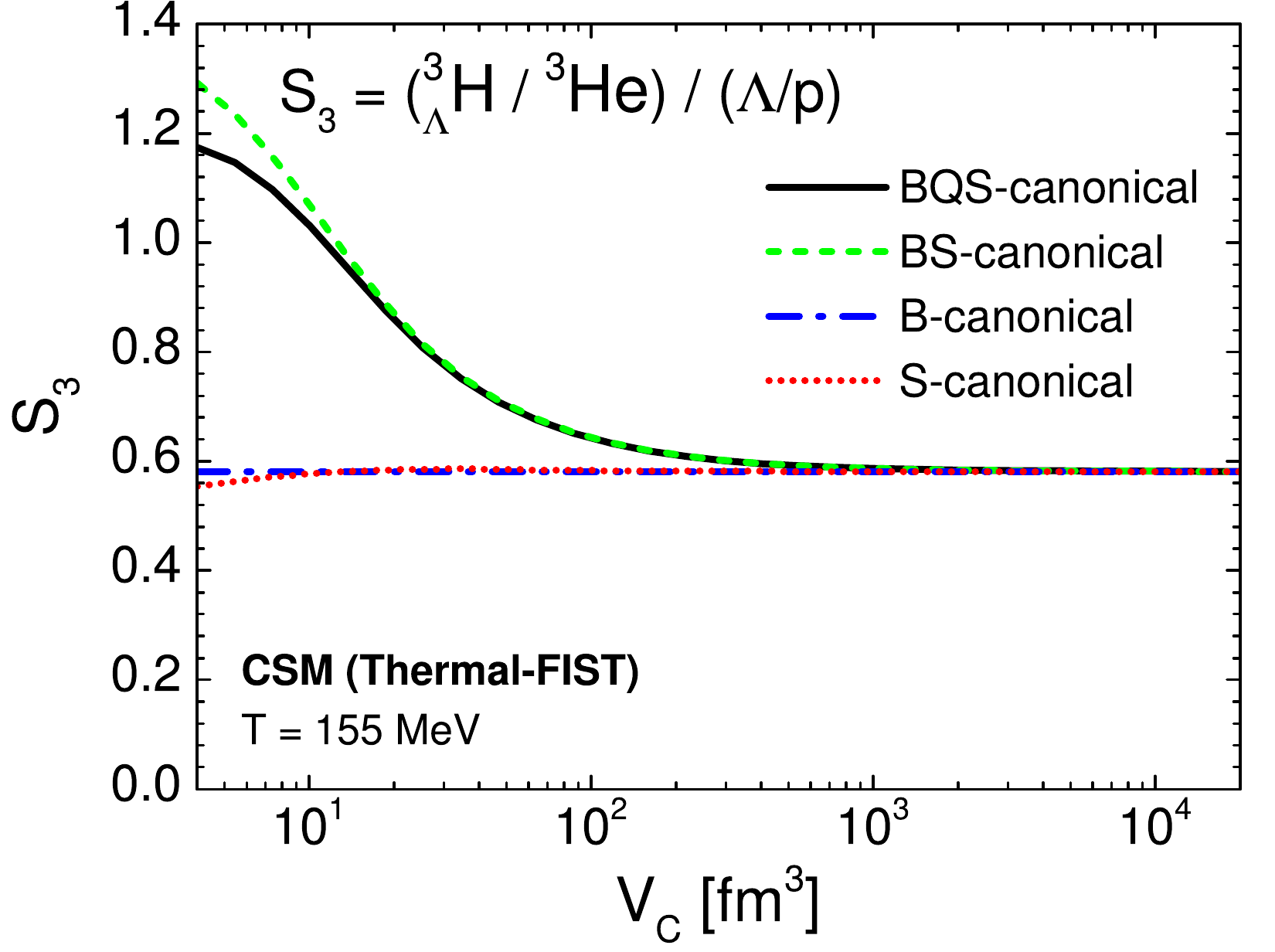}
\includegraphics[width=0.49\textwidth]{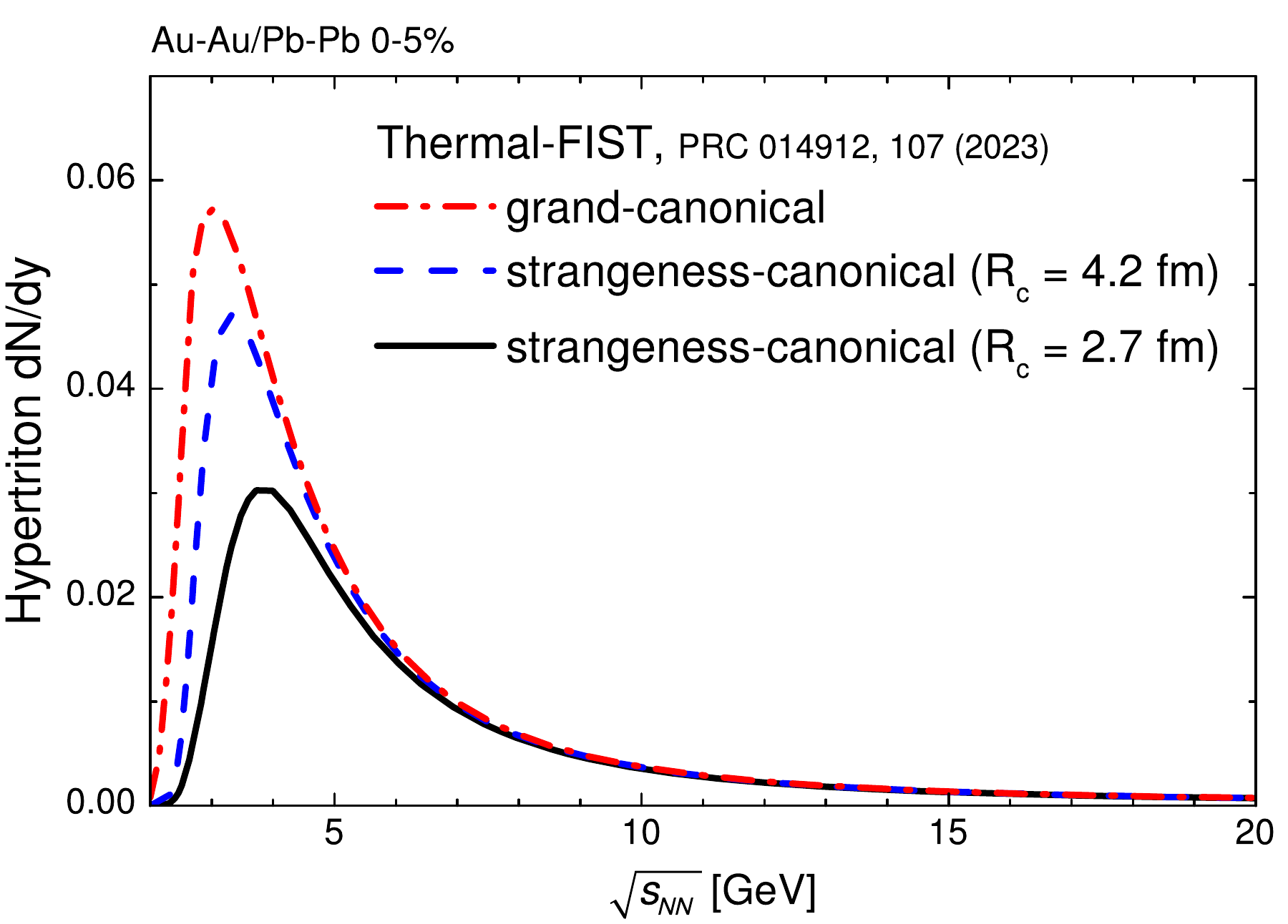}
\caption{
Left: Yields of d, $^3$He, $^3_\Lambda$H, and $^4$He as a function of the conservation volume $V_C$ in the CSM at $T=155$~MeV and vanishing conserved charges, normalized to their grand-canonical (infinite-volume) limits~\cite{Vovchenko:2018fiy}. Different lines indicate the charges treated canonically.
Right: Midrapidity hypertriton yield in $0--5\%$ central Au--Au/Pb--Pb collisions as a function of collision energy for $\sqrt{s_{\rm NN}}<20$~GeV~\cite{Reichert:2022mek}. Thermal-FIST calculations are shown for the grand-canonical ensemble (dash-dotted red) and the strangeness-canonical ensemble with $R_c=4.2$~fm (dashed blue) and $R_c=2.7$~fm (solid black).
}
\label{fig:CSMhyper}
\end{figure}

In this ratio, the fugacity factors cancel, as do, to a large extent, the canonical chemical factors entering Eq.~\eqref{eq:Ncanon}. The dependence of $S_3$ on $V_c$ at LHC energies is shown in the left panel of Fig.~\ref{fig:CSMhyper} for the $BQS$-, $BS$-, $B$-, and $S$-canonical versions of the CSM. The ratio is essentially independent of $V_c$ and coincides with the grand-canonical limit in the $B$- and $S$-canonical ensembles. In contrast, in the $BS$- and $BQS$-canonical ensembles, $S_3$ increases toward small $V_c$, corresponding to low multiplicities. This behavior arises from the interplay between exact baryon-number and strangeness conservation in small systems. In particular, the $\Lambda$ hyperon carries both baryon number and strangeness and is therefore simultaneously affected by both conservation laws.

At lower collision energies ($\sqrt{s_{\rm NN}} \lesssim 5$~GeV), canonical effects are qualitatively different, as exact strangeness conservation becomes important due to the reduced production of strange particles. The right panel of Fig.~\ref{fig:CSMhyper} shows the collision-energy dependence of the midrapidity $^3_\Lambda$H yield in $0--5\%$ central Au--Au/Pb--Pb collisions, calculated in the grand-canonical and strangeness-canonical ensembles~\cite{Reichert:2022mek}. The latter calculations use strangeness conservation radii of $R_c=4.2$ and $2.7$~fm, corresponding to $V_c=\frac{4\pi}{3}R_c^3$, as obtained from strangeness-canonical analyses of the $\phi/K^-$ and $\phi/\Xi^-$ ratios, respectively, in Au--Au collisions at $\sqrt{s_{\rm NN}}=3$~GeV~\cite{STAR:2021hyx}. The calculations predict a significant suppression of the $^3_\Lambda$H yield at $\sqrt{s_{\rm NN}}\lesssim5$~GeV, while the effect becomes negligible in central collisions at higher energies. Similar behavior is expected for other hypernuclei.

\subsubsection{Correlations and fluctuations}

Canonical conservation laws can significantly affect particle-number fluctuations and correlations in the HRG model, even in the large systems created in central heavy-ion collisions~\cite{Begun:2004gs,Vovchenko:2021kxx}. These effects also influence the moments of light-nuclei multiplicity distributions. As an illustration, we consider fluctuations and correlations of antideuterons and antiprotons, motivated by recent experimental measurements~\cite{ALICE:2022xiu,STAR:2023ebz}.

Consider an acceptance covering a fraction $\alpha$ of the full system, described by the HRG model with exact baryon-number conservation. In the thermodynamic limit, appropriate for central heavy-ion collisions, the second-order joint cumulants can be written as~\cite{Vovchenko:2020gne}
\begin{equation}
\kappa_{ij} = (1-\alpha) \kappa_{ij}^{\rm gce} + \alpha \kappa_{ij}^{\rm ce},
\end{equation}
where $\kappa_{ij}^{\rm gce}$ and $\kappa_{ij}^{\rm ce}$ denote the cumulants in the grand-canonical and canonical limits, respectively. The latter can be expressed in terms of grand-canonical cumulants as~\cite{Vovchenko:2020gne}
\begin{equation}
\kappa_{ij}^{\rm ce} = \kappa_{ij}^{\rm gce} - \frac{\kappa_{iB}^{\rm gce} \kappa_{Bj}^{\rm gce}}{\kappa_{BB}^{\rm gce}},
\end{equation}
where $B$ denotes the baryon number. In the ideal HRG, the grand-canonical cumulants are determined by the mean multiplicities:
\begin{align}
\kappa_{BB}^{\rm gce} &\simeq \langle B + \bar{B} \rangle,
\quad
\kappa_{\bar{p}\bar{p}}^{\rm gce} = -\kappa_{\bar{p}B}^{\rm gce} = \langle \bar{p} \rangle,
\quad
\kappa_{\bar{d}\bar{d}}^{\rm gce} = \langle \bar{d} \rangle,
\nonumber\
\kappa_{\bar{d}B}^{\rm gce} = -2 \langle \bar{d} \rangle,
\quad
\kappa_{\bar{p}\bar{d}}^{\rm gce} = 0.
\end{align}

Using these expressions, the scaled variance of the antiproton multiplicity is
\begin{align}
\label{eq:kappap}
\frac{\kappa_{\bar{p}\bar{p}}}{\langle \bar{p} \rangle}
= 1 - \alpha \frac{\langle \bar{p} \rangle}{\langle B + \bar{B} \rangle}
= 1 - \frac{\alpha_{\bar{p}}}{2},
\end{align}
where
$\alpha_{\bar{p}} = 2 \langle \bar{p} \rangle / \langle B + \bar{B} \rangle_c
= 2\alpha \langle \bar{p} \rangle / \langle B + \bar{B} \rangle$
and
$\langle B + \bar{B} \rangle_c = \langle B + \bar{B} \rangle/\alpha$
is the total mean number of baryons and antibaryons in the conservation volume $V_c$ (corresponding to $\alpha=1$). Similarly, for antideuterons one obtains
\begin{align}
\frac{\kappa_{\bar{d}\bar{d}}}{\langle \bar{d} \rangle}
= 1 - 4 \frac{\langle \bar{d} \rangle}{\langle B + \bar{B} \rangle_c}
= 1 - 2 \alpha_{\bar{p}} R_{\bar{d}\bar{p}},
\end{align}
where $R_{\bar{d}\bar{p}} = \langle \bar{d} \rangle / \langle \bar{p} \rangle$ is the antideuteron-to-antiproton ratio.

\begin{figure}[!hbt]
\centering
\includegraphics[width=0.45\linewidth]{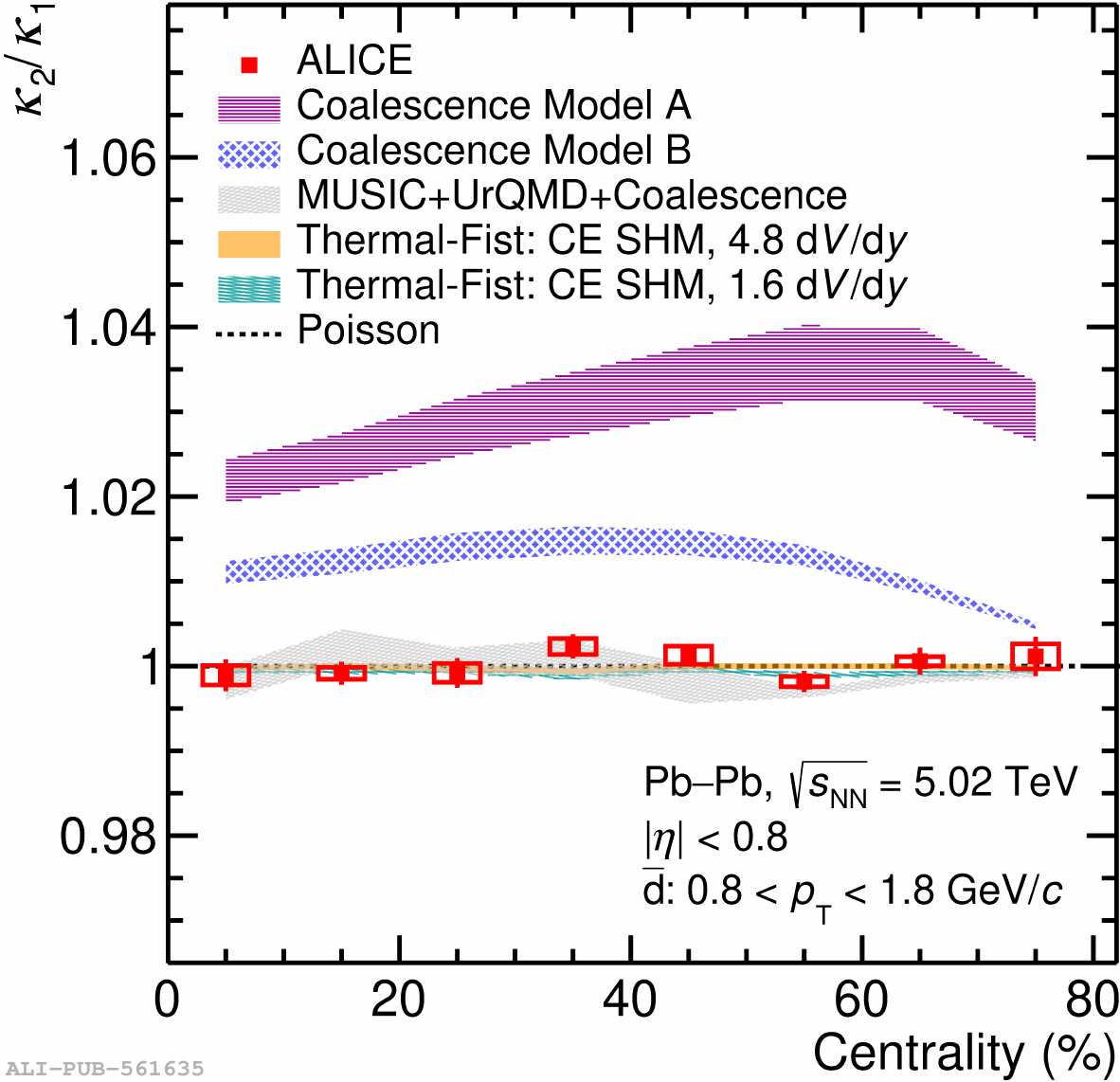}
\caption{Scaled variance of the antideuteron multiplicity as a function
of centrality in Pb--Pb collisions at $\sqrt{s_{\rm NN}}=5.02$~TeV.
Figure reproduced from Ref.~\cite{ALICE:2022xiu}.}
\label{fig:dflucs}
\end{figure}

The deviation of $\kappa_{\bar{d}\bar{d}}/\langle \bar{d} \rangle$ from unity is expected to be very small. At $\sqrt{s_{\rm NN}}=5.02$~TeV, ALICE measurements of net-proton cumulants~\cite{ALICE:2022xpf} and deuteron production~\cite{ALICE:2022veq} imply $\alpha_{\bar{p}}\lesssim0.06$ and $R_{\bar{d}\bar{p}}\lesssim0.004$, yielding $\kappa_{\bar{d}\bar{d}}/\langle \bar{d} \rangle \simeq 0.9995$. Such a small deviation from unity would be extremely challenging to measure. The available ALICE measurement is indeed consistent with unity~\cite{ALICE:2022xiu}, as shown in Fig.~\ref{fig:dflucs}.

For the antideuteron--antiproton correlation coefficient,
\begin{equation}
\rho_{\bar{d}\bar{p}}
= \frac{\kappa_{\bar{d}\bar{p}}}
{\sqrt{\kappa_{\bar{d}\bar{d}}}\sqrt{\kappa_{\bar{p}\bar{p}}}},
\end{equation}
one obtains
\begin{align}
\rho_{\bar{d}\bar{p}}
&= -2\frac{\langle \bar{d} \rangle \langle \bar{p} \rangle}
{\langle B + \bar{B} \rangle_c
\sqrt{\langle \bar{d} \rangle (1-2\alpha_{\bar{p}}R_{\bar{d}\bar{p}})}
\sqrt{\langle \bar{p} \rangle (1-\alpha_{\bar{p}}/2)}}
\nonumber\\
&= -\alpha_{\bar{p}}\sqrt{R_{\bar{d}\bar{p}}}
+ O(\alpha_{\bar{p}}^2).
\end{align}
  
This gives an estimated correlation coefficient of $\rho_{\bar{d}\bar{p}}\simeq-0.004$, consistent with the ALICE measurement~\cite{ALICE:2022xiu}.

Measurements of $\kappa_{\bar{p}\bar{p}}$, $\kappa_{\bar{d}\bar{d}}$, and $\rho_{\bar{d}\bar{p}}$ can therefore be used to test whether the observed fluctuations and correlations are fully explained by canonical conservation effects. Eliminating $\alpha_{\bar{p}}$ using Eq.~\eqref{eq:kappap} gives
\begin{align}
\rho_{\bar{d}\bar{p}}
\simeq 2\left(
\frac{\kappa_{\bar{p}\bar{p}}}{\langle \bar{p} \rangle} - 1
\right)
\sqrt{R_{\bar{d}\bar{p}}},
\end{align}
providing a direct relation that can be tested experimentally.

The acceptance dependence of $\kappa_{\bar{d}\bar{p}}$ or $\kappa_{\bar{p}\bar{p}}$ can be used to constrain the correlation volume $V_c$. Assuming that the mean antiproton multiplicity scales linearly with the acceptance window $\Delta\eta$ and that the mean (anti)baryon multiplicity scales with $V_c$, one obtains $\alpha_{\bar{p}}\propto\Delta\eta/V_c$, and therefore
\begin{align}
\rho_{\bar{d}\bar{p}} \propto -\frac{\Delta\eta}{V_c}.
\end{align}
Thus, $V_c$ determines the slope of the approximately linear dependence of the $\bar{d}$--$\bar{p}$ correlation coefficient on $\Delta\eta$.

\begin{figure}[!hbt]
\centering
\includegraphics[width=0.9\linewidth]{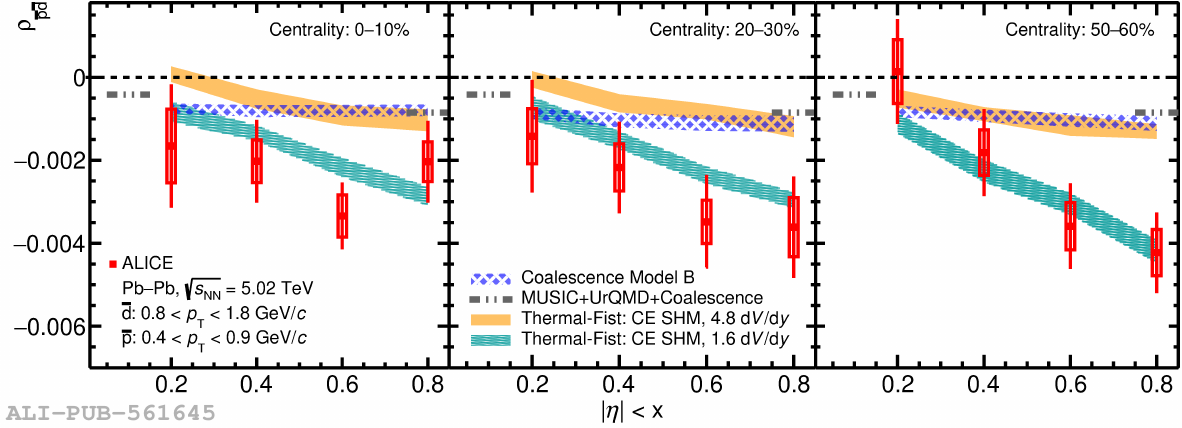}
\caption{Antideuteron--antiproton correlation coefficient
$\rho_{\bar{d}\bar{p}}$ as a function of the pseudorapidity acceptance
window $\Delta\eta$ for three centrality intervals in Pb--Pb collisions
at $\sqrt{s_{\rm NN}}=5.02$~TeV.
Figure reproduced from Ref.~\cite{ALICE:2022xiu}.}
\label{fig:dpcorr}
\end{figure}

The above derivation assumes a uniform thermal system and acceptances for
$\bar{d}$ and $\bar{p}$ such that the yield ratio $R_{\bar{d}\bar{p}}$
is the same as in the full conservation volume.
A more realistic analysis was performed by the ALICE Collaboration using
a blast-wave model and experimental kinematic cuts~\cite{ALICE:2022xiu}.
The measured scaled variance of the antideuteron multiplicity is shown
in Fig.~\ref{fig:dflucs}, while the antideuteron--antiproton correlation
coefficient as a function of the pseudorapidity acceptance window is
presented in Fig.~\ref{fig:dpcorr}.
The results largely confirm the above expectations and constrain the
conservation volume $V_c$ to approximately 1.6 units of rapidity.

The canonical thermal-model description of light nuclei can be further tested using other cumulants and particle correlations. For instance, assuming that net-$\Lambda$ fluctuations and their correlations with antideuterons are driven solely by baryon-number conservation, one obtains
\begin{align}
\rho_{\bar{d}\Delta\Lambda}
&= 2\alpha_{\Delta\Lambda}
\sqrt{\frac{\langle \bar{d} \rangle}
{\langle \Lambda + \bar{\Lambda} \rangle}}
+ O(\alpha_{\Delta\Lambda}^2)
\nonumber\\
&\simeq -2\left(
\frac{\kappa_2[\Lambda - \bar{\Lambda}]}
{\langle \Lambda + \bar{\Lambda} \rangle} - 1
\right)
\sqrt{\frac{\langle \bar{d} \rangle}
{\langle \Lambda + \bar{\Lambda} \rangle}}~.
\end{align}
  
This relation provides a direct experimental test of the canonical thermal-model description.

\section{Coalescence model}
\label{subsec:coalescence}

The coalescence model was introduced in the early 1960s to describe the deuteron momentum spectrum observed in high-energy proton--nucleus collisions. Based on a second-order perturbative treatment of proton--neutron reactions catalyzed by the nuclear mean field~\cite{Butler:1961pr}, it was found that the deuteron yield at a given momentum is proportional to the product of the proton and neutron yields evaluated at half the deuteron momentum~\cite{Butler:1963pp}. Equivalently, the deuteron-to-proton ratio can be related to the probability of finding a neutron within a small sphere around a proton in momentum space~\cite{Schwarzschild:1963zz}. Treating the radius of this coalescence sphere as a free parameter, the model was subsequently found to provide a good description of the momentum spectra of deuterons, tritons, helium-3, and helium-4 measured in high-energy heavy-ion collisions at the Bevalac~\cite{Gutbrod:1976zzr}.

The success of the coalescence model subsequently motivated its reformulation within the sudden approximation of quantum mechanics, in terms of the overlap between the nucleon phase-space distributions and the nucleus phase-space distributions~\cite{Bond:1977fd}, or equivalently between the multi-nucleon density matrix and the internal wave functions of the nuclei~\cite{Sato:1981ez,Remler:1981du,Mrowczynski:1987oid,Scheibl:1998tk,Zhao:2025glf}. It was also realized that the same proton--neutron final-state interaction underlies both deuteron production and proton--neutron correlations in heavy-ion collisions~\cite{Mrowczynski:1992gc,Blum:2019suo,Bellini:2020cbj}.

Furthermore, microscopic transport approaches based on multi-nucleon reactions have been developed to describe light-nucleus production in heavy-ion collisions and to investigate their connection to the coalescence picture~\cite{Remler:1982zz,Danielewicz:1991dh,Danielewicz:1992pei,Oh:2009gx,Sun:2022rjh}. These kinetic approaches are discussed in detail in Sec.~\ref{sec:kinetic cluster}.

Recently, the coalescence model has been widely used to describe the production of light (anti)nuclei and (anti)hypernuclei at RHIC and the LHC~\cite{Sun:2015jta,Sun:2015ulc,Coalescence4,Sun:2017ooe,Sun:2018mqq,Liu:2019nii,Sun:2020uoj,Sun:2022xjr,Liu:2024ygk,Sun:2022rjh,Mrowczynski:2020ugu,Zhao:2022xkz,Hillmann:2021zgj,Mahlein:2023fmx,Florkowski:2023uim}. It has been applied to explain the suppression of light-(hyper)nucleus production in small collision systems at the LHC~\cite{Liu:2019nii}, the softening of the hypertriton $p_T$ spectrum~\cite{Liu:2024ygk}, the enhanced deuteron coalescence parameter $B_2$ in energetic jets relative to the underlying event~\cite{ALICE:2020hjy,ALICE:2022ugx,Mrowczynski:2023hbn,ALICE:2026vqn}, and deuteron production in cosmic rays~\cite{Dal:2015sha,Blum:2017qnn}. The coalescence model has also been used to investigate the effects of baryon-density fluctuations and correlations on light-nucleus production, which may provide information relevant to the search for the QCD critical point~\cite{Sun:2017xrx,Sun:2018jhg,Sun:2020zxy}. Furthermore, it has been applied to studies of the spin polarization and alignment of particles with non-zero spin in non-central heavy-ion collisions~\cite{Liang:2004ph,Liang:2004xn,Sun:2024vpp}. Some of these applications and their underlying physics are discussed below.

\subsection{Suppressed production in small collision systems}

In the phase-space coalescence model~\cite{Scheibl:1998tk}, the formation probability of a light nucleus in heavy-ion collisions is determined by the overlap between the nucleon phase-space distributions in the emission source and the Wigner function of the light nucleus, obtained from the Wigner transform of its internal wave function~\cite{Chen:2003qj,Chen:2003ava}. The internal wave functions, and hence the spatial sizes, of light (hyper)nuclei therefore play a crucial role in their production. In particular, it has been shown that the finite sizes of light (hyper)nuclei lead to a pronounced suppression of their yields in small collision systems~\cite{Sun:2018mqq}, a phenomenon observed in various ALICE measurements at the LHC~\cite{ALICE:2017xrp,ALICE:2019fee,ALICE:2020foi,ALICE:2021ovi,ALICE:2021mfm,ALICE:2021sdc}. In the following, a simple coalescence model is used to illustrate the underlying physics of this phenomenon.

For deuteron production in heavy-ion collisions, the deuteron yield in the coalescence model is given by
\begin{align}
N_{\text{d}}
&=
g_{\text{d}}
\int \text{d}^3{\bf x}_1
\int \text{d}^3{\bf k}_1
\int \text{d}^3{\bf x}_2
\int \text{d}^3{\bf k}_2\,
f_n({\bf x}_1,{\bf k}_1)
\notag\\
&\quad\times
f_p({\bf x}_2,{\bf k}_2)
W_{\text{d}}\left(
\frac{{\bf x}_1-{\bf x}_2}{\sqrt{2}},
\frac{{\bf k}_1-{\bf k}_2}{\sqrt{2}}
\right).
\label{Eq1}
\end{align}

where $g_{\rm d}=3/4$ is the statistical factor for forming a spin-one deuteron from a spin-half proton and neutron, $f_{p,n}({\bf x},{\bf k})$ denote the proton and neutron phase-space distributions, and $W_{\rm d}({\bf x},{\bf k})$ is the Wigner function of the deuteron.

Neglecting collective flow, the nucleon phase-space distributions in a thermalized spherical fireball with kinetic freeze-out temperature $T_K$ and radius $R$ can be written as
\begin{eqnarray}
f_{p,n}({\bf x},{\bf k})
=\frac{N_{p,n}}{(2\pi)^3(mT_KR^2)^{3/2}}
\exp\left(-\frac{k^2}{2mT_K}-\frac{x^2}{2R^2}\right),
\end{eqnarray}
where $m$ is the nucleon mass. The distributions are normalized to the proton and neutron yields according to
\begin{eqnarray}
N_{p,n}=\int \text{d}^3{\bf x}\int \text{d}^3{\bf k},
f_{p,n}({\bf x},{\bf k}).
\end{eqnarray}

Approximating the internal deuteron wave function by that of a harmonic oscillator, as commonly done in coalescence calculations, its Wigner function takes the Gaussian form~\cite{Sun:2015jta,Sun:2015ulc,Zhu:2015voa},
\begin{eqnarray}
W_\text{d}({\bf x},{\bf k})
=8\exp\left(-\frac{x^2}{\sigma^2}\right)
\exp\left(-\sigma^2k^2\right),
\label{Eq2}
\end{eqnarray}
with the normalization
\begin{eqnarray}
\int \text{d}^3{\bf x}\int \text{d}^3{\bf k},
W_\text{d}({\bf x},{\bf k})=(2\pi)^3.
\end{eqnarray}

Introducing the center-of-mass and relative coordinates and momenta,
\begin{align}
{\bf X}
&=
\frac{{\bf x}_1+{\bf x}_2}{2},
\qquad
{\bf x}
=
\frac{{\bf x}_1-{\bf x}_2}{\sqrt{2}},
\notag\\
{\bf K}
&=
{\bf k}_1+{\bf k}_2,
\qquad
{\bf k}
=
\frac{{\bf k}_1-{\bf k}_2}{\sqrt{2}},
\label{Eq3}
\end{align}
which give 
\begin{align}
x_1^2+x_2^2
&=
2X^2+x^2,
\notag\\
k_1^2+k_2^2
&=
\frac{K^2}{2}+k^2,
\end{align}
the integrals in Eq.~(\ref{Eq1}) can then be evaluated analytically, yielding
\begin{align}
N_{\text{d}}
&=
\frac{8g_{\rm d}N_pN_n}{(2\pi)^6(mT_KR^2)^3}
\int \text{d}^3{\bf X}\,
e^{-X^2/R^2}
\int \text{d}^3{\bf x}\,
e^{-\left(\frac{1}{\sigma^2}+\frac{1}{2R^2}\right)x^2}
\notag\\
&\quad\times
\int \text{d}^3{\bf K}\,
e^{-K^2/(4mT_K)}
\int \text{d}^3{\bf k}\,
e^{-\left(\sigma^2+\frac{1}{2mT_K}\right)k^2}
\notag\\
&=
\frac{3N_nN_p}{4(mT_KR^2)^{3/2}}\,
\frac{1}
{\left(1+\frac{1}{2mT_K\sigma^2}\right)^{3/2}}
\frac{1}
{\left(1+\frac{\sigma^2}{2R^2}\right)^{3/2}}
\notag\\
&\approx
\frac{3N_nN_p}{4(mT_KR^2)^{3/2}}\,
\frac{1}
{\left(1+\frac{\sigma^2}{2R^2}\right)^{3/2}}.
\label{Eq4}
\end{align}
where, in the last step, the condition $mT_K\sigma^2\gg1$ has been used.

The parameter $\sigma$ in Eq.~(\ref{Eq2}) is related to the root-mean-square matter radius of the deuteron, $r_{\rm d}=1.96~{\rm fm}$~\cite{Ropke:2008qk}, through
\begin{eqnarray}
\sigma=\sqrt{\frac{4}{3}},r_{\rm d}\approx2.26~{\rm fm}.
\end{eqnarray}
For a typical nucleon kinetic freeze-out temperature of $T_K\sim100$ MeV, the condition $mT_K\gg1/\sigma^2$ or equivalently $mT_K\sigma^2\gg1$, i.e., the thermal wavelength of the nucleon is much smaller than the size of the deuteron, is well satisfied, justifying the approximation used in the final expression of Eq.~(\ref{Eq4}).

Similarly, the hypertriton (\hyt) yield is given by
\begin{eqnarray}
N_{^3_\Lambda\mathrm{H}}
&\approx&
\frac{N_nN_pN_\Lambda}{4(mT_KR^2)^3}
\left(\frac{2m+m_\Lambda}{3m_\Lambda}\right)^{3/2}
\frac{1}{\left(1+\frac{\sigma_1^2}{2R^2}\right)^{3/2}}
\frac{1}{\left(1+\frac{\sigma_2^2}{2R^2}\right)^{3/2}},
\label{Eq5}
\end{eqnarray}
where $\sigma_1$ and $\sigma_2$ characterize the hypertriton Wigner function,

\begin{eqnarray}
W_\mathrm{ht}
&=&
8^2
\exp\left(-\frac{x^2}{\sigma_1^2}-\sigma_1^2 k_x^2\right)
\exp\left(-\frac{\lambda^2}{\sigma_2^2}-\sigma_2^2 k_\lambda^2\right),
\label{Eq:wig}
\end{eqnarray}
defined in terms of the Jacobi relative coordinates and momenta
\begin{align}
{\bf x}
&=
\frac{{\bf x}_1-{\bf x}_2}{\sqrt{2}},
\qquad
{\boldsymbol\lambda}
=
\sqrt{\frac{2}{3}}
\left(
\frac{m_1{\bf x}_1+m_2{\bf x}_2}{m_1+m_2}
-{\bf x}_3
\right),
\notag\\
{\bf k}_x
&=
\sqrt{2}\,
\frac{m_2{\bf k}_1-m_1{\bf k}_2}{m_1+m_2},
\qquad
{\bf k}_\lambda
=
\sqrt{\frac{3}{2}}\,
\frac{m_3({\bf k}_1+{\bf k}_2)-(m_1+m_2){\bf k}_3}
{m_1+m_2+m_3}.
\label{Eq:coordinate}
\end{align}

Here, $m_1=m_n$, $m_2=m_p$, and $m_3=m_\Lambda$ denote the neutron, proton, and $\Lambda$ masses, respectively, while ${\bf x}_i$ and ${\bf k}_i$ denote their corresponding spatial coordinates and momenta. In Eq.~(\ref{Eq5}), $m=m_n\simeq m_p$ has been assumed. The mass-dependent factor
\begin{eqnarray}
\left(\frac{2m+m_\Lambda}{3m_\Lambda}\right)^{3/2}\approx0.85
\end{eqnarray}
was neglected in Ref.~\cite{Sun:2018mqq}.

Under the simplifying assumption that the nucleons and the $\Lambda$ hyperon are equally distributed within the hypertriton (see, however, the discussion below), one has $\sigma_1=\sigma_2=\sqrt{\langle r_{^3_\Lambda\mathrm{H}}^2\rangle}=4.9$ fm, which leads to~\cite{Sun:2018mqq}
\begin{eqnarray}
\frac{N_{^3_\Lambda\mathrm{H}}}{N_\Lambda}
\approx
\frac{0.85\times7.1\times10^{-6}}
{\left[1+\left(\frac{3.46~\mathrm{fm}}{R}\right)^2\right]^3}.
\notag
\end{eqnarray}

This expression relies on an oversimplified description of the hypertriton wave function. In realistic descriptions of the hypertriton, the $\Lambda$, neutron, and proton are not equally distributed within the bound state~\cite{Hildenbrand:2019sgp}. The above equation should therefore not be interpreted as a quantitative prediction of the coalescence model, but rather as an illustration of the sensitivity of coalescence calculations to the assumed nuclear wave function. This distinction is particularly relevant because the expression has also been compared directly with experimental data~\cite{hypppB}, which may lead to its interpretation as a quantitative coalescence prediction.

A more realistic description of the hypertriton wave function can be obtained by exploiting its loosely bound deuteron--$\Lambda$ structure. In this case, $\sigma_1=2.26$ fm characterizes the internal structure of the deuteron core, while $\sigma_2$ can be inferred from the $\Lambda$ separation energy $B_\Lambda$ according to

\begin{equation}
\sigma_2 =
\left[
\frac{2.15}{\sqrt{B_\Lambda/\mathrm{MeV}}}
+1.23
\right]\mathrm{fm}.
\end{equation}

where $B_\Lambda$ is expressed in MeV and $\sigma_2$ in fm. For $B_\Lambda=0.17$ MeV~\cite{Chen:2023mel}, this gives $\sigma_2=6.4$ fm, corresponding to a root-mean-square separation between the deuteron core and the $\Lambda$ hyperon of $\sqrt{\langle r_{\Lambda d}^2\rangle}=9.6$ fm. The hypertriton-to-$\Lambda$ yield ratio then becomes
\begin{eqnarray}
\label{eq:HT2b}
\frac{N_{^3_\Lambda\mathrm{H}}}{N_\Lambda}
\approx
\frac{0.85\times7.1\times10^{-6}}
{\left[1+\left(\frac{1.6~\mathrm{fm}}{R}\right)^2\right]^{3/2}
\left[1+\left(\frac{4.5~\mathrm{fm}}{R}\right)^2\right]^{3/2}}.
\end{eqnarray}

Equation~(\ref{eq:HT2b}) should not be interpreted as a two-body coalescence model in which deuterons are assumed to be present at kinetic freeze-out. Rather, the deuteron--$\Lambda$ picture introduced above provides a simplified description of the internal three-body hypertriton wave function. A full coalescence calculation using a three-body hypertriton wave function, without assuming the presence of deuterons at freeze-out, was presented in Ref.~\cite{Bellini:2020cbj,Zhao:2025glf}. Within a Gaussian approximation of the corresponding Wigner functions, the result of that calculation reduces to the form of Eq.~(\ref{eq:HT2b}).

\subsection{Softening of hypertriton transverse momentum spectrum}

A recent study has shown that the halo structure of the hypertriton, characterized by a large $\Lambda-d$ distance of approximately $10$ fm, leads not only to a suppression of the \hyt~yield but also to a softening of its transverse momentum ($p_T$) spectrum~\cite{Liu:2024ygk}. This effect can be illustrated within the blast-wave model, which assumes both the hypertriton and helium-3 to be in thermal equilibrium at the kinetic freeze-out temperature $T_{\rm kin}$, with an invariant momentum spectrum of the form~\cite{Cooper:1974mv,Schnedermann:1993ws}:
\begin{equation}
f_\text{BL}\equiv E\frac{d^3N}{dp^3} \propto \int m_T I_0\left(\frac{p_T\sinh(\rho)}{T_\text{kin}}\right) K_1\left(\frac{m_T\cosh(\rho)}{T_\text{kin}}\right)r,\text{d}r.
\label{eq:bw_pt}
\end{equation}
Here, $I_0$ and $K_1$ are modified Bessel functions, $m_T=\sqrt{m^2+p_T^2}$ is the transverse mass, and $\rho = \tanh^{-1}[(r/R_0)^n\beta_s]$ is the flow rapidity profile. The quantities $r$, $R_0$, $\beta_s$, and $n$ denote the radial distance in the transverse plane, the transverse radius of the fireball, the transverse expansion velocity at its surface, and the exponent of the velocity profile, respectively. For large $m_T$, $f_\text{BL}$ takes the form $f_\text{BL} \propto \text{exp}(-m_T/T_\text{eff})$, where $T_\text{eff}=T_{\rm kin}\sqrt{\frac{1+\langle\beta_T\rangle}{1-\langle\beta_T\rangle}}$ is the effective (blue-shifted) temperature. The ratio of the hypertriton to helium-3 spectra at large $m_T$ is then given by
\begin{equation}
\frac{f_\text{BL}^\text{ht}}{f_\text{BL}^\text{he3}} \propto \text{exp}\left(-\frac{m^2_\text{ht}-m^2_\text{he3}}{\left(\sqrt{m^2_\text{ht}+p_T^2}+\sqrt{m^2_\text{he3}+p_T^2}\right) T_\text{eff}}\right),
\end{equation}
which increases with increasing $p_T$ because radial flow boosts the heavier hypertriton to higher transverse momentum than the lighter helium-3.

In the coalescence model, on the other hand, the ratio of the hypertriton to helium-3 $p_T$ spectra is approximately given by~\cite{Scheibl:1998tk,Bellini:2018epz,Blum:2019suo}
\begin{equation}
\frac{f_\text{COAL}^\text{ht}}{f_\text{COAL}^\text{he3}}\approx \frac{f_\text{BL}^\text{ht}}{f_\text{BL}^\text{he3}}
\frac{\left[1+\frac{\sigma_\text{he3}^2}{2R^2(p_T)}\right]^{3}}{\left[1+\frac{\sigma_1^2}{2R^2(p_T)}\right]^{3/2}\left[1+\frac{\sigma_2^2}{2R^2(p_T)}\right]^{3/2}}.
\label{eq:qs}
\end{equation}
where $R(p_T)$ is the Gaussian homogeneity radius of the nucleon and $\Lambda$ emission source for hypertritons and helium-3 nuclei produced at transverse momentum $p_T$. The last factor, hereafter denoted by $\mathcal{C}$, accounts for quantum corrections associated with the hypertriton wave function~\cite{Scheibl:1998tk,Sun:2018mqq}. Since $R(p_T)$ decreases with increasing $p_T$~\cite{ALICE:2011kmy,ALICE:2015hvw,ALICE:2023zbh} and $\sigma_2\gg \sigma_\text{he3}\sim \sigma_1$, the quantum correction factor $\mathcal{C}$ increasingly suppresses hypertriton production relative to helium-3 at larger $p_T$.

\begin{figure*}[!t]
\centering
\includegraphics[width=0.95\linewidth]{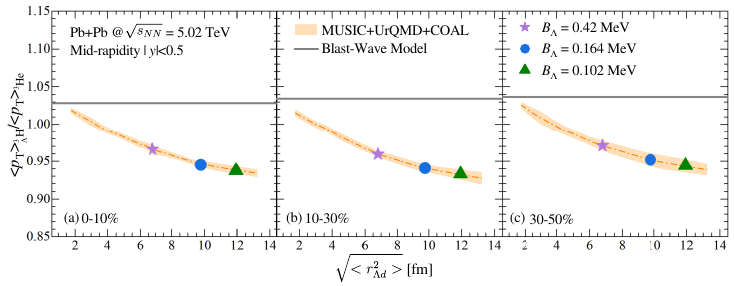}
\caption{Ratio of the mean transverse momenta of hypertritons and helium-3 nuclei, $\langle p_T\rangle_{^3_\Lambda\text{H}}/\langle p_T\rangle_{^3\text{He}}$, as a function of the $\Lambda-d$ distance in Pb+Pb collisions at $\sqrt{s_{NN}} = 5.02$ TeV for centralities of 0--10\% (a), 10--30\% (b), and 30--50\% (c). The solid lines show the blast-wave model predictions, while the shaded bands show the coalescence model predictions. The solid symbols indicate coalescence results obtained for different values of the hypertriton $\Lambda$ separation energy $B_\Lambda$. Figure taken from Ref.~\cite{Liu:2024ygk}.}
\label{fig:average_pt}
\end{figure*}

Figure~\ref{fig:average_pt} shows the ratio $\langle p_T\rangle_{^3_\Lambda\text{H}}/\langle p_T\rangle_{^3\text{He}}$ of the hypertriton to helium-3 mean transverse momenta as a function of the $\Lambda-d$ distance. The shaded bands represent the coalescence model predictions, while the solid symbols correspond to results obtained for different values of the hypertriton $\Lambda$ separation energy $B_\Lambda$. In the coalescence model, the mean-$p_T$ ratio is smaller than unity and decreases with increasing $\Lambda-d$ distance, or equivalently with decreasing $B_\Lambda$, with a variation of order 10\%. In contrast, the blast-wave model predictions, shown by the solid lines, yield a ratio slightly above unity, ranging from approximately 1.03 for 0--10\% central collisions to about 1.04 for 30--50\% central collisions.

This difference between the coalescence and blast-wave predictions for the hypertriton-to-helium-3 mean-$p_T$ ratio provides a complementary observable to the suppression of hypertriton production in peripheral collisions. The latter can be described both by coalescence calculations incorporating the hypertriton wave function and by thermal models with a canonical correlation volume~\cite{Vovchenko:2018fiy}. By contrast, a softening of the hypertriton $p_T$ spectrum and the corresponding reduction of its mean transverse momentum relative to helium-3 would provide a more direct discriminator between the two production scenarios, despite the relatively small magnitude of the predicted effect.

Such quantum-mechanical softening of the (anti)hypertriton spectrum is a general consequence of incorporating the extended hypertriton wave function in coalescence approaches based on final-state interactions. The effect can be tested experimentally in high-energy nuclear collisions at different collision energies and in different collision systems, providing further insight into the production mechanism of (anti)hypernuclei and potentially constraining the $\Lambda$--nucleon interaction.

\subsection{Nucleon-number scaling of light-nuclei collective flow}

In the coalescence model, the small binding energies of light (hyper-)nuclei require the coalescing nucleons to have approximately equal velocities. This suggests that the azimuthal-angle distribution of a nucleus with mass number $A$ can be factorized as $F_A(\phi)\propto f(\phi)^A$, leading to scaling relations for the anisotropic flow coefficients~\cite{Kolb:2004gi}. For deuterons, this results in the following approximate relation~\cite{Molnar:2003ff}:
\begin{equation}
v_{2(3),\mathrm{d}}(p_{\rm T}) \approx \frac{2 v_{2(3),\mathrm{p}} (p_{\rm {T}}/2)}{1+2{v^2_{2(3),\rm{p}}} (p_{\mathrm {T}}/2)},
\label{eq:v2coal}
\end{equation}
which reduces to the naive scaling relation $v_{2(3),\mathrm{d}}(p_{\rm T}) = 2 v_{2(3),\mathrm{p}}(p_{\rm {T}}/2)$ when $v_{2(3)}$ is small.

\begin{figure}[!htb]
\begin{center}
\includegraphics[width=0.49\textwidth]{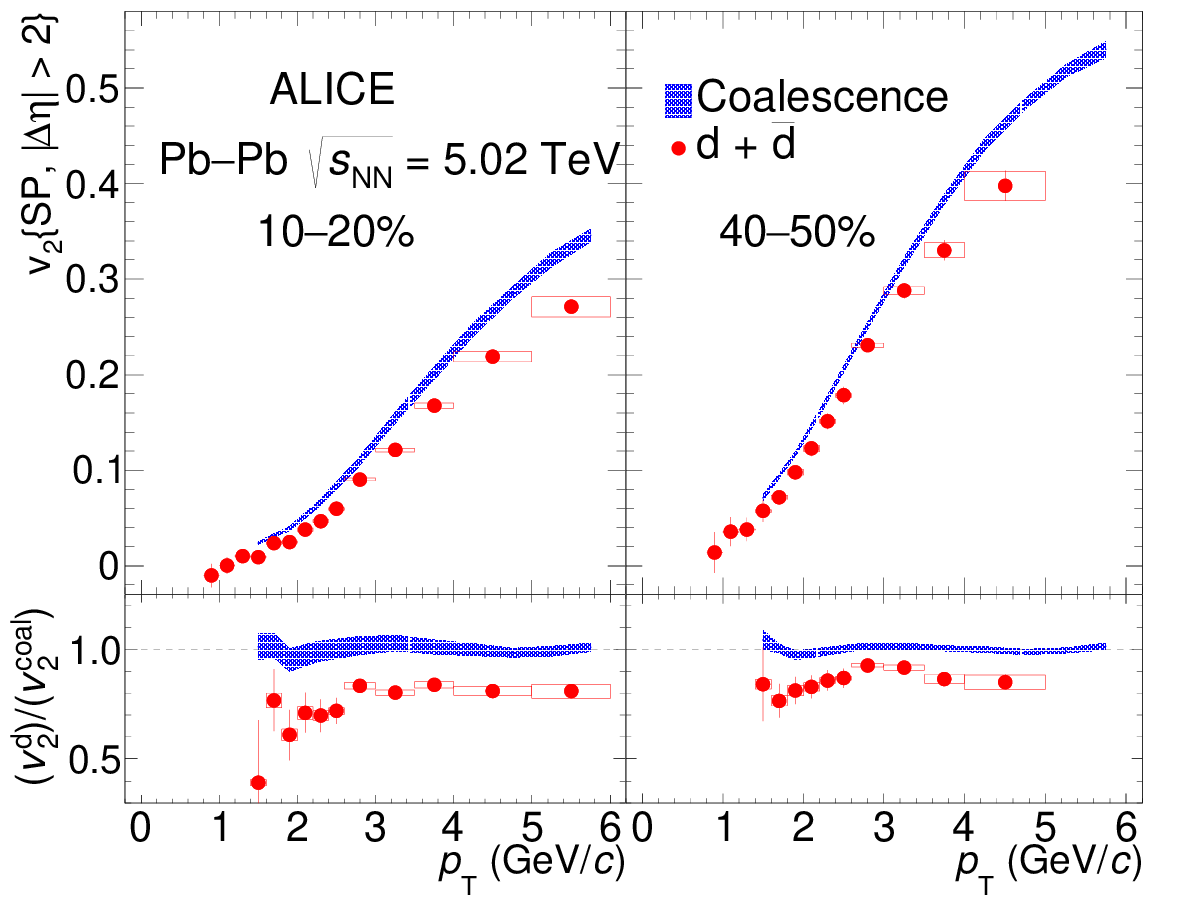}
\includegraphics[width=0.49\textwidth]{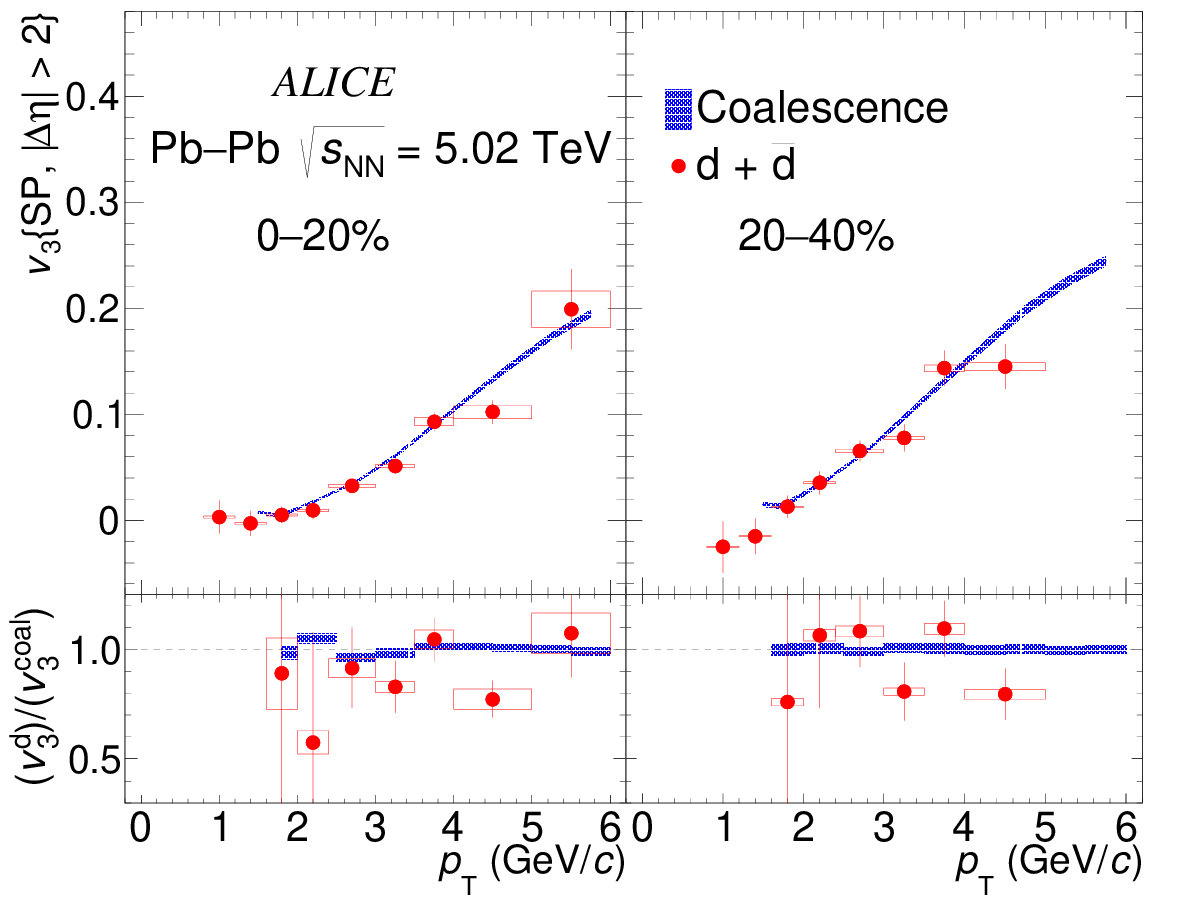}
\caption{Measured deuteron $v_2$ and $v_3$ (red circles) compared with expectations from a simple coalescence model assuming nucleons with equal velocities (Eq.~(\ref{eq:v2coal}); blue shaded bands) for different centrality intervals. The left panels show $v_2$ in the 10--20$\%$ and 40--50$\%$ centrality intervals, while the right panels show $v_3$ in the 0--20$\%$ and 20--40$\%$ centrality intervals. The bottom panels show the ratios of the measured $v_2$ and $v_3$ values to the corresponding coalescence-model expectations. Vertical bars and boxes represent statistical and systematic uncertainties, respectively. The horizontal line at unity is shown to guide the eye. Figure taken from Ref.~\cite{ALICE:2020chv}.}
\label{fig:CoalComparisonv2v3}
\end{center}
\end{figure}

The left panels of Fig.~\ref{fig:CoalComparisonv2v3} compare the measured deuteron elliptic flow $v_2$ in the 10--20\% and 40--50\% centrality intervals with the coalescence-model expectations obtained from Eq.~(\ref{eq:v2coal}) using the measured proton $v_2$. Similarly, the right panels compare the measured deuteron triangular flow $v_3$ in the 0--20\% and 20--40\% centrality intervals with the corresponding coalescence-model expectations~\cite{ALICE:2020chv}. The coalescence model overestimates the deuteron $v_2$ by about 20$\%$--30$\%$ in central collisions, while it provides a better description of the data in semi-peripheral collisions. For deuteron $v_3$, the coalescence model shows somewhat better agreement with the measurements; however, the large statistical uncertainties prevent firm conclusions. 

\subsection{Deuteron production in jets} 
\label{dinjets}

In the coalescence model, in addition to the light-nucleus wave function, the properties of the emission source play a crucial role in determining light-nucleus production in heavy-ion collisions. An extreme example of such an emission source is provided by energetic jets produced in hard parton scatterings in high-energy collisions, such as $pp$ and $pA$ collisions. Particles within a jet cone are highly collimated, which can enhance the probability of deuteron formation through the coalescence of a proton and a neutron. Recently, a strong enhancement of deuteron production within jets has been observed in proton-proton collisions at $\sqrt{s} = 13$ TeV at the LHC~\cite{ALICE:2020hjy,ALICE:2022ugx}. This enhancement has been attributed to the reduced phase-space separation between nucleons within jets~\cite{Mrowczynski:2023hbn,ALICE:2022ugx}.

The effect of jets on deuteron production can be illustrated by considering a highly simplified momentum distribution of nucleons within a jet. Using spherical coordinates with the $z$ axis aligned with the direction of the leading particle, the momentum distribution of jet protons can be written as
\be
\label{distribution-N}
\frac{d N_p}{d^3{\bf p}} = N_p \frac{e^{-\alpha p}}{\pi\alpha^3}
\frac{\Theta(\cos\theta - \cos\theta_c) }{1 - \cos\theta_c} ,
\ee
where $\alpha$ is a positive real parameter, $\Theta(x)$ is the step function, and $\theta_c$ is the maximum zenithal angle defining the jet cone. For small angles, $\theta_c$ expressed in radians approximately corresponds to the jet radius $R$. For typical values of $R=0.2$--$0.5$, this gives $\theta_c \simeq 11^\circ$--$29^\circ$. The momentum distribution in Eq.~(\ref{distribution-N}) is normalized to the total number of protons, $N_p$, within the jet cone.

The deuteron yield is then given by ~\cite{Mrowczynski:2023hbn}
\be
\label{cross-D-model}
\frac{d N_D^{\rm jet}}{d^3{\bf p}} =  {\cal A} \, 
\bigg(\frac{dN_p^{\rm jet}}{d^3\big(\frac{1}{2}{\bf p}\big)} \bigg)^2={\cal A} \, N_p^2 \frac{e^{-\alpha p} }{\pi^2 \, \alpha^6} \, 
\frac{\Theta(\cos\theta - \cos\theta_c) }{(1 - \cos\theta_c)^2} ,
\ee
where the deuteron formation rate is
\be
\label{d-form-rate}
{\cal A} = \frac{3}{4} (2\pi)^3 \int d^3r \, D ({\bf r})\, |\phi_d({\bf r})|^2 ,
\ee
with the source function $D ({\bf r})$ normalized to unity distribution of relative space-time positions of the neutron and proton at the kinetic freeze-out and $\phi _d({\bf r})$ being the deuteron wave function of relative motion. The factor $\frac{3}{4}$ reflects the fact that the deuterons come from the neutron-proton pairs in the spin triplet state. The total yield of jet deuterons is integrated out as
\be
\label{cross-D-model-total}
N_D^{\rm jet} = \int d^3p \, \frac{d N_D^{\rm jet}}{d^3{\bf p}} 
=\frac{2{\cal A} \,N_p^2}{\pi \, \alpha^3 (1 - \cos\theta_c)} .
\ee
As one sees, the deuteron yield grows when $\theta_c$ decreases, that is when the jet becomes more and more collimated. Reducing $\theta_c$ from $180^0$, when the momentum distribution is isotropic, to $\theta_c= 20^0$ the deuteron yield increases by the factor of 33.  

Since the function $|\phi _d({\bf r})|^2$ changes slowly around ${\bf r} = 0$ as compared to $D ({\bf r})$ when $r \rightarrow 0$, the value of $|\phi _d({\bf r})|^2$ at ${\bf r} = 0$ can be pulled out of the integral (Eq.(\ref{d-form-rate})). Thus, one finds
\be
{\cal A}^{\rm max} \approx \frac{3}{4} (2\pi)^3 |\phi _d({\bf r}=0)|^2 \int d^3r \, D ({\bf r}) 
= \frac{3}{4} (2\pi)^3 |\phi _d({\bf r}=0)|^2.
\ee
The last equality holds because the source function is normalized to unity. Using the Hulth\'en wave function, one finally finds  
\be
\label{d-form-rate-max}
{\cal A}^{\rm max} = 3\pi^2 \alpha \beta (\alpha + \beta ) \approx  20.2~{\rm fm}^{-3}.
\ee
with $\alpha = 0.23$ fm$^{-1}$ and $\beta = 1.61$ fm$^{-1}$ \cite{Mattiello:1996gq}. The result is independent of the adopted parameterization of the source function as it fully depends on the wave function at ${\bf r}=0$. One can calculate the coalescence parameter ${\cal B}_2$ by ${\cal B}_2 = 2 {\cal A}/m$ where $m$ is the nucleon mass.  The maximal value of ${\cal B_2}$ corresponding to ${\cal A}^{\rm max}$ given in Eq.~(\ref{d-form-rate-max}) is ${\cal B_2}^{\rm max} \approx 0.33~{\rm GeV}^2$.
This value is not too far from the experimentally found values of  approximately ${\cal B}_2 = 0.13 \pm 0.07~{\rm GeV}^2 $  and $0.33~{\rm GeV}^2$ extracted, respectively, from jet-associated deuterons produced in $pp$ collisions at $\sqrt{s}=13$ TeV and $p-Pb$ collisions at $\sqrt{s_{NN}}=5.02$ TeV by the ALICE Collaboration~\cite{ALICE:2022ugx} at the LHC, which are about a factor 10 and 15 larger than those expected in deuteron production from thermal nucleons in the collisions. 

A more quantitative study based on the AMPT model~\cite{Lin:2004en} has shown, however, that the order of magnitude enhancement in the deuteron coalescence parameters from the ALICE experiments mainly comes from the coalescence of nucleons inside the jet with the underlying-event nucleons~\cite{Feng:2024tmc}, instead of from the coalescence of nucleons only inside the jet, which dominates the production of deuterons at high momentum.

\begin{figure}[!ht]
\centering
\includegraphics[width=0.7\linewidth]{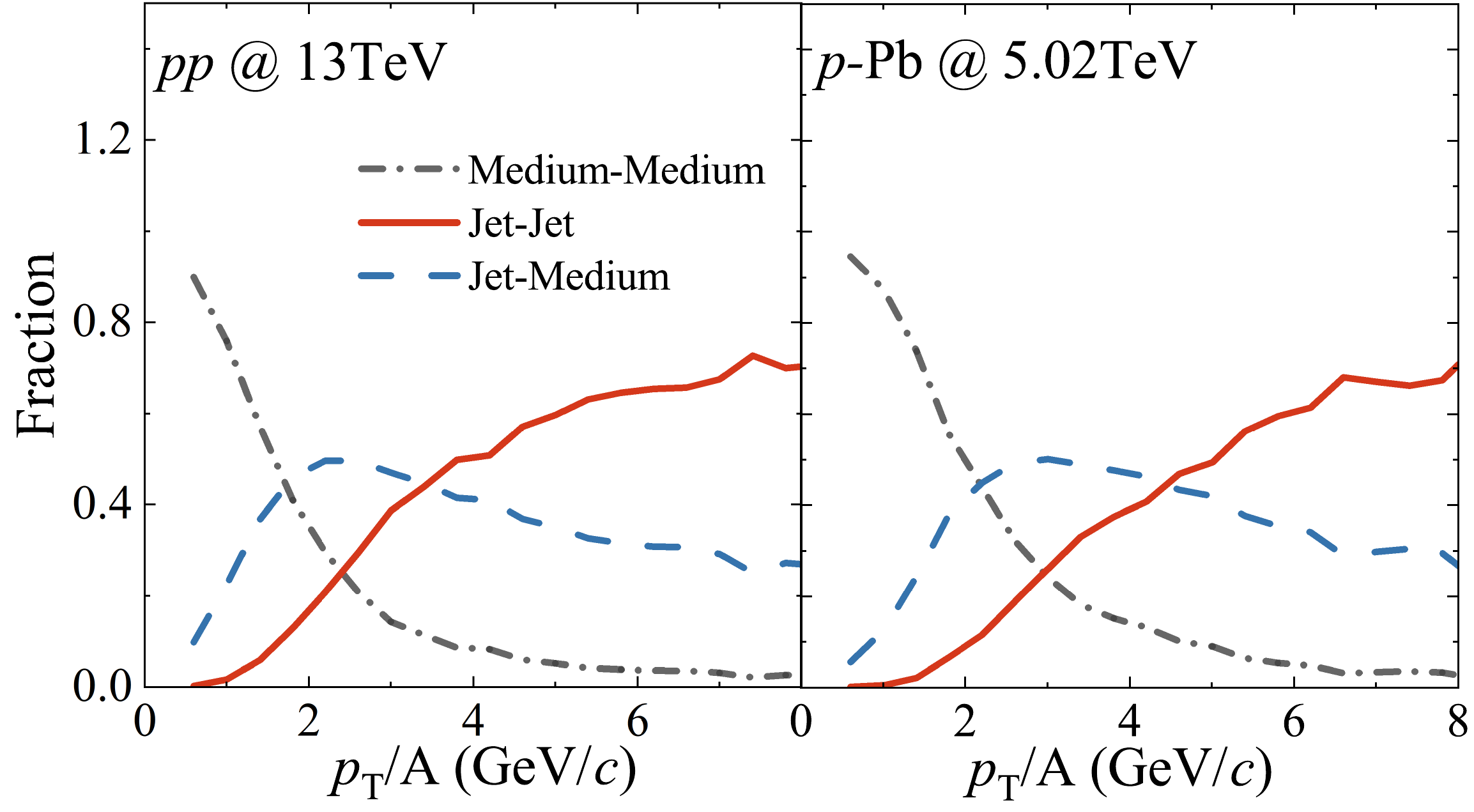}
  \caption{Fractions of deuterons produced by Jet-Jet nucleon coalescence (solid line), Medium-Medium nucleon coalescence (dash-dotted line), and Jet-Medium nucleon coalescence (dashed line) as a function of transverse momentum. Figure is taken from Ref.~\cite{Feng:2024tmc}.}
  \label{pic:fraction}
\end{figure}

Figure~\ref{pic:fraction} shows the fractions of deuterons originating from three different coalescence contributions: Jet--Jet, where both nucleons originate from the jet; Jet--Medium, where one nucleon originates from the jet and the other from the underlying event (UE); and Medium--Medium, where both nucleons originate from the UE. At $p_{\rm T}/A < 1.5$ GeV/$c$, the Jet--Jet contribution is much smaller than the Jet--Medium contribution, which in turn is much smaller than the Medium--Medium contribution. At $p_{\rm T}/A > 4$ GeV/$c$, however, Jet--Jet coalescence becomes dominant. In this momentum region, both the deuteron-to-proton yield ratio, $\mathrm{d}/p$, and the coalescence parameter $B_2$ are significantly larger along the jet direction than in the direction perpendicular to the jet axis, as a consequence of the strong collinear correlations among nucleons produced through jet fragmentation. Measurements of jet-associated deuteron production in relativistic nuclear collisions therefore provide a novel probe of the phase-space structure of nucleons within jets.

Future measurements of light-nuclei production in jets reconstructed with the anti-$k_{\rm T}$ algorithm will provide an important test of this coalescence picture. The technique introduced by the CDF Collaboration, currently employed
in these measurements, consists of identifying the leading particle,
defined as the particle with the highest transverse momentum
($p_{\rm T}$) in the event, requiring $p_{\rm T} > 5$ GeV/$c$,
and dividing the azimuthal plane into three regions relative to its
direction: \textit{toward} ($|\Delta\varphi| < 60^\circ$),
\textit{transverse} ($60^\circ < |\Delta\varphi| < 120^\circ$),
and \textit{away} ($|\Delta\varphi| > 120^\circ$).
Compared with this approach, the anti-$k_{\rm T}$ algorithm is expected to enhance the relative contribution from nucleons associated with the jet, thereby suppressing the Jet--Medium and, even more strongly, the Medium--Medium contributions. If the observed enhancement originates from the interplay of these different coalescence components, the resulting change in their relative contributions should lead to a reduced difference between the $B_2$ values measured in the jet and UE regions, with a characteristic $p_{\rm T}$ dependence reflecting the transition from Medium--Medium and Jet--Medium to Jet--Jet dominated coalescence.

\subsection{Effects of density fluctuations and correlations}

In the coalescence model, the production of light nuclei in heavy-ion collisions naturally reflects multi-nucleon correlations. Light nuclei can therefore serve as sensitive probes of baryon density fluctuations and correlations, which are expected to be enhanced in the vicinity of the QCD critical point~\cite{Sun:2017xrx,Sun:2018jhg,Sun:2020pjz,Sun:2020zxy}.

By taking into account density fluctuations and correlations in the emission source created in high-energy heavy-ion collisions, the deuteron yield from the coalescence of proton--neutron pairs can be written as~\cite{Sun:2020zxy}
\begin{eqnarray}
N_\text{d}
&\approx&N_\text{d}^{(0)}\left[1+C_\text{np}+\frac{\lambda}{\sigma_d}G\left(\frac{\xi}{\sigma_d}\right)\right],
\label{Eq:d04}
\end{eqnarray}
where
$N_\text{d}^{(0)}=\frac{3}{2^{1/2}}\left(\frac{2\pi}{mT}\right)^{3/2} N_{\rm p}\langle\rho_{\rm n}\rangle$
denotes the deuteron yield obtained in the conventional coalescence model without density fluctuations and correlations in the emission source, with $N_{\rm p}$ being the number of protons in the source. The term
$C_\text{np}=\langle\delta\rho_{\rm n}({\bf x})\delta\rho_{\rm p}({\bf x})\rangle /(\langle\rho_{\rm n}\rangle\langle\rho_{\rm p}\rangle)$, introduced in Refs.~\cite{Sun:2017xrx,Sun:2018jhg,Sun:2020pjz}, quantifies the correlation between neutron and proton density fluctuations, where $\langle\cdots\rangle$ denotes an average over coordinate space~\cite{Sun:2017xrx,Sun:2018jhg}. The parameter $\xi$ is the correlation length, while $\lambda$ characterizes the strength of the long-range nucleon--nucleon correlations and depends on the nucleon--nucleon interaction as well as on the critical temperature and baryon chemical potential of the QCD critical point.

The function $G$ describes the contribution from long-range neutron--proton correlations and is given by
\begin{equation}
G(z)=\sqrt{\frac{2}{\pi}}-\frac{1}{z}e^{\frac{1}{2z^2}}\text{erfc}\left(\frac{1}{\sqrt{2}z}\right),
\end{equation}
where $z=\xi/\sigma_d$ and $\text{erfc}(z)$ denotes the complementary error function. As the correlation length $\xi$ increases, $G(z)$ increases monotonically and approaches the limiting value $\sqrt{2/\pi}$ for $z\gg1$, corresponding to $\xi\gg\sigma_d$. Therefore, even though the correlation length diverges at the critical point, its contribution to the deuteron yield remains finite, and the deuteron yield itself does not diverge.

Similarly, the triton yield can be expressed as
\begin{eqnarray}
N_\text{t} &\approx&N_{\rm t}^{(0)}\left[1+\Delta\rho_n+2C_\text{np}+\frac{3 \lambda}{ \sigma_t}G\left(\frac{\xi}{\sigma_t}\right)+\mathcal O(G^2)\right],
\label{Eq:t02}
\end{eqnarray}
where $N_{\rm t}^{(0)}= \frac{3^{3/2}}{4}\left(\frac{2\pi}{mT}\right)^{3} N_{\rm p}\langle\rho_{\rm n}\rangle^2$ denotes the triton yield in the absence of nucleon density fluctuations and correlations in the emission source, and $\Delta\rho_n=\langle\delta\rho_n(x)^2\rangle/\langle\rho_n\rangle^2\ge 0$ quantifies the relative neutron density fluctuation~\cite{Sun:2017xrx,Sun:2018jhg,Sun:2020pjz}.

By considering the yield ratio \pdt, which can be expressed as the double ratio of $N_{\rm t}/N_{\rm d}$ and $N_{\rm d}/N_{\rm p}$, the temperature- and proton-number-dependent prefactors in Eqs.~(\ref{Eq:d04}) and~(\ref{Eq:t02}) can be eliminated. Neglecting, for simplicity, the difference between $\sigma_d$ and $\sigma_t$ by taking $\sigma\approx\sigma_\text{d}\approx\sigma_\text{t}$, and retaining only terms linear in $G$, the yield ratio \pdt~can be approximated as
\begin{eqnarray}
\frac{N_\text{t}N_\text{p}}{N_\text{d}^2} &\approx& \frac{1}{2\sqrt{3}}\left[1+\Delta \rho_n + \frac{\lambda}{\sigma}G\left(\frac{\xi}{\sigma}\right)\right].
\label{Eq:pdt}
\end{eqnarray}

Equation~(\ref{Eq:pdt}) shows that the yield ratio \pdt~is sensitive both to neutron density fluctuations, characterized by $\Delta\rho_n$~\cite{Sun:2017xrx,Sun:2018jhg}, and to nucleon density correlations, characterized by the correlation length $\xi$. While $\Delta\rho_n$ vanishes in a homogeneous hadronic medium, the correlation length $\xi$ diverges in the vicinity of the critical point. In contrast, a first-order phase transition involving the coexistence of QGP and hadronic matter can generate large spatial density inhomogeneities~\cite{Steinheimer:2012gc,Herold:2013qda,Li:2016uvu,Sun:2020pjz}, leading to non-vanishing density fluctuations even when the correlation length within each phase remains significantly smaller than that expected near the critical point. Consequently, both a first-order phase transition and the presence of a critical point can lead to an enhancement of the yield ratio \pdt, provided that the associated fluctuations and correlations survive the subsequent hadronic evolution.

The \pdt~ratio has a natural spatial resolution scale of approximately 2~fm, comparable to the correlation length $\xi\simeq2$--$3$~fm that could develop in heavy-ion collisions~\cite{Berdnikov:1999ph}. This makes the \pdt~ratio a potentially sensitive probe of critical phenomena in the QCD phase diagram.

The above result can be generalized to yield ratios involving the heavier $^4$He ($\alpha$) nucleus~\cite{Sun:2017xrx,Shuryak:2018lgd,Shuryak:2019ikv}. One example is
\begin{eqnarray}
\frac{N_\alpha N_\text{p}}{N_{^3\text{He}}N_\text{d}} \approx  \frac{2\sqrt{2}}{9\sqrt{3}}\left[1+C_\text{np} + \Delta \rho_p  + \frac{2\lambda}{\sigma}G\left(\frac{\xi}{\sigma}\right)\right]. \label{Eq:ratio1}
\end{eqnarray}
Compared with the yield ratio \pdt~in Eq.~(\ref{Eq:pdt}), the above ratio exhibits a dependence on the correlation length $\xi$ that is twice as strong. Although the production yield of $\alpha$ particles is small in heavy-ion collisions at very high energies, due to the large penalty factor $e^{-A(m-\mu_B)/T}$ for the production of a nucleus with $A$ nucleons when the baryon chemical potential is small~\cite{Agakishiev:2011ib}, it is expected to increase toward lower collision energies as $\mu_B$ increases. Such collisions probe regions of the QCD phase diagram at larger baryon chemical potential, where the QCD critical point may be located~\cite{Shuryak:2020yrs,Vovchenko:2020dmv}.

\subsection{Wigner function Coalescence using ToMCCA}
\label{ss:tomcca}

ToMCCA (Toy Monte-Carlo Coalescence Afterburner)~\cite{Mahlein:2024pur} is a simplified, high-performance coalescence model for small collision systems based on the Wigner-function formalism. It enables event-by-event coalescence studies, ranging from high-precision predictions of absolute light-nucleus yields to studies of event-by-event number fluctuations.

Within this framework, the pair-by-pair coalescence probability for the formation of a cluster with $A=N$ can be calculated as a function of the source size $\sigma$, which characterizes the spatial separation of the nucleons, and their relative momenta $q_i$:

\begin{equation}
    \mathcal{P}_N(\sigma,q_1,...,q_{N-1})=\int d^3r_1...\int d^3r_{N-1} H(\Vec{r}_1,...,\Vec{r}_{N-1};\sigma)\mathcal{D}(q_1,...,q_{N-1},r_1,...,r_{N-1})
    \label{eq:CoalescenceProbabilityWigner}
\end{equation}

where $H(\Vec{r}_1,...,\Vec{r}_{N-1};\sigma)$ is the source function and $\mathcal{D}(q_1,...,q_{N-1},r_1,...,r_{N-1})$ denotes the Wigner function of the $N$-nucleon cluster,

\begin{equation}
\begin{aligned}
\mathcal{D}(q_1,...,q_{N-1},r_1,...,r_{N-1})
=&\int d^3\xi_1...\int d^3\xi_{N-1}
\psi^*(r_1+\xi_1/2,...,r_{N-1}+\xi_{N-1}/2)
\\
&\times\psi(r_1-\xi_1/2,...,r_{N-1}-\xi_{N-1}/2)
\\
&\times\exp(i(q_1\xi_1+....+q_{N-1}\xi_{N-1}))
\end{aligned}
\label{eq:WignerFunction}
\end{equation}

and $r_i$ and $q_i$ are the relative positions and momenta expressed in Jacobi coordinates. An important feature of this formulation is that the Wigner function depends only on the wave function of the final-state nucleus.

For deuteron production in this model, the deuteron momentum spectrum
is given by
\begin{equation}
\frac{d^3N_d}{d^3P_d}
=
\frac{S}{(2\pi)^6}
\int d^3r_1\,d^3r_2\,d^3q\,
\mathcal{D}(\vec{q},\vec{r}_1-\vec{r}_2)
W_{\mathrm{pn}}\left(
\frac{\vec{P}_d}{2}+\vec{q},
\frac{\vec{P}_d}{2}-\vec{q},
\vec{r}_1,\vec{r}_2
\right),
\label{eq:DeuteronSpectrum}
\end{equation}
where $\mathcal{D}$ is the deuteron Wigner function,
$\vec{q}$ is the relative momentum of the proton--neutron pair,
and $W_{\mathrm{pn}}$ denotes their phase-space distribution.
The factor $S$ accounts for the statistical spin weight.

\begin{figure}[!hbt]
     \centering
     \includegraphics[width=0.75\linewidth]{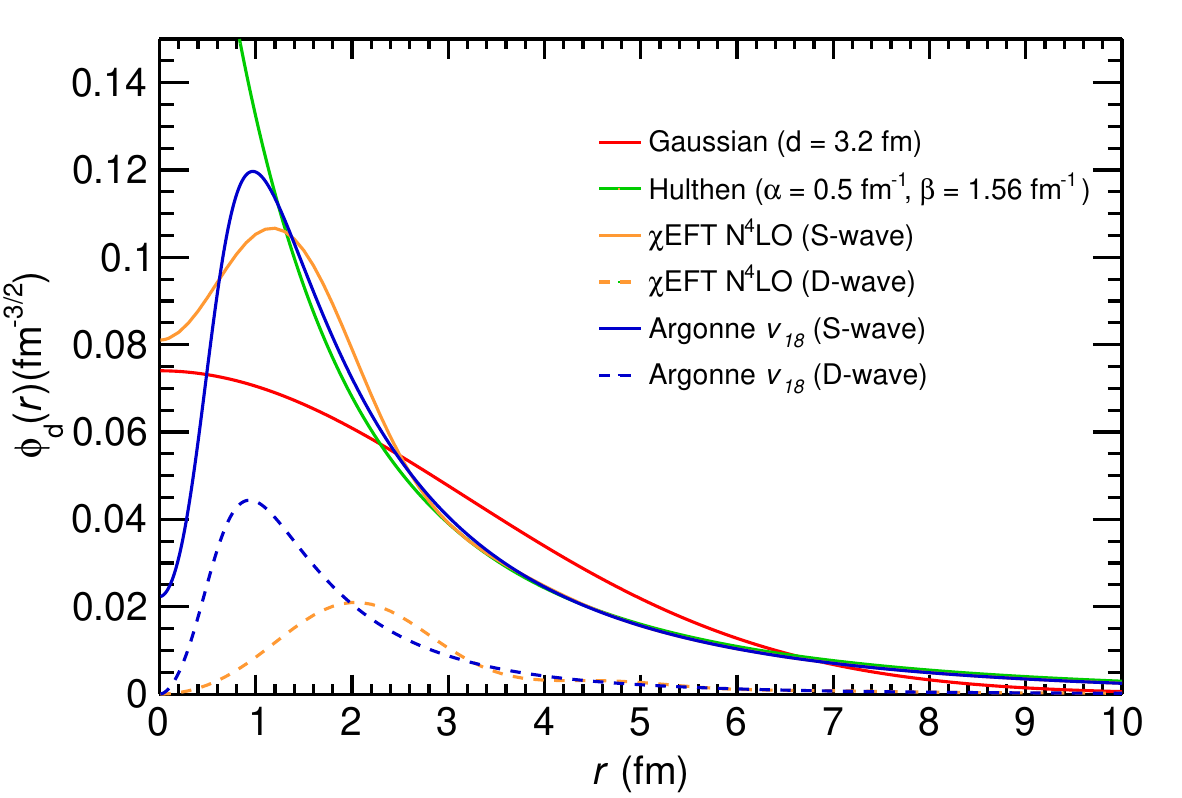}
     \caption{Deuteron wave function from various models.}
     \label{fig:deuteronWavefunctions}
\end{figure}

Several choices for the deuteron wave function have historically been employed in coalescence calculations. AS shown in Fig.~\ref{fig:deuteronWavefunctions}, these include a Gaussian wave function; the Hulthén wave function~\cite{Mahlein:2023fmx,Kachelriess:2019taq}, derived from a phenomenological potential and constrained by low-energy nucleon--nucleon scattering; the Argonne $v_{18}$ wave function, obtained from a high-precision phenomenological nucleon--nucleon potential fitted to scattering data; and wave functions derived within chiral effective field theory ($\chi$EFT). Among these approaches, the Argonne $v_{18}$ and $\chi$EFT wave functions provide the most realistic descriptions of the deuteron structure.

The choice of the nuclear wave function has been shown to have a negligible impact on coalescence predictions in heavy-ion collisions~\cite{Bellini:2020cbj}. In small collision systems, such as pp collisions, however, different choices can lead to substantially different results~\cite{Mahlein:2023fmx}. ToMCCA predictions for deuteron production are based on the Argonne $v_{18}$ wave function.

As an event-by-event coalescence model, ToMCCA generates individual pp collisions. For simplicity and computational efficiency, only nucleons are explicitly generated, while all other particles are neglected. This approximation is justified by the much lower particle multiplicity in pp collisions compared with heavy-ion collisions, which reduces the probability that other particles significantly affect the coalescence process. Once the numbers of protons and neutrons in an event are determined, each nucleon is assigned a three-dimensional momentum, including the relevant angular correlations, as well as a spatial position relative to the other nucleons. Strong space--momentum correlations are not explicitly included, except through the dependence of the source size on the transverse momentum of the particles. This approximation is motivated by the absence of a strongly expanding medium in pp collisions and, consequently, by the much weaker collective space--momentum correlations compared with heavy-ion collisions.

\begin{figure}
    \centering
    \includegraphics[width=0.9\linewidth]{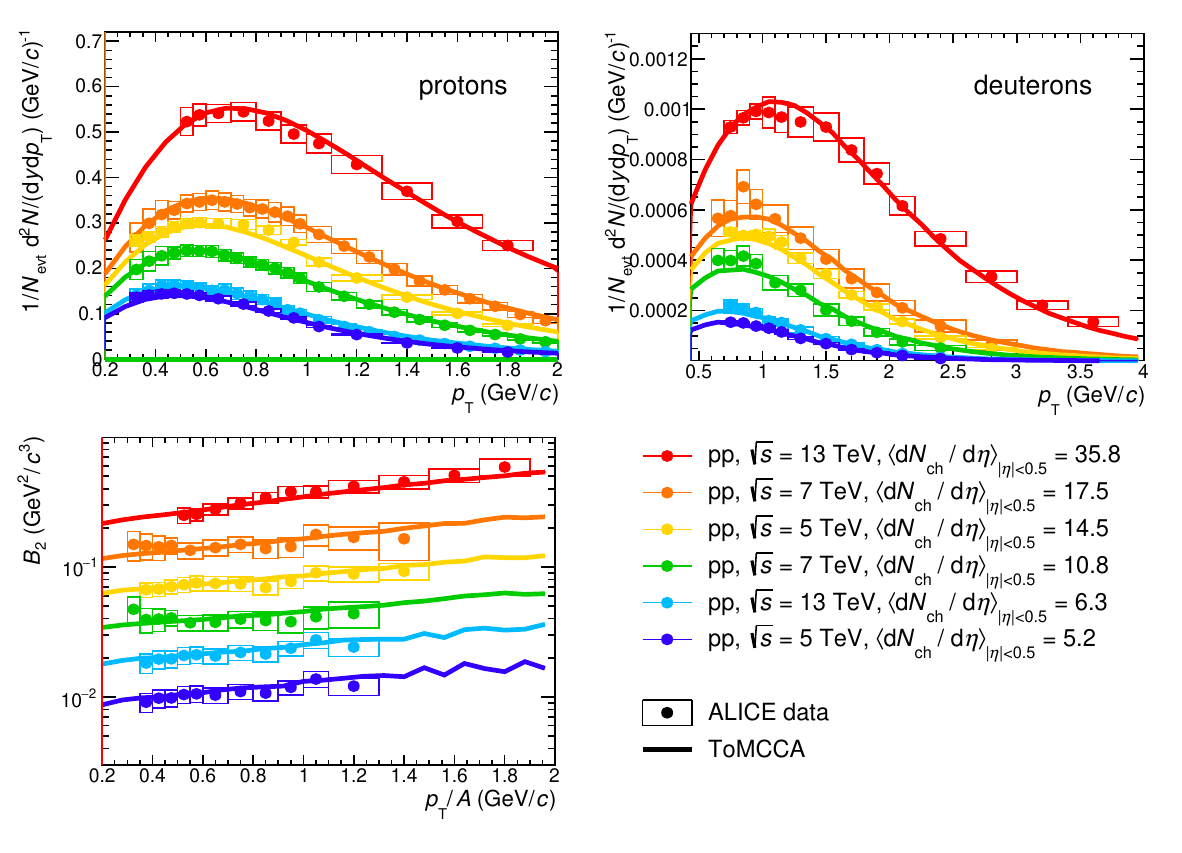}
    \caption{ToMCCA results on proton and deuteron spectra as well as the coalescence factor $B_2$ for various multiplicity classes and collision energies at the LHC.}
    \label{fig:ToMCCADeuteronResults}
\end{figure}

\begin{figure}
    \centering
    \includegraphics[width=0.7\linewidth]{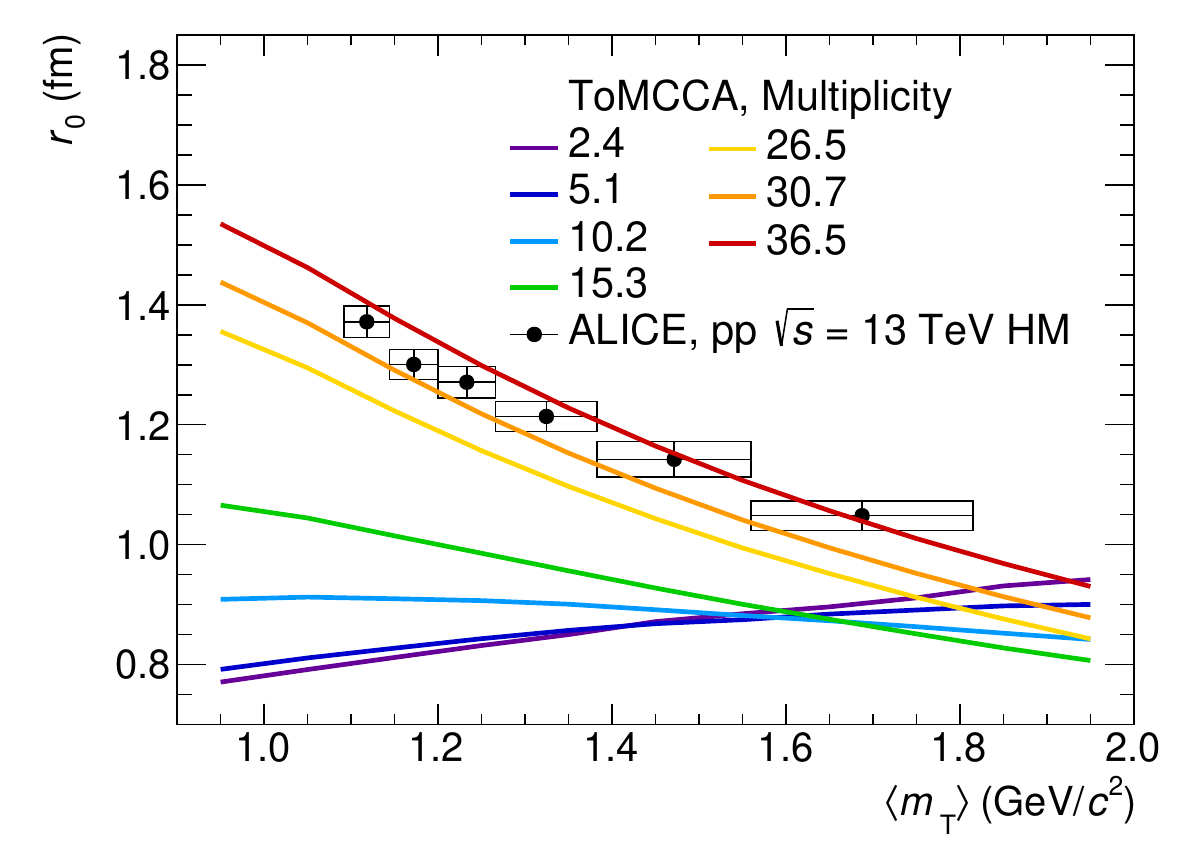}
    \caption{The fitted source size using ToMCCA. For reference, the measurement by ALICE has a multiplicity of $\sim$ 31 charged particles.}
    \label{fig:SourceSizeToMCCA}
\end{figure}

Once the momenta and relative positions of all nucleons have been assigned, the coalescence probability in Eq.~(\ref{eq:CoalescenceProbabilityWigner}) is evaluated and the resulting clusters are counted while retaining their full momentum information. ToMCCA has so far been tuned to high-energy pp collisions at $\sqrt{s}=5$--$13$ TeV measured by ALICE. Within this energy range, no significant collision-energy dependence has been observed; instead, the system exhibits an approximate scaling with the event multiplicity~\cite{Mahlein:2024pur}. ToMCCA provides a good description of the available measurements at LHC energies, as shown in Fig.~\ref{fig:ToMCCADeuteronResults}. In addition, ToMCCA can be used to determine the source size in pp collisions as a function of multiplicity, which represents an important input for coalescence calculations, as illustrated in Fig.~\ref{fig:SourceSizeToMCCA}.

\subsection{The coalescence-correlations relation}
\label{sss:ccr}

Coalescence is sometimes considered as a plug-in for kinetic theory (see discussion in the introduction of Sec.~\ref{sec:kinetic cluster}).
This perspective is consistent with the historical development that formulated cluster formation using asymptotic-state scattering theory methods (e.g.~\cite{Butler:1963pp}).

Viewed this way, both kinetic theory in general and its coalescence plug-in in particular suffer from the same problem: the asymptotic condition does not apply to nucleons in the high-excitation state (HXS) of high-energy hadronic collisions; namely, these nucleons are not asymptotic ``in'' states.

However, some predictions extend beyond the asymptotic condition. Correlation studies (see Sec.~\ref{subsec:correlations}) aim to measure a source function for hadrons~\cite{Schwarzschild:1963zz,Yano:1978gk,Csernai:1986qf}, corresponding to a density-matrix description of the HXS~\cite{Lednicky:1981su,Lednicky:2005tb}. Correlations are interpreted as final-state interactions that enter the projection of the source function onto the asymptotic ``out'' states at the detector.

Coalescence lives in the same framework: bound-cluster final states are an unavoidable consequence of the same final-state interactions. This relationship permits model-independent predictions that connect cluster formation and correlation functions, dubbed the coalescence-correlations relation in~\cite{Blum:2019suo} and recognized early on for concrete HXS models~\cite{Mrowczynski:1987oid,Mrowczynski:1992gc,Scheibl:1998tk}. From this perspective, coalescence predictions do not require, and are more general than, kinetic-theory models of the HXS.

The coalescence-correlations relation is the underlying idea behind methods such as ToMCCA (Sec.~\ref{ss:tomcca}). A test of coalescence as the origin of nuclei, independent of the validity of kinetic theory or other model assumptions, can be attempted via direct application of the same source parametrization used to fit pair correlations in the coalescence computation~\cite{ALICE:2021mfm}.

\section{Kinetic cluster production}
\label{sec:kinetic cluster}

Transport approaches assume that there is no sudden hadronization at a given temperature and chemical potential, but that interactions among partons and later hadrons continue until, in the expanding system, the distance between the hadrons becomes larger than $r = \sqrt{\sigma/\pi}$, at which point collisions cease automatically and the system freezes out. Here, $\sigma$ is the largest cross section among the hadrons in the expanding system. The existence of a long hadronic expansion phase is supported by detailed studies of resonances, which decay and are recreated during the expansion \cite{Knospe:2015nva}. Transport approaches show a broad distribution of freeze-out times, i.e., the times when the baryons undergo their last collisions with other hadrons.

The study of cluster formation in transport approaches is a challenging task on which many groups are working intensively. As solutions of the many-body Schr"odinger equation, clusters are quantum states with exact binding energies. Transport theories cannot presently be formulated at the level of quantum states. They describe the approximate time evolution of one-body or many-body phase-space densities and are therefore semiclassical approaches. How to extract quantum states such as clusters from these time-evolved phase-space densities is a yet unsolved problem, and therefore, at present, only more or less phenomenological approaches are possible.

Three processes have been advanced to study theoretically, within these transport approaches, the production of deuterons (and sometimes of $A=3$ and $A=4$ clusters) in heavy-ion collisions at midrapidity.
\begin{itemize}
\item Formation by coalescence. Here, it is assumed that nucleons form a deuteron if, at the time when the last of the cluster nucleons undergoes its last collision, the distances between the cluster nucleons in phase space are smaller than ${\bf r}_0$ and ${\bf p}_0$~\cite{Sombun:2018yqh,Butler:1963pp}. These parameters are usually chosen to describe the multiplicities. In more advanced formulations of the coalescence model, the sharp cutoff in phase space is replaced by a Wigner density of the deuteron~\cite{Scheibl:1998tk,Zhu:2015voa}. The coalescence model can be extended to the formation of (hyper)clusters with $A=3$, if additional approximations are made. A description of the coalescence model was given in Sec.~\ref{subsec:coalescence}.

\item Formation by collisions. Deuterons can be produced by three-body collisions such as $NNN \to dN$ or $NN\pi\to d\pi$ \cite{Coci:2023daq,Oliinychenko:2018ugs}. Because the inverse cross sections have been measured, the rate for deuteron formation can be calculated assuming that the matrix element depends only on the center-of-mass energy, $\sqrt{s}$, of the collisions, an assumption that may be considered severe in view of the angular distribution of the inverse reaction. This approach faces the problem that the cross section is calculated under the assumption that the deuteron, with an rms radius of 2.1 fm, does not, as a final-state particle, interact with any other strongly interacting hadron. In heavy-ion collisions at freeze-out, we usually find several strongly interacting particles within a sphere with a radius of 2.1 fm, a fact that calls the kinetic approximation into question. Only at the very end of the expansion is this no longer the case, but if one limits oneself to this situation, the measured deuteron yield cannot be reproduced by assuming that deuterons are produced by three-body collisions \cite{Coci:2023daq}. We call this approach kinetic deuteron production.

\item Formation by potential. During the whole heavy-ion reaction, baryons interact through potentials, which can be calculated in the Brueckner G-matrix approach \cite{Gaitanos:2001hv} or using chiral effective theories \cite{Holt:2016pjb}, and which are attractive at low density and repulsive at densities well above normal nuclear matter density. In nuclear matter below saturation density, the nucleons cluster. The potential used in the simulation program reproduces the binding energies of larger clusters, as given by the Weizsäcker mass formula. It is therefore natural that, during the expansion, nucleons form clusters under the influence of this potential. This approach allows clusters of all sizes, and not only deuterons, to be identified \cite{Aichelin:2019tnk} and has the advantage that it relies completely on the ingredients already present in the transport approach, such as potentials and cross sections, whose validity can be controlled by observables that are not related to clusters, and that there is no need to switch to another model, as in the coalescence approach.\footnote{On the other hand, coalescence allows model-independent predictions (that is, predictions that do not require any detailed transport model) concerning relations between phase-space correlations of emitted baryons and cluster yields~\cite{Blum:2019suo,Bellini:2020cbj}. These predictions can be understood as consistency constraints that transport models should satisfy.}

\end{itemize}

In each transport approach, at least one of the three cluster identification prescriptions is employed to calculate cluster observables. There have been efforts to search for observables for which the three approaches give different predictions \cite{Kireyeu:2023spj,Kireyeu:2024hjo}. Indeed, such observables exist, but the available experimental results do not yet allow the most suitable approach to be identified.

\subsection{Kinetic production using URQMD}
\label{subsec:kinetic_production_urqmd}

The UrQMD (Ultra-relativistic Quantum Molecular Dynamics) approach is based on the classical propagation of hadrons in phase space \cite{Bass:1998ca,Bleicher:1999xi}. Collisions are modelled via resonance excitation and decays at moderate energies, supplemented by string degrees of freedom at high collision energies. As a QMD-type model, it includes $n$-particle correlations, which makes it suitable for calculating cluster yields. At beam energies below 20 GeV, different potential interactions (classical soft and hard Skyrme-type potentials or even density-dependent potentials that mimic a phase transition to the QGP) can be employed \cite{Steinheimer:2022gqb}.

The production of light fragments in UrQMD is realized either by a coalescence mechanism that is local in space and time for each individual light cluster or through bulk fragmentation using a statistical multifragmentation model. The latter is especially suited for low collision energies and for the production of heavy clusters \cite{Hillmann:2021zgj}.

The coalescence process is realized using the space-time point of the last interaction of each individual nucleon \cite{Mattiello:1996gq,Sombun:2018yqh}. One then boosts into the two-particle rest frame of all proton-neutron pairs (in the case of deuteron formation) and propagates the earlier frozen-out nucleon to the time of the later nucleon. The spatial and momentum distances can then be compared, since both nucleons are evaluated at the same time in their two-particle rest frame. If the coalescence criterion is fulfilled, this pair is clustered into a deuteron (with a probability given by the spin/isospin factor, in our case 3/8) and removed from further analysis. This procedure is followed until all pairs have been analyzed. Different choices are possible for the coalescence criterion: one can either employ a probability density extracted from a Wigner representation of the deuteron wave function or use a box approximation to the Wigner function of the cluster. We have chosen the latter approach because it was shown to yield results similar to those obtained with the Wigner density in the case of the Hulthén wave function \cite{Nagle:1996vp}.

A critical remark: We further believe that the Wigner-function approach may suffer from (potentially small but) uncontrolled problems due to the non-orthogonality of the states emerging from classical simulations. Let us briefly elaborate on this using the example of a pure helium state to make this statement clearer. Assuming that one starts from a pure helium wave function, one can perform a Wigner transformation to obtain a phase-space representation of this helium state. It is then clear that a projection of the helium state onto the Wigner density of the deuteron should yield a vanishing overlap. However, this is not what a transport simulation will provide. A transport simulation will provide a truly classical phase-space distribution, i.e., it will generate not a Wigner representation of the helium wave function but a Husimi representation (i.e., a Gaussian-smeared Wigner density \cite{Colom_s_2015}) in phase space. Unfortunately, the overlap of the Husimi representation of the helium state with the Wigner representation of the deuteron wave function will not vanish. Instead, the projection will create deuterons, tritons, and helium states and will not obey baryon-number conservation due to the loss of orthogonality of the states. This problem is avoided in the box-coalescence procedure.

Using the coalescence model described above then allows us to calculate the production of clusters from the same underlying events as those used for all other investigations, e.g., pion HBT or resonance studies.

\begin{figure}[!hbt]
\centering
	\includegraphics[width=0.9\textwidth]{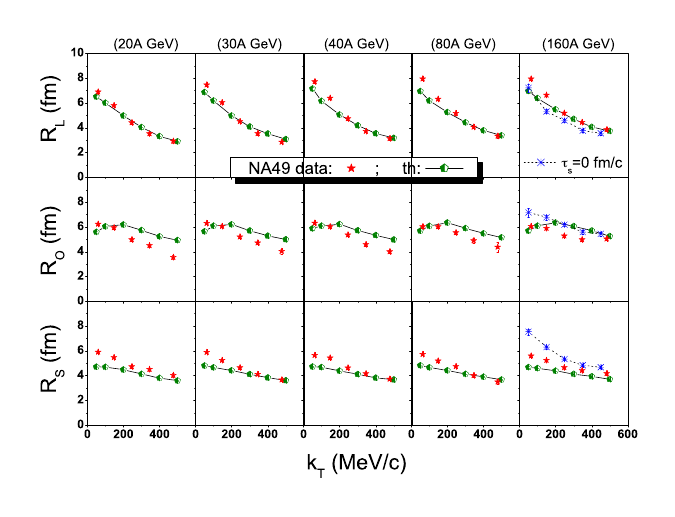}
\caption{Transverse-momentum dependence of the HBT radii at midrapidity in central Pb+Pb collisions for various beam energies ranging from $E_{\rm lab}$ = 20 to 160 GeV, compared with preliminary NA49 data. Figure taken from \cite{Li:2006gb}.}

\label{fig:urqmdhbt}
\end{figure}

Let us start with the lifetime of the hadronic system as seen in pion HBT calculations. Figure \ref{fig:urqmdhbt} shows the HBT radii calculated with UrQMD in comparison with the measured radii in Pb+Pb reactions in the SPS energy range. One observes generally good agreement between the data and the calculations. However, the outward radii are systematically larger than the data, indicating a slightly broader decoupling-time distribution in the model than in the data. Nevertheless, the discrepancies are small and indicate that the overall lifetime of the fireball in the simulation is not far from that inferred from the experimental data.

\subsubsection{System size dependence at the LHC and final state annihilations}

An interesting aspect of light-nuclei production at the LHC is that it can be used to constrain the degree of baryon annihilation during the hadronic phase of the collisions. This is of particular interest, as a detailed description of the annihilation and regeneration processes is difficult \cite{Becattini:2012sq,Rapp:2000gy,Cassing:2001ds,Greiner:2002nc,Garcia-Montero:2021haa}. On the other hand, the impact of annihilation on final-state hadronic multiplicities \cite{Steinheimer:2012rd} and proton-number fluctuations can be essential \cite{Savchuk:2021aog}.

As light nuclei are formed after kinetic freeze-out, their multiplicities are sensitive to the phase-space densities of nucleons at kinetic freeze-out, raised to the power $A$, where $A$ is the mass number of the corresponding nucleus. Any annihilation will reduce this density and thus have an increasingly strong effect on nuclei with increasing mass number $A$~\cite{Sombun:2018yqh}.

\begin{figure}[!hbt]
\centering
	\includegraphics[width=0.7\textwidth]{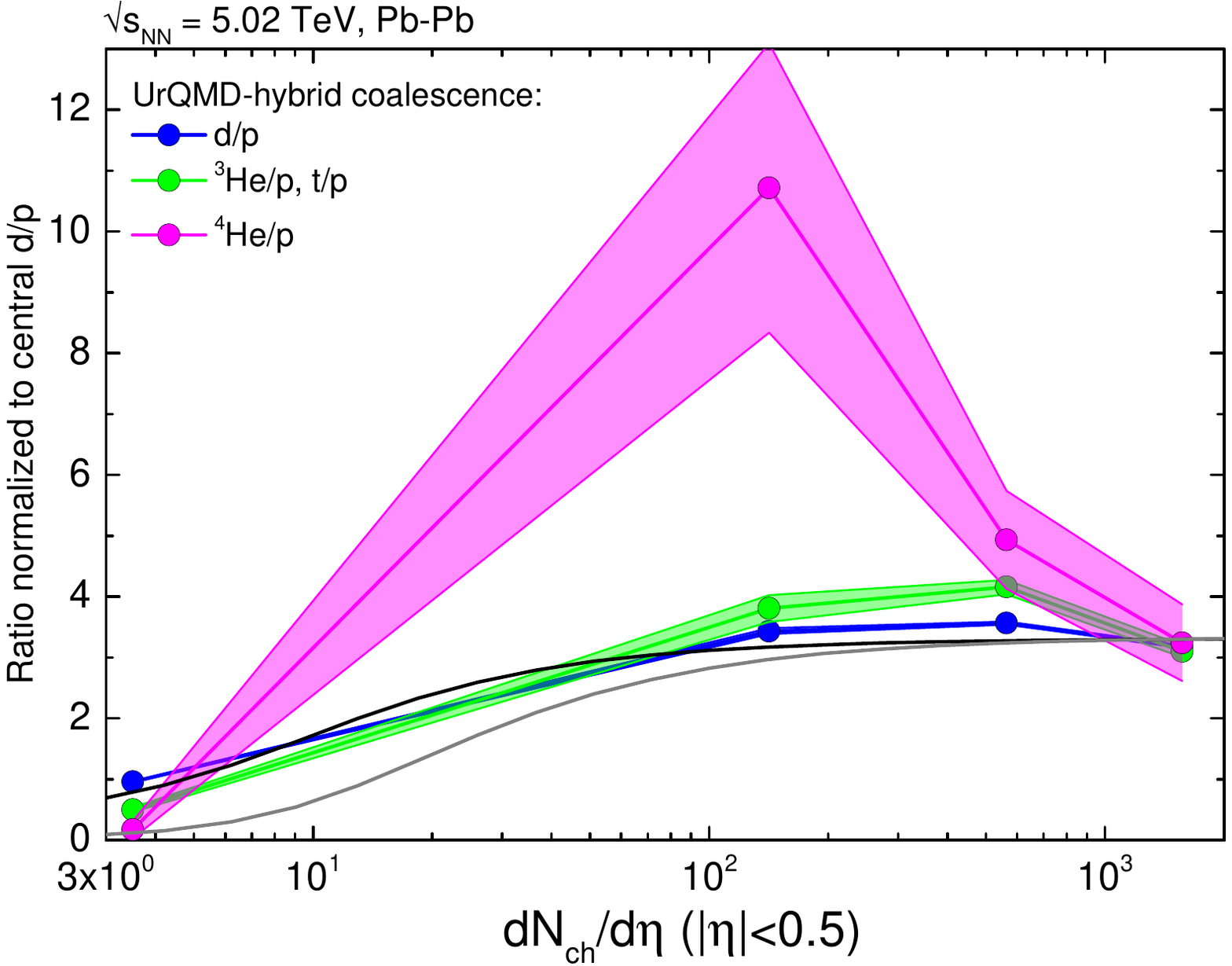}
\caption{System-size dependence of different light-nuclei midrapidity ratios in AuAu collisions at the LHC. All curves are normalized to the central $d/p$ value. The colored curves with symbols depict the UrQMD simulations including annihilation, while the black and grey lines show results from the Thermal-FIST package \cite{Reichert:2022mek}.}
\label{fig:LHC_urqmd_anni}
\end{figure}

Figure \ref{fig:LHC_urqmd_anni} shows the system-size dependence of different light-nuclei ratios, normalized to the value of $d/p$ in central Pb-Pb collisions, at the LHC. As one can see, nuclei with larger mass number $A$ show a stronger suppression of the midrapidity yield. They are very sensitive to the degree of annihilation and regeneration in the hadronic phase.

\subsubsection{Hypernuclei from coalescence in the STAR BES}

\begin{figure}[!hbt]
\centering
	\includegraphics[width=0.7\textwidth]{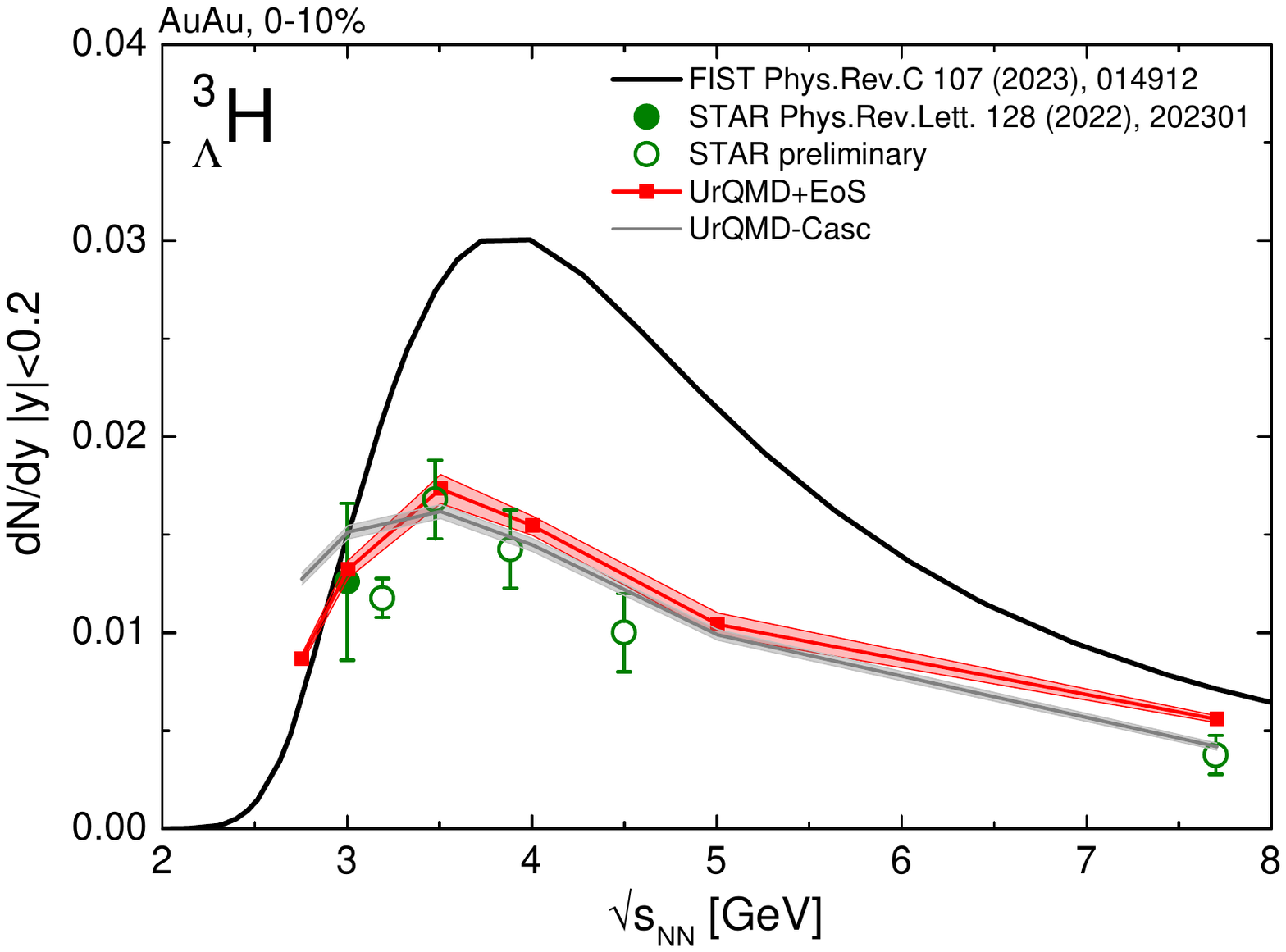}
\caption{Beam-energy dependence of the hypertriton $dN/dy$ at midrapidity for central AuAu collisions from the UrQMD model (red curve) and the Thermal-FIST package (black curve), compared with preliminary data from STAR (green symbols). {\bf{(VV: I think the Thermal-FIST curve here is for 0--5\%, not 0--10\%.)}} The grey line indicates results from the UrQMD cascade model, which overestimates strangeness production compared with UrQMD simulations including a density-dependent QMD potential \cite{Steinheimer:2022gqb}.}
\label{fig:hyper_star_urqmd}
\end{figure}

Hypernuclei production in heavy-ion collisions in the low- to intermediate-beam-energy range is of twofold interest. First, it has been discussed as being sensitive to the correlations between strangeness and baryon number \cite{Zhang:2009ba,Steinheimer:2012tb}, and second, it may shed light on hyperon-nucleon interactions in dense systems. Especially at beam energies close to the elementary production threshold of hyperons, the production yield in transport simulations depends on the equation of state. As the introduction of baryonic interactions modifies the correlations of baryons at the freeze-out surface, a refit of the coalescence parameters is necessary. Using new data from the STAR Collaboration at $3 A$ GeV to perform this refit, one can use UrQMD and the coalescence model to predict the hypernuclei yields for the entire STAR-BES energy range. The results are presented in Fig.~\ref{fig:hyper_star_urqmd} as a red curve. As a reference, we also show predictions from the Thermal-FIST thermal model implementation~\cite{Steinheimer:2012tb} and simulations of UrQMD in cascade mode (without a high-density EoS). The transport simulations, with parameters fixed at one beam energy, are able to describe the preliminary data over the entire BES energy range. In addition, effects of the EoS on strangeness production are visible mainly at the lowest beam energy. This shows that the coalescence model is able to reproduce hypernuclei multiplicities well and does not suffer from the additional uncertainty associated with having to constrain a freeze-out volume, as the thermal fit does.

\subsection{Kinetic production using SMASH}
\label{subsec:kinetic_production_smash}
%Responsible: Hannah Elfner

The hadronic transport approach SMASH (Simulating Many Accelerated Strongly-interacting Hadrons) \cite{SMASH:2016zqf} is based on the relativistic Boltzmann equation and can be applied to non-equilibrium stages of heavy-ion collisions. The degrees of freedom in SMASH are hadrons and resonances (up to 2.3 GeV), with properties taken from the PDG. For the resonances, we assume vacuum Breit-Wigner distributions for the spectral functions. In the model, light nuclei can be formed either by coalescence or through multi-particle collisions. As we present here only results obtained using multi-particle reactions, we provide a more detailed description of them. Besides elastic interactions, the most important reactions for light nuclei are pion- or nucleon-catalysis reactions of the following form: $\text{d}\pi\leftrightarrow\text{NN}$, $\text{dX}\leftrightarrow\text{npX}$, $^{3}\text{He}\text{X}\leftrightarrow\text{nppX}$, $\text{tX}\leftrightarrow\text{nnpX}$, and $^{3}_{\Lambda}\text{H}\text{X}\leftrightarrow\text{np}\Lambda\text{X}$. Here, X can be either a pion or a nucleon (n, p). The multi-particle reactions are realized using a stochastic collision criterion, on which more details can be found in Ref.~\cite{Staudenmaier:2021lrg}. This allows us to perform the multi-particle reactions directly instead of using the prescription employed in Refs.~\cite{Oliinychenko:2018ugs} and \cite{Oliinychenko:2020znl}, where an intermediate dibaryon resonance $d'$ was necessary. In this work, SMASH version 3.0 is used \cite{dmytro_oliinychenko_2023_7870822}, and the results, as well as further details, can be found in Ref.~\cite{Ege:2024vls}.

Light nuclei are produced dynamically during the afterburner stage of a hybrid approach. For the hot and dense stage, hadronic transport is not applicable, so the system is treated hydrodynamically. The (3+1)D relativistic viscous hydrodynamic equations are solved using the MUSIC code \cite{Schenke:2010nt,Schenke:2010rr,Paquet:2015lta}. When the system becomes more dilute, at an energy density of $\epsilon=0.26$ GeV/fm$^{3}$, we switch to a hadronic transport treatment. Subsequently, hadrons are sampled on the hypersurface, and the hadronic afterburner is performed, including rescattering and resonance decays. SMASH is applied to perform this stage, during which light-nuclei formation takes place. Results for the light-nuclei multiplicities are presented in the following sections.

\begin{figure}
    \centering
    \includegraphics[width=0.7\textwidth]{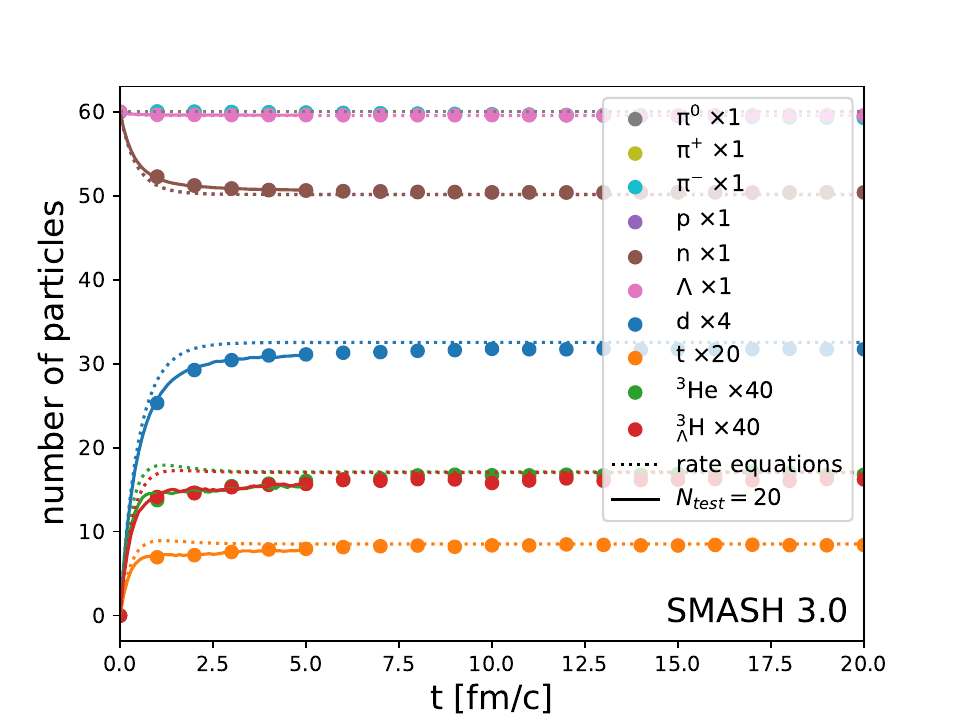}
\caption{Particle multiplicities as a function of time in a box of $(10\text{fm})^{3}$ at $T=150$ MeV in SMASH. The calculations were performed with 10 test particles, and the first 5 fm were also calculated with 20 test particles. The dotted curves show the analytical solutions from rate equations.}
    \label{fig:boxtest_smash}
\end{figure}

In Fig.~\ref{fig:boxtest_smash}, one can find the multiplicities of different particle species in a box of $(10\text{fm})^{3}$ at $T=150$ MeV as a function of time. For comparison, the analytical solutions from rate equations \cite{Pan:2014caa} are shown as well (dotted lines). This is done to check whether the numbers of particles as a function of time are well reproduced by the transport approach using the stochastic criterion. Here, it is important to note that the analytical calculation works only in a box and only if no cross sections other than those for nuclei formation are included. Thus, in a realistic scenario, the full transport calculation is necessary. As one can see in the plot, the equilibrium multiplicities are correctly reproduced. We also checked the numerical sensitivity by changing the number of test particles from 10 to 20 (solid lines). Since there was no significant difference, we decided to retain the former numerical setup for the following calculations.

\begin{figure}
    \centering
    \includegraphics[width=\textwidth]{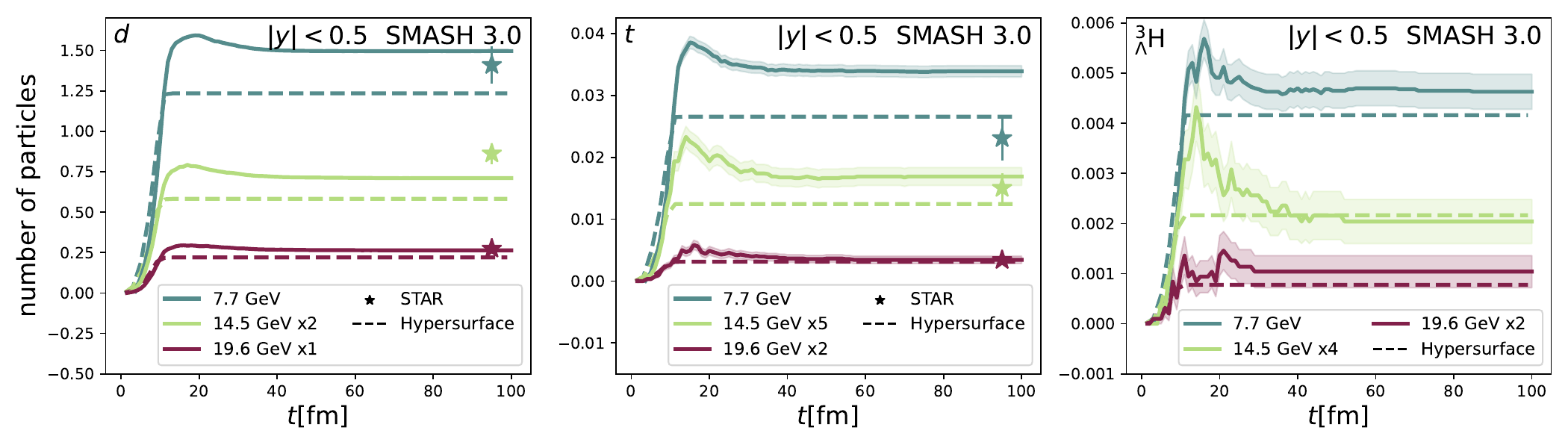}
    \caption{Light-nuclei multiplicities at different energies in central Au--Au collisions. Data points are taken from Refs.~\cite{STAR:2019sjh,STAR:2022hbp}.}
    \label{fig:mult_smash}
\end{figure}

For the main part of this work, the production of light nuclei is studied in 0--10\% central collisions of two gold nuclei at lower RHIC-BES energies (7.7 GeV up to 19.6 GeV). In Fig.~\ref{fig:mult_smash}, the numbers of deuterons, tritons, and hypertritons as a function of time for different energies at midrapidity are presented. For each energy, we compare two scenarios. The solid curve represents the full afterburner calculation. This means that the nuclei experience hadronic rescattering and decays during the afterburner stage. The dashed curve represents the number of nuclei sampled from the hypersurface without subsequent rescattering. For this reason, this curve rises only at the beginning and remains constant after particlization. This comparison allows us to directly assess the impact of the afterburner on the multiplicities, given that the numbers of particles change significantly. Looking at the left plot, one can see that the dynamic production of deuterons leads to a higher number of nuclei compared with the hydrodynamic curve for all energies. Since the full afterburner calculation describes the data points better, it becomes evident that hadronic rescattering is necessary in this case. For the tritons in the middle plot, one can observe similar behaviour. Only at $\sqrt{s}_{\text{NN}}=19.6$ GeV does the afterburner multiplicity reach the multiplicity on the Cooper-Frye hypersurface. The data points are well reproduced at the higher energies but show some deviation at 7.7 GeV, which is the system with the largest baryon density. The hypertriton multiplicities in the right plot also exceed the hydrodynamic curves. At later times, the multiplicities equilibrate to the hydrodynamic curves within the error bars.

\subsection{Cluster production using PHQMD}
\label{subsec:kinetic_production_phqmd}

\subsubsection{PHQMD}

Because clusters are correlations between nucleons, the starting point of a transport theory that studies cluster formation should not be a mean-field approach such as the Boltzmann-Uehling-Uhlenbeck (BUU) approach. It should start from an $N$-body Hamiltonian and study the trajectories of the nucleons under the influence of mutual two- or many-body interactions. Such an approach allows one to follow correlations and fluctuations.

To comply with this requirement, we have developed over the last few years the Parton-Hadron-Quantum-Molecular Dynamics (PHQMD) approach \cite{Aichelin:2019tnk,Glassel:2021rod,Kireyeu:2022qmv,Coci:2023daq,Kireyeu:2024woo}, which now allows us to study all three cluster-production mechanisms within the same transport framework and to identify how the different propositions for cluster formation are reflected in cluster observables. PHQMD is a microscopic $N$-body transport model that combines the characteristics of baryon propagation from the Quantum Molecular Dynamics (QMD) model~\cite{Aichelin:1991xy,Aichelin:1987ti,Aichelin:1988me,Hartnack:1997ez} with the dynamical properties and interactions, in and out of equilibrium, of hadronic and partonic degrees of freedom from the Parton-Hadron-String-Dynamics (PHSD) approach~\cite{Cassing:2008sv,Cassing:2008nn,Cassing:2009vt,Bratkovskaya:2011wp,Linnyk:2015rco}. We refer to these papers for details and discuss here only the essential ingredients necessary to understand the differences from the other approaches used to study cluster production.

If one does not apply a Boltzmann equation, which describes the time evolution of the one-body Wigner density, but wants to study the time evolution of an $n$-body quantum system, the first question to address is how one can describe the time evolution of such a system. The Dirac-Frenkel-McLachlan approach \cite{raab:2000,broeck:1988}, extensively used in quantum chemistry, provides the framework for such a time evolution. In this approach, which can be considered an extension of the Ritz variational principle to time-dependent problems, a test wave function with time-dependent parameters has to be chosen. The variational principle
\be
\delta \int_{t_1}^{t_2} dt <\psi(t)|i\frac{d}{dt}-H|\psi(t)> = 0
\label{eq:DF}
\ee
then provides the time evolution of the parameters of the wave function. Equation~\ref{eq:DF} requires two ingredients: the $n$-body wave function $\psi(t)$ and the $n$-body Hamiltonian $H$. This approach has also been applied in nuclear physics to other QMD-like models \cite{Feldmeier:1989st,Aichelin:1991xy,Ono:1992uy,Hartnack:1997ez}.

This approach provides the time derivatives of the centers of the wave functions, ${\bf r}_{i0}$ and ${\bf p}_{i0}$. Being an $n$-body approach, it conserves the correlations in the system and does not suppress fluctuations, as mean-field calculations do.

In the PHQMD approach, we follow the ansatz of the IQMD approach and choose Gaussian wave functions in coordinate space for the nucleons, centered around ${\bf r}_{i0}$, which, after a Wigner transformation, yield the Wigner density
\begin{equation} \label{fdefinition}
f ({\bf r}_i, {\bf p}_i,{\bf r}_{i0},{\bf p}_{i0},t)
= \frac{1}{\pi^3 \hbar^3 }
{\rm e}^{-\frac{2}{L} ({\bf r}_i - {\bf r}_{i0} (t) )^2 }
{\rm e}^{-\frac{L}{2\hbar^2} ({\bf p}_i - {\bf p}_{i0} (t) )^2}.
\end{equation}
The Gaussian width $L$ is taken to be $L=8.66$~fm$^2$. We use the $\hbar = c = 1$ convention in the following. The $n$-body Wigner density is then the product of the one-body Wigner densities of all nucleons:
\begin{equation}
f ({\bf r}_1, {\bf p}_1,{\bf r}_{10},{\bf p}_{10},....,t)=
\prod_{i=1}^n f ({\bf r}_i, {\bf p}_i,{\bf r}_{i0},{\bf p}_{i0},t).
\end{equation}
The next step is to fix the $n$-body Hamiltonian $H$. In PHQMD, it is the sum of the Hamiltonians of the nucleons, composed of kinetic and two-body potential energies:
\begin{equation}
H = \sum_i H_i  = \sum_i  (T_i+\sum_{j\neq i}  V_{i,j}).
\end{equation}
The interaction between the nucleons has two parts: a local Skyrme-type interaction and a Coulomb interaction,

\begin{eqnarray}
&&\phantom{a}\hspace*{-5mm}
V_{i,j}= V({\bf r}_i, {\bf r}_j,{\bf r}_{i0},{\bf r}_{j0},t)
= V_{\rm Skyrme}+ V_{\rm Coul}  \label{EP} \\
&& \phantom{a}\hspace*{-5mm}
=\frac{1}{2} t_1 \delta ({\bf r}_i - {\bf r}_j)
+ \frac{1}{\gamma+1}t_2 \delta ({\bf r}_i - {\bf r}_j)
\, \rho^{\gamma-1}({\bf r}_i,{\bf r}_j,
{\bf r}_{i0},{\bf r}_{j0},t) \nonumber \\
&&\phantom{a}\hspace*{-5mm}
+\frac{1}{2}\frac{Z_i Z_j e^2}{|{\bf r}_i-{\bf r}_j|} ,
\nonumber
\end{eqnarray}

with the density
$\rho({\bf r}_i,{\bf r}_j,{\bf r}_{i0},{\bf r}_{j0},t)$
defined as
\begin{eqnarray}
&&\rho({\bf r}_i,{\bf r}_j,{\bf r}_{i0},{\bf r}_{j0},t)
=\nonumber \\
&=&C\frac{1}{2}\Big[
\sum_{j,i\neq j}
\Big(\frac{1}{\pi L}\Big)^{3/2}
e^{-\frac{1}{L}({\bf r}_i-{\bf r}_j-{\bf r}_{i0}(t)+{\bf r}_{j0}(t))^2}
\nonumber \\
&+&
\sum_{i,i\neq j}
\Big(\frac{1}{\pi L}\Big)^{3/2}
e^{-\frac{1}{L}({\bf r}_i-{\bf r}_j-{\bf r}_{i0}(t)+{\bf r}_{j0}(t))^2}
\Big].
\label{dens1}
\end{eqnarray}
where $C$ is a correction factor discussed in Ref.~\cite{Aichelin:2019tnk}.

We now have to calculate the expectation value of the Hamiltonian, which enters Eq.~\ref{eq:DF}. It is finally given by

\begin{eqnarray}
\langle H \rangle &=& \langle T \rangle + \langle V \rangle
\label{ham} \\
&=& \sum_i \big(\sqrt{p_{i0}^2+m^2}-m\big) \nonumber \\
&+& \sum_{i} \langle V_{\rm Skyrme}({\bf r}_{i0},t)
+V_{\rm Coul}({\bf r}_{i0},t)\rangle .
\nonumber
\end{eqnarray}

For the Skyrme potential, we use the approximate form (for a discussion, we refer to Ref.~\cite{Aichelin:2019tnk})

\begin{equation} \label{eosinf} \langle V_{\rm Skyrme}({\bf r_{i0}},t) \rangle ,=, \alpha
\left(\frac{\rho_{int} ({\bf r}_{i0},t)}{\rho_0}\right) +
\beta \left(\frac{\rho_{\rm int} ({\bf r}_{i0},t)}{\rho_0}\right)^{\gamma}.
\end{equation}

In the above, $\rho_{\rm int}({\bf r_{i0}})$ is the sum of the Gaussian single-particle densities of all baryons $j\ne i$ at ${\bf r_{i0}}$, but with twice the width $(L\to 2L)$. The expectation value of the Coulomb interaction can also be calculated analytically.

With our assumption that the wave functions have a Gaussian form and that their width is time independent, one obtains two equations for the time evolution of the centroids of the Gaussian single-particle Wigner densities. These equations resemble the equations of motion of a classical particle with phase-space coordinates ${\bf r}_{i0}$ and ${\bf p}_{i0}$ \cite{Aichelin:1991xy}.

The time-evolution equations
\begin{equation}
\dot{r_{i0}}=\frac{\partial\langle H \rangle}{\partial p_{i0}} \qquad
\dot{p_{i0}}=-\frac{\partial \langle H \rangle}{\partial r_{i0}} \quad .
\label{prop}
\end{equation}

look rather similar to those obtained in the BUU approach, but they are actually quite different. Whereas in the BUU approach the potential in $H$ is a local mean field, in QMD, $H$ contains all interactions among all nucleons.

In infinite matter, the parameters $\alpha(t_1,t_2)$, $\beta(t_1,t_2)$, and $\gamma$ can be related to the nuclear equation of state (EOS). Two of the three parameters are determined by the condition that the energy per nucleon has a minimum of $\frac{E}{A}=-16$~MeV at $\rho_0$, the ground-state density of nuclear matter. The third one is usually expressed in terms of the compression modulus $K$ of nuclear matter, the inverse of the compressibility $\chi = \frac{1}{V}\frac{d V}{d P}$, where $P$ is the pressure in the system of volume $V$. Here, we employ $K = 380$~MeV, which is usually called a hard EOS.

\subsubsection{Cluster formation by potential}
The very same nucleon-nucleon potential that governs the time evolution of the whole heavy-ion reaction (with the exception of a possible QGP phase) produces clusters at the end of the expansion of the system. This reflects the $n$-body character of our approach and is absent in mean-field calculations, where the potential is not strong enough to bind together the test particles, which represent a fraction of a nucleon, when the system expands.

We call this {\it dynamical} cluster formation because the formation of clusters does not occur at a given time but is a continuous process due to the interactions among the baryons. In comparison with other approaches, cluster formation by potential does not require the use of an additional model, such as a coalescence model or a statistical model, on top of the baryon dynamics to determine clusters.

As explained in Ref.~\cite{Aichelin:2019tnk}, cluster identification or recognition is realized in PHQMD using the Minimum Spanning Tree (MST) procedure \cite{Aichelin:1991xy}. In the MST algorithm, only coordinate-space information and the binding energy are used to define clusters: two nucleons are considered to be part of a cluster if their distance is less than $r_0 = 4$~fm in the cluster rest system obtained by a Lorentz transformation from the computational frame. Nucleons that are more distant no longer experience an attractive interaction. Nucleons with a large relative momentum do not remain together for a long time. Consequently, additional cuts in momentum space change the cluster distribution only slightly \cite{Kireyeu:2021igi}. In addition, clusters must have a finite binding energy.

\begin{figure*}[ht]
\centering
\includegraphics[width=\textwidth]{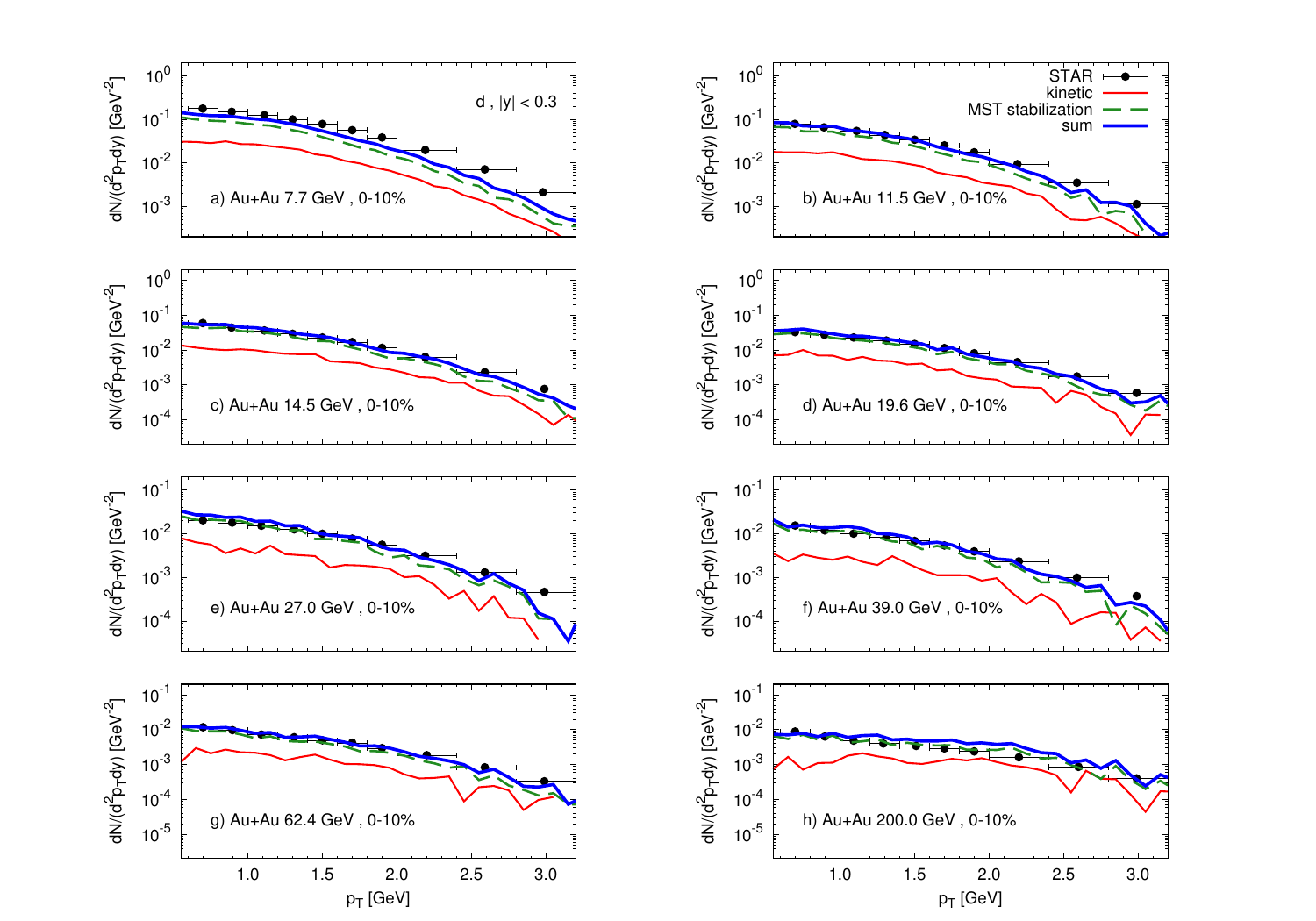}
\caption{\label{fig:BESpTd} (color online) Transverse momentum distributions of deuterons in central Au+Au collisions at all RHIC BES energies, from $\sqrt{s_{NN}}=7.7$ GeV (a) to $200$ GeV (h). The points are the experimental data from the STAR Collaboration~\cite{STAR:2019sjh}. The lines correspond to the PHQMD results for the different production mechanisms: kinetic with finite size (solid red), potential from advanced MST identification (dashed green), and the sum of all contributions (thick solid blue). The PHQMD $p_T$ spectra are shown for the same midrapidity interval, $|y|<0.3$, as the STAR measurements.}
\end{figure*}

PHQMD has been used to study heavy-ion reactions from $E_{\rm kin} = 1.5$ AGeV up to $\sqrt{s} = 200$ GeV. The results, which show good agreement with experiment for many observables, can be found in Refs.~\cite{Aichelin:2019tnk,Glassel:2021rod,Kireyeu:2022qmv,Coci:2023daq,Kireyeu:2024woo}. Here we show just one example in Fig.~\ref{fig:BESpTd}, namely the beam-energy dependence of the deuteron $p_T$ spectra measured in the RHIC Beam Energy Scan.

Instead, we concentrate here on the results that are important for the question of how clusters are produced in heavy-ion reactions.

\subsubsection{Clusters are produced from very low to very high beam energies}

\begin{figure}[ht!]
\centering
\includegraphics[width=0.8\textwidth]{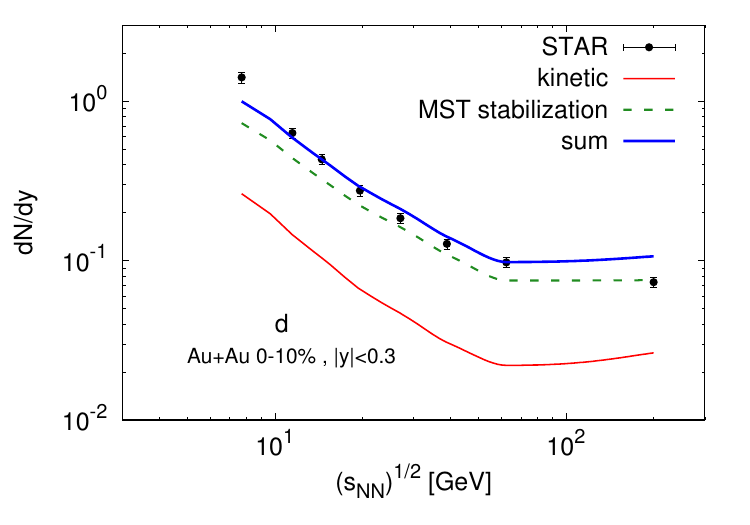}
\caption{\label{fig:BESexcid} (color online) The midrapidity ($|y|<0.3$) excitation function of $dN/dy$ for deuterons as a function of $\sqrt{s_{NN}}$ for $0--10\%$ central Au+Au collisions, compared with experimental data from the STAR Collaboration~\cite{STAR:2019sjh}. The different lines indicate the different contributions to deuteron production: kinetic production with modelling of finite-size effects in coordinate and momentum space (solid red), potential from MST with stabilization, i.e., advanced MST (dashed green), and their sum (blue). The rapidity interval of the PHQMD results is the same as that measured by STAR.}
\end{figure}

The first observation is that cluster production at midrapidity is a very common process in heavy ion collisions, from beam energies around $E_{\rm kin}=1.5$ AGeV up to $\sqrt{s}=200$ GeV and even beyond, where PHQMD calculations are presently too time consuming to obtain spectra. The cluster multiplicities, their transverse momentum spectra (see Fig.~\ref{fig:BESpTd}), and the ratios of clusters and anticlusters all evolve very smoothly with beam energy. In Fig.~\ref{fig:BESexcid} we show, as an example, the excitation function of the deuteron multiplicity, although the same behavior is observed for many other observables. Such s smooth collision energy dependence is surprising because the environment of the clusters at midrapidity changes dramatically with the collision energy.

At $E_{\rm kin}=1.5$ GeV, the environment is completely baryonic; at higher energies, mesons dominate; and at still higher energies, a QGP is formed. In view of this observation, an exclusive study of cluster production at the highest energies is not a useful approach. One should exploit the already available data over the entire beam-energy range in which clusters are produced at midrapidity.

\subsubsection{Transport approaches show a wide freeze out time distribution}

\begin{figure}[ht!]
\centering
\includegraphics[scale=0.49]{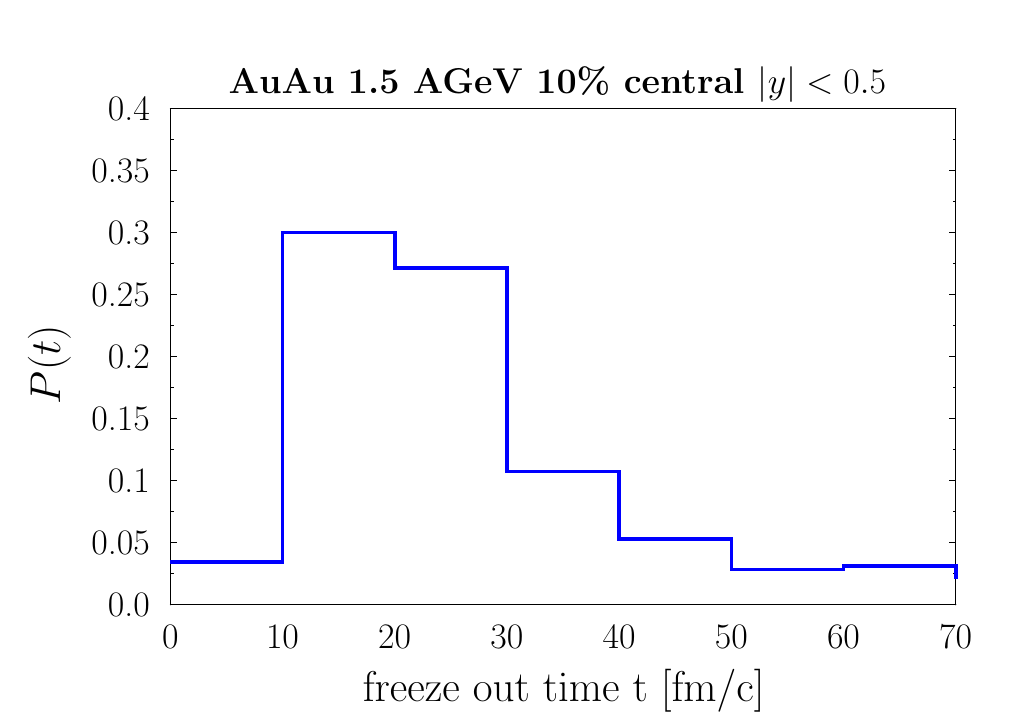}
\includegraphics[scale=0.49]{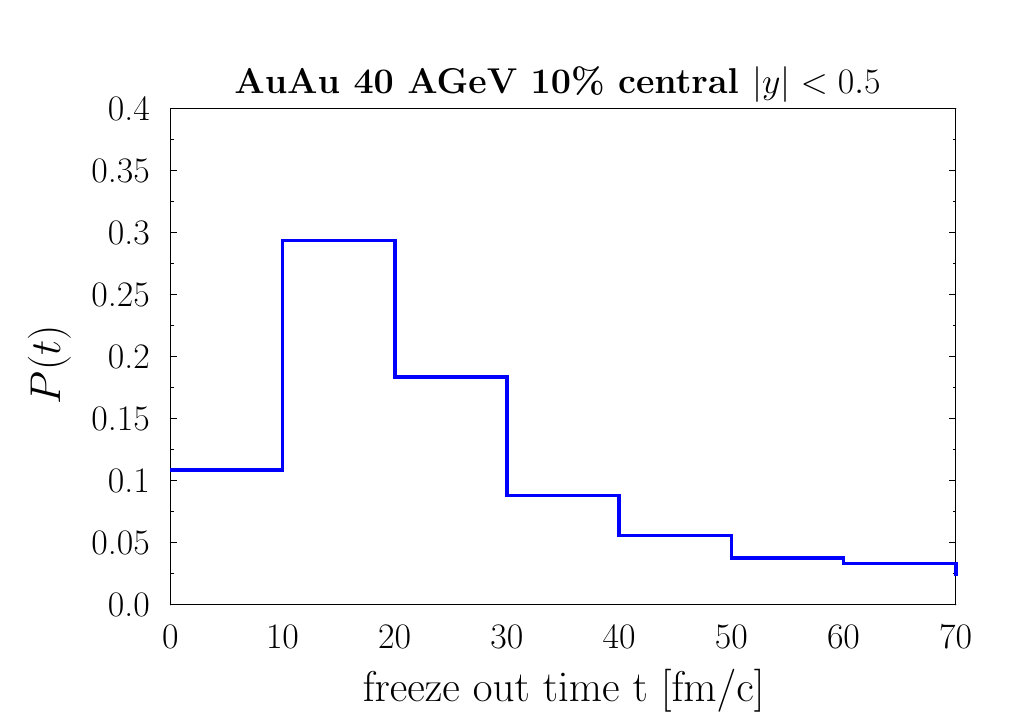}
\caption{\label{fig:frz} Normalized distribution of the freeze-out times of baryons (nucleons and hyperons), i.e., the times of the last hadronic collisions of the baryons observed at midrapidity, $|y|<0.5$, for 10\% central Au+Au collisions at beam energies of $E_{\rm kin} = 1.5$~$A$GeV (left) and $E_{\rm kin} = 40$~$A$GeV (right).}
\end{figure}

Transport approaches, which describe not only cluster spectra but also many single-particle observables, show a very broad distribution of freeze-out times, i.e., the times at which the baryons undergo their last collisional interaction. This is shown in Fig.~\ref{fig:frz}, which displays this distribution for two beam energies and for particles emitted at midrapidity. Until freeze-out, many collisions among the hadrons take place, and this raises the question of how clusters at midrapidity, with binding energies of around or less than 2 MeV and root-mean-square radii of 2 fm or more, can survive the expansion of the system. Collisions with other particles should destroy them on a very short timescale. Regeneration during the expansion would change their multiplicity, reflecting the local temperature rather than that at chemical freeze-out.

\subsubsection{The surviving clusters have a different expansion dynamics than free nucleons}

\begin{figure}[ht!]
\centering
\includegraphics[scale=0.32]{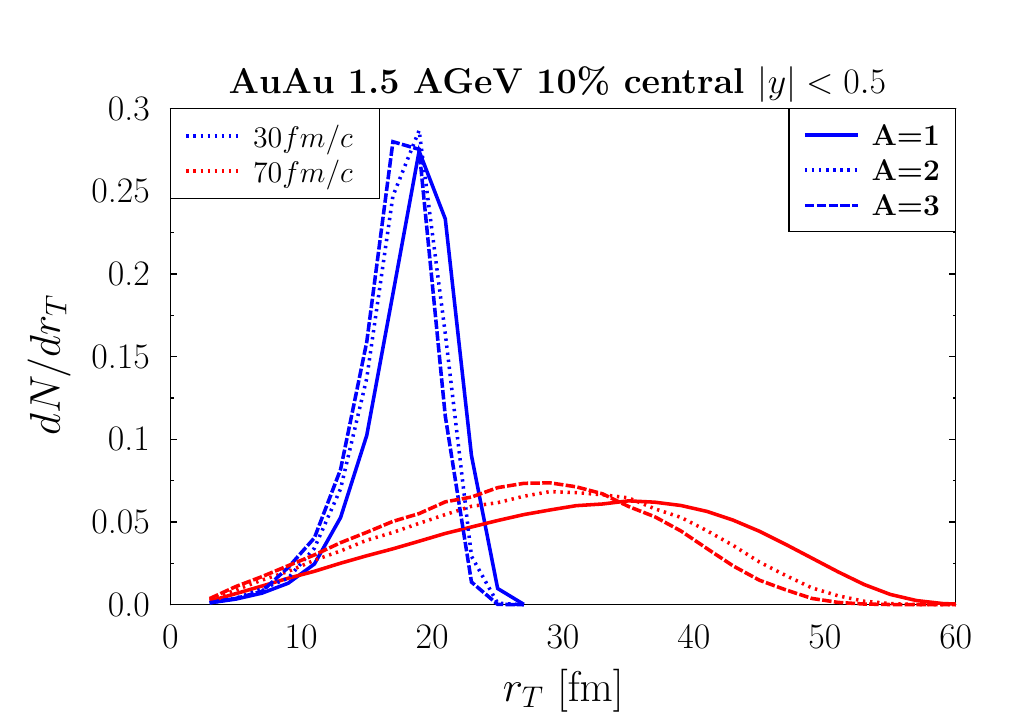}
\includegraphics[scale=0.32]{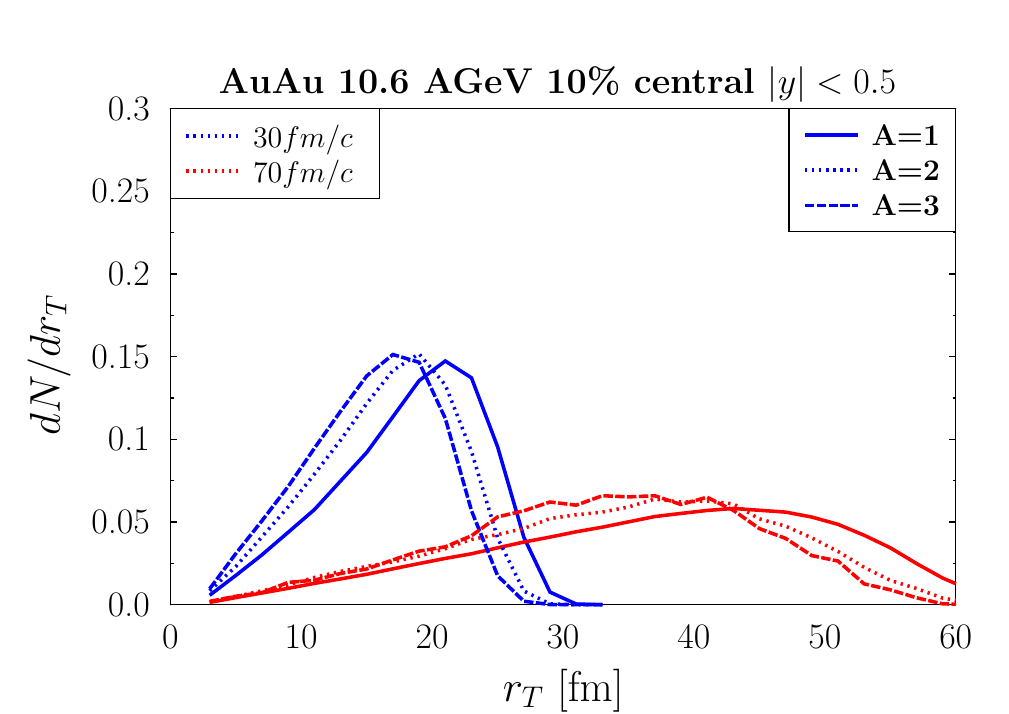}
\includegraphics[scale=0.32]{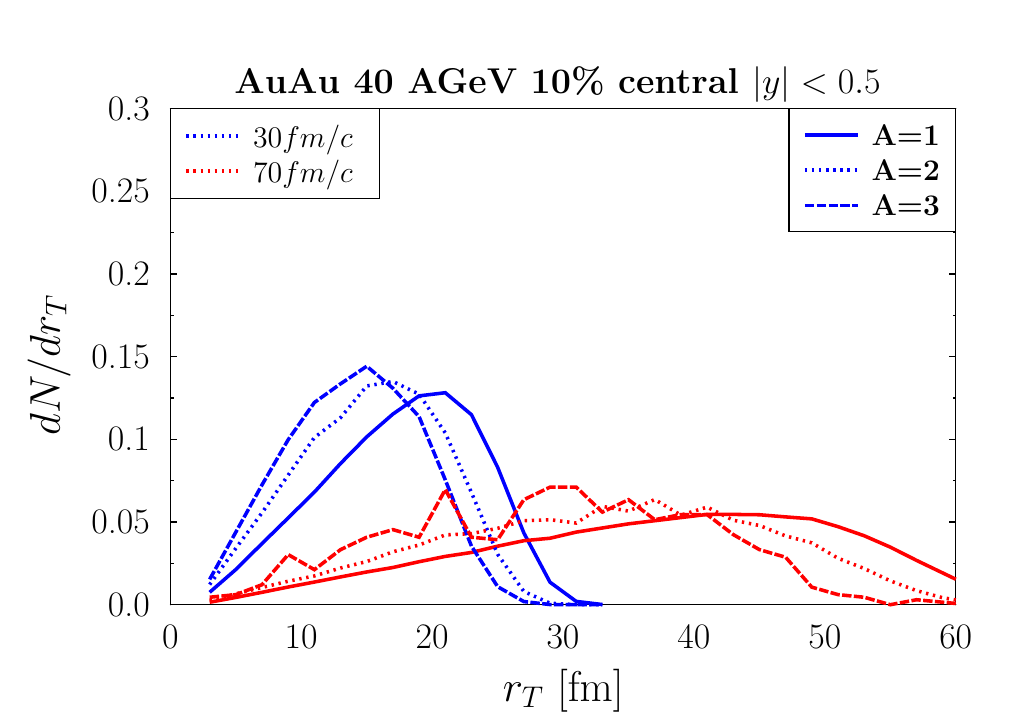}
\caption{\label{fig:RT} Normalized distribution of the transverse distance of nucleons observed at midrapidity ($|y|<0.5$) from the reaction center at times of 30~fm/$c$ (blue lines) and 70~fm/$c$ (red lines) for 10\% central Au+Au collisions at beam energies of $E_{\rm kin} = 1.5$~$A$GeV (left), $E_{\rm kin} = 10$~$A$GeV (middle), and $E_{\rm kin} = 40$~$A$GeV (right). The dotted lines show the distributions of nucleons belonging to clusters of size $A=2$, the dashed lines those belonging to $A=3$ clusters, and the solid lines the distributions of free nucleons ($A=1$) \cite{Glassel:2021rod}.}
\end{figure}

One step towards understanding how clusters can survive the expanding fireball is the investigation of their spatial distributions. Such an investigation is possible in transport approaches, where the positions of all free and bound nucleons are known. We study these distributions in Fig.~\ref{fig:RT}, where we display the distributions of the transverse radius for nucleons that are free or bound in $A=2$ and $A=3$ clusters. This is shown at two different times, 30~fm/$c$ (blue lines) and 70~fm/$c$ (red lines). We see that, at these times, the clusters are closer to the reaction center than the free nucleons, with $A=3$ clusters being even closer than $A=2$ clusters. This means that, during the expansion, the clusters are not at the same location as the free nucleons but trail behind them. This is the reason why they can survive.

Whether clusters with the same velocity as the free nucleons can survive or not is beyond the scope of our analysis. This result is not limited to clusters produced by the potential interactions among the nucleons. The same behaviour is observed for clusters formed by coalescence \cite{Kireyeu:2022qmv}.

\subsubsection{Can the mechanism for deuteron production be identified?}

\begin{figure}[t]
\centering
\includegraphics[width=0.5\textwidth]{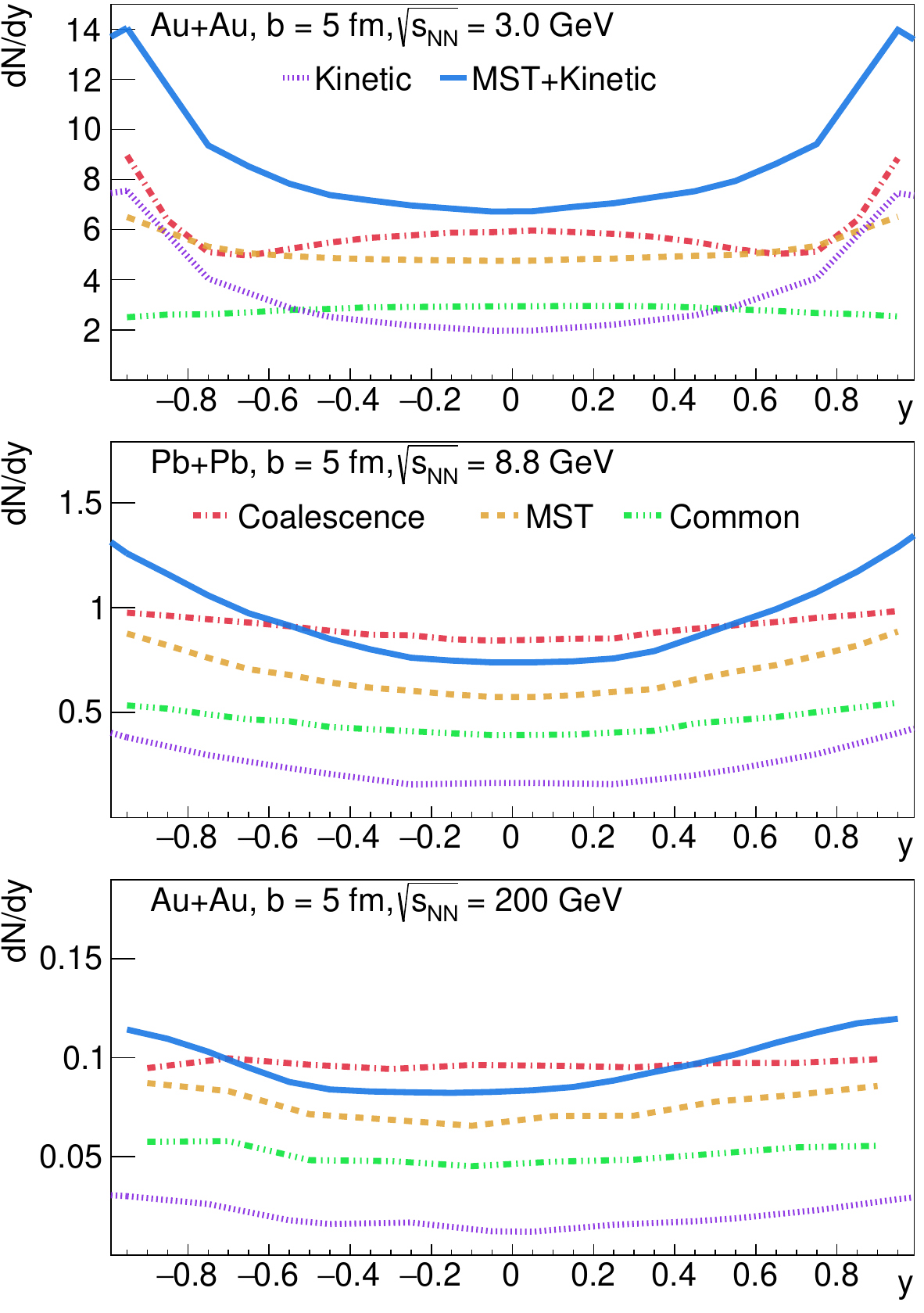}
\caption{Rapidity distributions of deuterons in central collisions at center-of-mass energies of $\sqrt{s_{NN}}=3$ GeV (top), $\sqrt{s_{NN}}=8.8$ GeV (middle), and $\sqrt{s_{NN}}=200$ GeV (bottom). The orange dashed lines show the results of the aMST with $E_B<0$ for `potential' deuterons, the dash-dotted red lines those for coalescence deuterons, and the dotted purple lines those for `kinetic' deuterons produced by collisions. The solid blue lines show the sum of aMST and kinetic deuterons. The green dash-triple-dotted lines indicate the rapidity distributions of those `potential' deuterons, identified by the aMST, which are at the same time `coalescence' deuterons without applying the factor 3/8 \cite{Kireyeu:2024woo}.}
\label{fig:dndy}
\end{figure}

  The fact that the three mechanisms for deuteron production, which were discussed at the beginning of this section, yield reasonable results when compared with experimental data is, of course, not very satisfactory. This motivates the search for observables that may be able to discriminate among them. It turns out that the rapidity distribution of deuterons at energies around $\sqrt{s}=3$ GeV is rather different in the coalescence model compared with the potential model \cite{Kireyeu:2024woo}, as shown in Fig.~\ref{fig:dndy}. If we compare the dashed orange and dash-dotted red lines, we clearly see a difference in shape, which is large enough to potentially allow experimental discrimination.

Experiments have been performed at the very same energies, but the $dN/dy$ distribution has not been measured precisely enough or, at $\sqrt{s}=3$ GeV, the problem is the lower $p_T$ cut of about 1.2 GeV for deuterons. A detailed investigation shows that the difference between the two models is most pronounced at low $p_T$ values, where neither model agrees with the blast-wave fit that has been applied to the experimental data to extrapolate to lower $p_T$ values in order to obtain the $dN/dy$ distribution. Thus, at present, no definitive conclusion can be obtained, but the potential for experimental discrimination still exists~\cite{Kireyeu:2024woo}.

\subsubsection{Hypernuclei are produced in the same way as non strange clusters}

\begin{figure}
\includegraphics[width=0.33\textwidth]{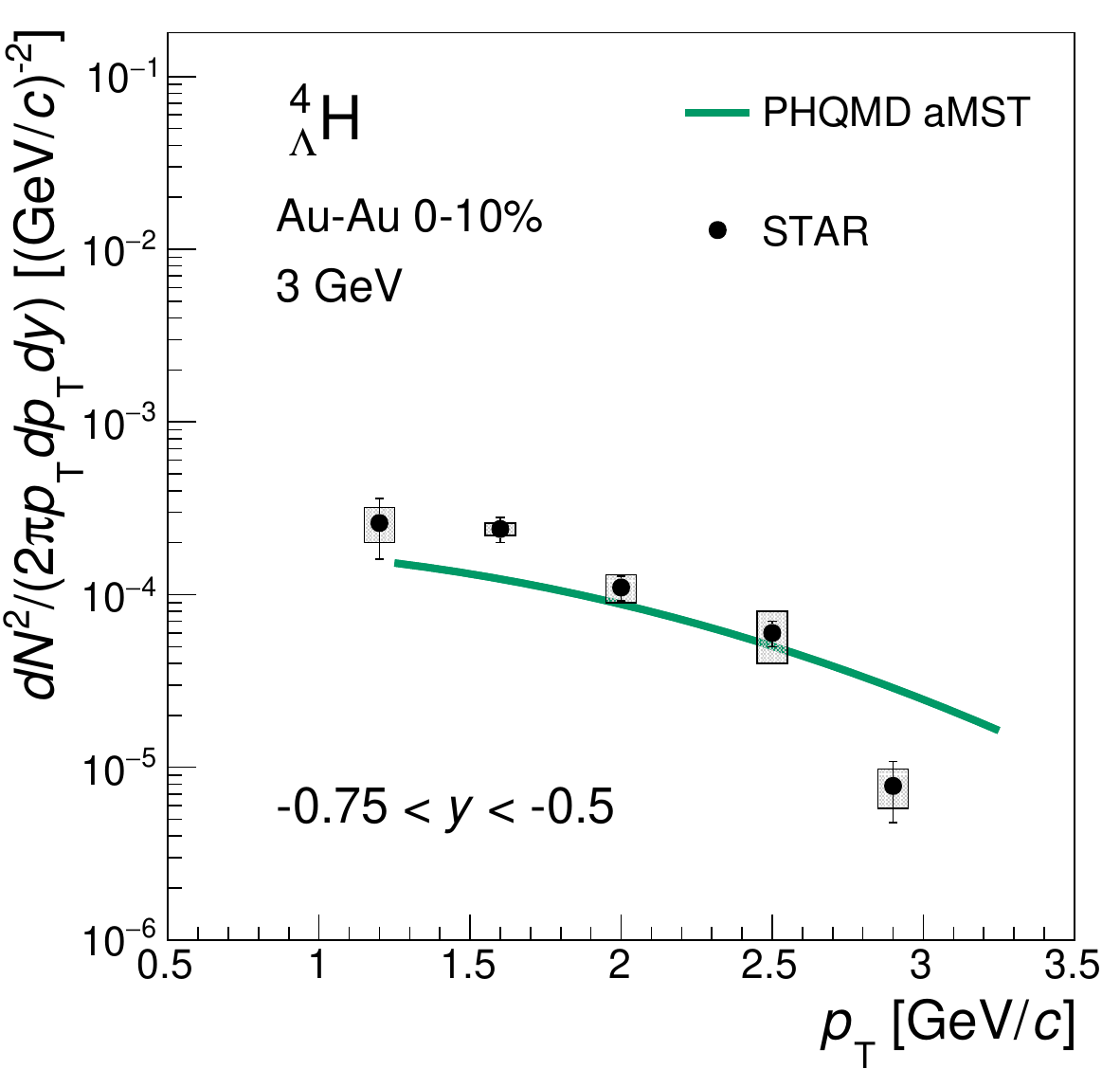}
\includegraphics[width=0.33\textwidth]{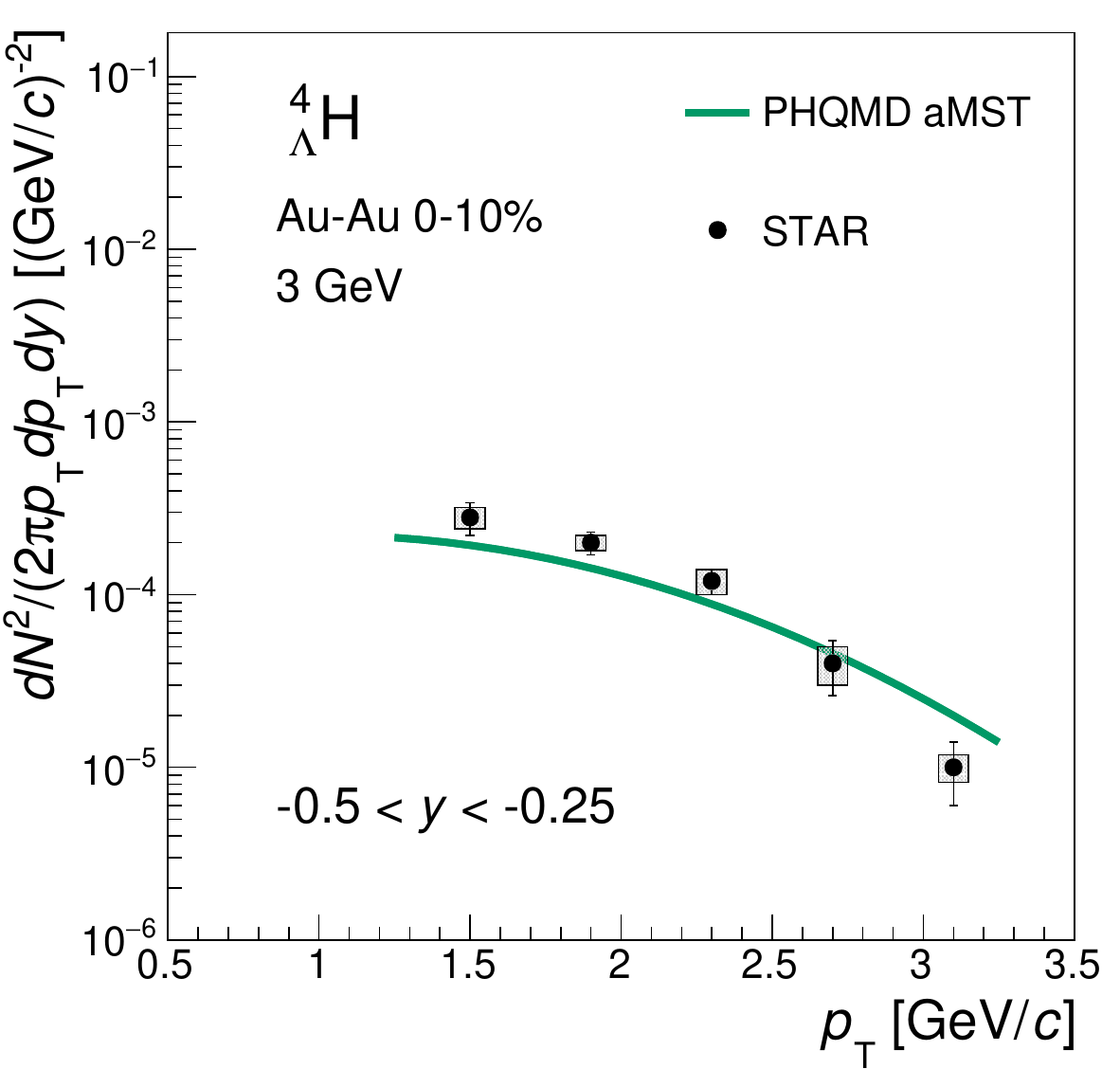}
\includegraphics[width=0.33\textwidth]{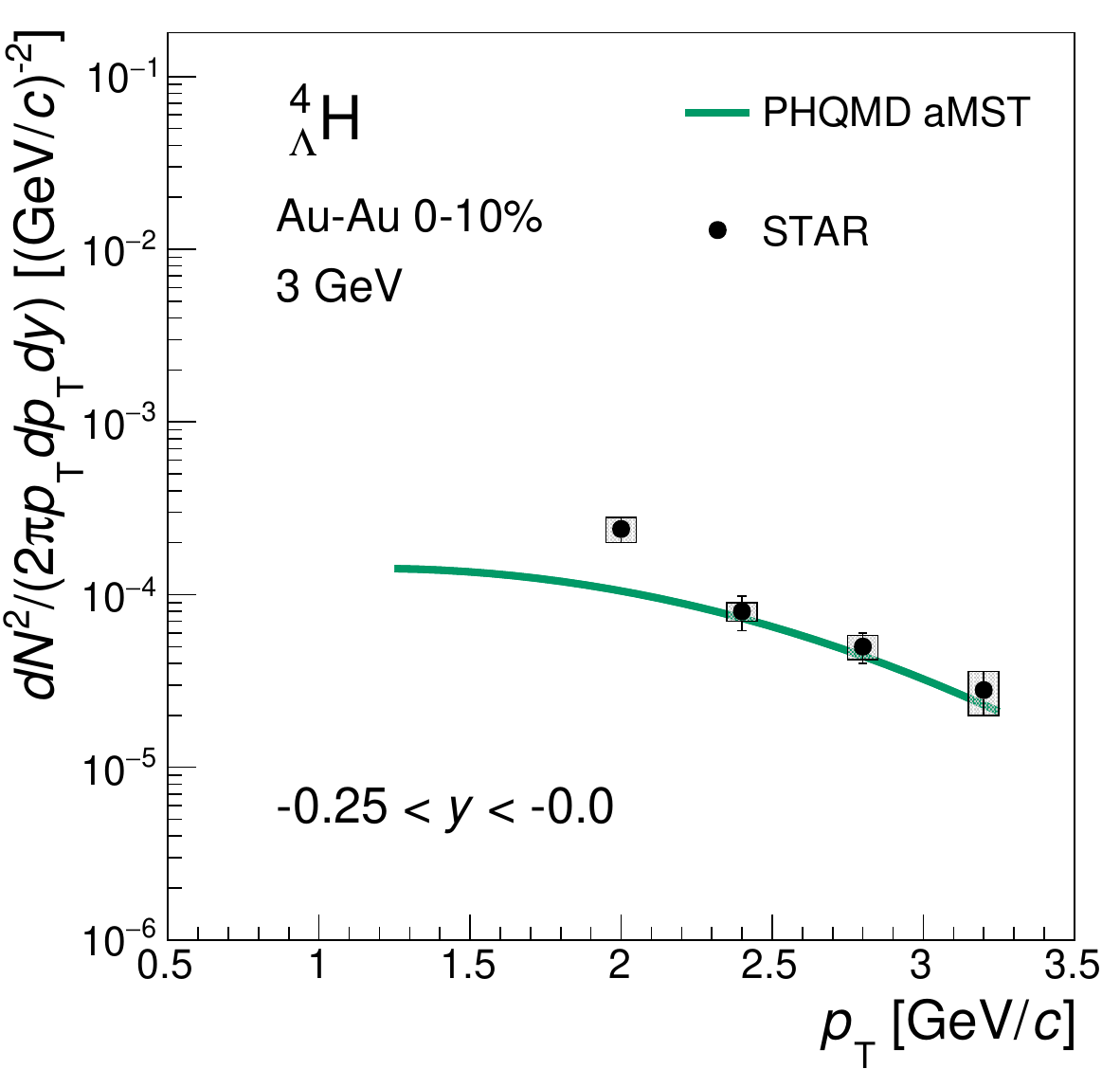}
\caption{\label{fig:H4L_pT_3}
Transverse momentum distributions of $^{4}_{\Lambda}$H for different rapidity intervals, as indicated in the legends, in central Au+Au collisions at $\sqrt{s_{NN}}=3$~GeV.
The filled circles indicate the preliminary experimental data from the STAR Collaboration \cite{Abdallah:2021ab}. The PHQMD results are represented by the green lines.}
\end{figure}

In PHQMD, hyperons interact through a potential that has 2/3 of the strength of the nucleon-nucleon potential. When compared with the Beam Energy Scan data, as shown in Fig.~\ref{fig:H4L_pT_3}, we obtain, as for non-strange clusters, a good description of the hypernucleus $p_T$ spectra. There is, of course, still room for improvement, but it seems that hypernucleus production is not fundamentally different from the production of non-strange clusters.

\section{Key questions emerging from the RRTF}
\label{sec:rrtf_discussion}

The discussions during the RRTF highlighted several conceptual and experimental questions that are central to developing a consistent description of light-nuclei and hypernuclei production. Particular attention was devoted to the definition of the source relevant for coalescence, the formation time of loosely bound states, the connection between femtoscopic correlations and bound-state formation, and the identification of observables sufficiently sensitive to discriminate among different production mechanisms. The main questions and insights emerging from these discussions are summarized below.

\subsection{Source size and formation of loosely bound states}
\label{ss_looselybs}

A central question concerns the source size that should be used in coalescence calculations. Source radii entering these calculations are commonly constrained using femtoscopic measurements. However, femtoscopy probes the homogeneity region of the emitting source, and the extracted radius depends on the particle species, transverse mass, collision system, and collision energy. A systematic determination of proton--proton correlation radii as a function of $m_{\mathrm{T}}$ across collision systems and energies would therefore provide particularly valuable input for coalescence calculations. In this context, proton--proton rather than pion--pion radii are expected to provide a more direct characterization of the nucleon source relevant for nuclear formation.

The source size relevant for coalescence in jets also remains to be established. This is particularly important for interpreting measurements of light nuclei in jet-associated regions, where the spatial characteristics of the nucleon-emitting source may differ substantially from those of the underlying event or minimum-bias collisions. A direct experimental determination of the source size in jets, for example through femtoscopic correlations of jet-associated nucleons, would therefore be highly valuable. Such a measurement is, however, experimentally challenging because of the limited pair statistics, the definition and reconstruction of the jet-associated source, possible contamination from the underlying event, and non-femtoscopic correlations associated with jet fragmentation and momentum conservation. These effects need to be carefully controlled before a femtoscopic radius can be interpreted as the source size relevant for coalescence in jets.

The appropriate source size may also depend on the nuclear state under consideration. This issue becomes particularly important for the hypertriton, whose exceptionally small $\Lambda$ separation energy implies a spatial extent considerably larger than that of ordinary light nuclei. It is therefore not evident that the same effective source description used for deuterons or $^3$He can be directly applied to hypertriton formation. More generally, the interplay between the size of the emitting source and the spatial structure of the final bound state is one of the key ingredients controlling the coalescence probability.

Closely related to the question of the source size is the meaning of the ``formation time'' of a loosely bound state. Different notions of formation time may enter the discussion and should be clearly distinguished. One may refer to the time at which the constituents become sufficiently close in phase space to form a bound-state candidate, or to the time required for the corresponding wave function to evolve into the asymptotic configuration of the observed nucleus. For weakly bound systems, these time scales need not coincide. One picture discussed during the RRTF is that a compact state carrying the mass and quantum numbers of the final nucleus may initially be produced and subsequently evolve into the spatially extended wave function of the physical bound state. A more explicit theoretical description of this evolution, the associated time scales, and their relation to the binding energy and spatial structure of the nucleus would be highly desirable.

The sensitivity of coalescence predictions to the assumed nuclear wave function provides an additional aspect of this problem. Predictions obtained using realistic wave functions based on nucleon--nucleon interactions, such as Argonne~v18 or chiral effective field theory, can differ from those based on simplified Gaussian wave functions. This indicates that coalescence observables may be sensitive not only to the characteristic size and binding energy of the nucleus, but also to the detailed shape of its wave function. Once the size and binding energy of a state are experimentally constrained, such sensitivity could potentially provide access to additional information on its internal structure.

\subsection{Few-body correlations and their connection to coalescence}

The relation between femtoscopic correlations and the formation of loosely bound states was another major topic of discussion. Proton--deuteron correlations provide a particularly instructive example. Treating the deuteron as a point-like particle in a two-body $p$--d calculation neglects effects associated with its composite nature, including Pauli blocking and short-range interactions involving its constituent nucleons. A three-body treatment may therefore be required for a consistent description of the correlation function, particularly in small collision systems, where the sensitivity to short-distance dynamics is enhanced.

In a three-body description, the asymptotic $p$--d final state emerges dynamically from the underlying three-nucleon system. This picture is closely connected to the notion of formation time discussed above: the individual nucleons may already be well defined while the asymptotic deuteron wave function has not yet fully developed. Comparisons between calculations based on realistic nucleon--nucleon interactions and those employing simplified deuteron wave functions may therefore provide information on the role of short-range dynamics and the internal structure of the deuteron in the measured correlation function.

The situation may be different for systems such as kaon--deuteron, where the external kaon is distinguishable from the nucleons forming the deuteron. In this case, the formation of the deuteron may approximately factorize from its interaction with the kaon, potentially making a two-body treatment more appropriate. Measurements of $K$--d correlations nevertheless raised additional questions during the RRTF. In particular, source radii extracted in Pb--Pb collisions appeared unexpectedly large. Possible contributions from strong decays producing kaons, correlations among the fitted scattering parameters, and underestimated uncertainties were discussed as potential explanations. These observations underline the importance of establishing a consistent treatment of composite particles and their sources in femtoscopic analyses.

Similar considerations apply to $\Lambda$--d correlations, where the presence of the hypertriton bound state provides an additional connection between few-body scattering, femtoscopic correlations, and coalescence. A systematic comparison of two- and three-body approaches, together with measurements across different collision systems and source sizes, would help clarify how the internal structure and formation dynamics of loosely bound states manifest themselves in correlation observables.

\subsection{Model discrimination and sensitive observables}

A recurring conclusion of the RRTF discussions was that reproducing integrated nuclear yields alone is generally insufficient to identify the underlying production mechanism. Statistical hadronization and coalescence approaches can both describe several existing measurements, despite being based on substantially different physical pictures. Identifying observables for which these approaches lead to quantitatively distinct predictions is therefore essential for discriminating between different production scenarios.

One example is the yield ratio

\begin{equation}
\frac{N_{\mathrm{t}}N_{\mathrm{p}}}{N_{\mathrm{d}}^2},
\end{equation}

which reduces part of the dependence on bulk thermodynamic quantities and has been proposed as a probe of density fluctuations and of the nuclear production mechanism. Both thermal and coalescence calculations predict a relatively weak multiplicity dependence in sufficiently large systems, while their predicted limiting values differ by approximately a factor of two. 
For smaller system sizes, coalescence calculations additionally predict a characteristic dependence arising from the interplay between the nucleon source size and the nuclear wave function. More precise measurements of this ratio over a broad multiplicity range would therefore provide a sensitive test of the different production scenarios.

The collision-energy dependence of nuclear yields provides another useful benchmark for model discrimination. During the RRTF, comparisons of triton production over a broad range of $\sqrt{s_{\mathrm{NN}}}$ revealed a systematic tension between the measured yields and some model calculations, with predictions exceeding the data by approximately a factor of two. Extending such comparisons in a consistent manner to deuterons, $^3$He, $^4$He, hypertritons, ${}^{4}_{\Lambda}$H, and ${}^{4}_{\Lambda}$He was identified as an important next step. Such a systematic study would help establish whether the observed discrepancies are specific to individual species or point to more general ingredients missing from the theoretical descriptions.

Canonical conservation effects constitute an additional source of model sensitivity in small systems. In particular, the multiplicity dependence of the $S_3$ ratio can be affected differently by the exact conservation of baryon number and strangeness. The discussions highlighted the importance of treating these conservation laws consistently when comparing statistical-model calculations, particularly at low multiplicity. More generally, comparisons among different implementations of the statistical model should clearly specify the treatment of canonical conservation, excluded-volume effects, hadronic interactions, and possible deviations from chemical equilibrium.

The role of the hadronic phase was also emphasized. Annihilation, regeneration, and rescattering can modify nuclear yields and momentum distributions after chemical freeze-out. Deviations of measured flow observables from hydrodynamic expectations, particularly outside the most central collisions, may indicate an increasing relevance of non-equilibrium effects. In small systems, nuclear production associated with jets introduces an additional non-equilibrium component that should be taken into account when interpreting inclusive yields and when comparing them with models based on an equilibrated source.

Anisotropic flow measurements provide another potentially sensitive discriminator between production mechanisms. In coalescence, the formation probability of a spatially extended (hyper)nucleus depends on the overlap of its wave function with the homogeneity region of the constituent nucleons. Since the size and shape of this region vary with the azimuthal angle relative to the event plane, this dependence can be reflected in the differential anisotropic flow of the produced (hyper)nuclei. Statistical production does not generally incorporate the spatial extent of the nuclear wave function in the same manner, and the two approaches may therefore lead to different predictions for $v_2$ and higher-order flow coefficients.

The first measurement of hypertriton elliptic flow, together with a more precise and differential measurement of the $^3$He elliptic flow, in Pb--Pb collisions at $\sqrt{s_{\mathrm{NN}}}=5.36$~TeV \cite{ALICE:2026hca} provides an important experimental test of this picture. Despite the much larger spatial extent and significantly smaller binding energy of the hypertriton compared with $^3$He, their measured $v_2$ values are compatible within the current uncertainties.

The measurements were compared with two model calculations based on different assumptions for nuclear production. The first employs a hybrid MUSIC+UrQMD description followed by a coalescence afterburner, while in the second nuclei are sampled directly from the Cooper--Frye hypersurface and subsequently propagated through UrQMD. The two approaches predict similar elliptic flow at low transverse momentum, $p_{\rm T}\lesssim3$~GeV/$c$, but increasingly diverge towards higher $p_{\rm T}$. In this region, the calculation including the coalescence afterburner is consistent with the measurements, whereas the calculation based on nuclei sampled at the hydrodynamic freeze-out underestimates the observed elliptic flow.

The compatibility of the hypertriton and $^3$He elliptic flow within the current uncertainties suggests that the large difference in their spatial sizes does not translate into a correspondingly large difference in their collective flow. At the same time, the increasing sensitivity to the assumed production mechanism at higher $p_{\rm T}$ makes differential flow measurements a promising observable for discriminating among different descriptions of nuclear formation. More precise measurements, including measurements of higher-order flow harmonics, together with quantitative comparisons with different production models, could therefore provide further constraints on the formation dynamics of loosely bound states. For such comparisons to be meaningful, the underlying source should be constrained independently by hadron spectra and flow observables rather than tuned to reproduce the nuclear measurements themselves.

\subsection{Experimental and theoretical priorities}

Several concrete measurements and theoretical developments were identified during the RRTF as particularly valuable for future progress. A systematic measurement of proton--proton femtoscopic source radii as a function of $m_{\mathrm{T}}$, collision system, and collision energy would provide improved experimental constraints for coalescence calculations. Establishing the source size relevant for nuclear production in jets represents another important experimental challenge. Femtoscopic measurements involving jet-associated nucleons could provide valuable information in this direction, although the limited pair statistics, underlying-event contamination, and non-femtoscopic correlations associated with jet fragmentation need to be carefully controlled. Few-body correlation measurements, in particular $p$--d, $d$--$d$, $\Lambda$--d, and $K$--d correlations, should be accompanied by theoretical treatments that consistently account for the composite nature of the deuteron and, where required, the full three-body dynamics.

Reducing experimental uncertainties associated with the interaction of hypernuclei with detector material is another important priority. For the hypertriton, absorption in detector material constitutes a significant source of systematic uncertainty. Dedicated measurements or improved constraints on the corresponding interaction cross sections could therefore substantially improve the precision of hypertriton measurements and may also provide information related to its unusually large spatial extent.

On the theory side, a more precise definition of bound-state formation time and a dynamical description of the evolution of the nuclear wave function would help establish connections among coalescence, transport, and femtoscopic approaches. The role of realistic nuclear wave functions should also be investigated systematically, particularly in small collision systems, where the sensitivity to their detailed structure is expected to be enhanced.

Finally, systematic comparisons of measured yields and yield ratios with different theoretical approaches over the full available range of collision energies and system sizes were identified as an important outcome of the RRTF. Presenting deuteron, triton, $^3$He, $^4$He, hypertriton, ${}^{4}_{\Lambda}$H, and ${}^{4}_{\Lambda}$He measurements within a common framework would provide a stringent test of whether the different production models can simultaneously describe the hierarchy of nuclear states and its evolution with collision energy and system size.

\section{Acknowledgments}
\label{Acknoledgments}

We thank the ExtreMe Matter Institute EMMI at GSI, Darmstadt, for support in the framework of an EMMI Rapid Reaction Task Force meeting during which this work was initiated.
BT thanks the Czech Science Foundation for the support under grant No.~22-25026S.
HZ thanks the National Science Centre, Poland, grants No: 2021/41/B/ST2/02409 and 2020/38/E/ST2/00019. 
KR acknowledges the support by the National Science Centre (NCN), Poland, under OPUS Grant No.~2022/45/B/ST2/01527, and the Polish Ministry of Science and Higher Education.
CP acknowledges support from the French National Agency of Research (ANR), under grant ANR-24-CE31-5571 (project  ROAD\_4\_EIC).

% Bibliography
\bibliographystyle{utphys}
\bibliography{bibliography}
\end{document}